\documentclass[twocolumn,showpacs,amsmath,amssymb,aps,pre,superscriptaddress,longbibliography]{revtex4-1}

\usepackage{graphicx}
\usepackage{dcolumn}
\usepackage{bm}
\usepackage{placeins}
\usepackage[english=american]{csquotes}

\begin{document}

\newcommand{\kb}{k_\text{B}}

\title{Heavy-tailed phase-space distributions beyond Boltzmann-Gibbs and equipartition: Statistics of confined cold atoms}

\author{Andreas Dechant}
\affiliation{Department of Physics and Institute of Nanotechnology and Advanced Materials, Bar Ilan University, Ramat-Gan 52900, Israel}
\affiliation{Department of Physics \#1, Graduate School of Science, Kyoto University, Kyoto 606-8502, Japan}

\author{Shalom Tzvi Shafier}
\affiliation{Department of Physics and Institute of Nanotechnology and Advanced Materials, Bar Ilan University, Ramat-Gan 52900, Israel}

\author{David A. Kessler}
\affiliation{Department of Physics and Institute of Nanotechnology and Advanced Materials, Bar Ilan University, Ramat-Gan 52900, Israel}

\author{Eli Barkai}
\affiliation{Department of Physics and Institute of Nanotechnology and Advanced Materials, Bar Ilan University, Ramat-Gan 52900, Israel}

\begin{abstract}
The Boltzmann-Gibbs density, a central result of equilibrium statistical mechanics, relates the energy of a system in contact with a thermal bath to its equilibrium statistics. 
This relation is lost for non-thermal systems such as cold atoms in optical lattices, where the heat bath is replaced by the laser beams of the lattice. 
We investigate in detail the stationary phase-space probability for Sisyphus cooling under harmonic confinement. 
In particular, we elucidate whether the total energy of the system still describes its stationary state statistics.
We find that this is true for the center part of the phase-space density for deep lattices, where the Boltzmann-Gibbs density provides an approximate description.
The relation between energy and statistics also persists for strong confinement and in the limit of high energies, where the system becomes underdamped.
However, the phase-space density now exhibits heavy power-law tails.
In all three cases we find expressions for the leading order phase-space density and corrections which break the equivalence of probability and energy and violate energy equipartition.
The non-equilibrium nature of the steady state is confounded by explicit violations of detailed balance.
We complement these analytical results with numerical simulations to map out the intricate structure of the phase-space density.
\end{abstract}


\maketitle

\section{Introduction \label{sec:introduction}}
For a particle in contact with a thermal bath, the stationary phase-space probability density is the Boltzmann-Gibbs density \cite{lan80}, which is given by $P_\text{BG}(x,p) \propto e^{-H(x,p)/(\kb T)}$, where $H(x,p)$ is the Hamiltonian in terms of the position $x$ and momentum $p$ of the particle and $T$ is the temperature of the heat bath. 
This relates the system's equilibrium statistics (the phase-space density) to its energy (the Hamiltonian) : In thermal equilibrium, states that have the same energy occur with the same probability.
If we replace the thermal bath by a non-thermal one, the system is generally not in equilibrium and the connection between probability and energy is lost.

An important example of such a non-thermal system is laser cooling \cite{coh90,cas90,bar94,met99}.
This term describes a multitude of techniques that are used to cool atoms or small particles in laboratories all over the world.
The common feature of all these techniques is that the surrounding heat bath is replaced with the light field of the laser, which generally does not constitute a thermal equilibrium bath.
Nevertheless, it is often convenient to assign an effective temperature, in term of the average kinetic energy, to this optical bath and thus to the atom or particle and treat its statistics as if the bath were thermal.
This effective temperature can be very low, which, along with the ease of tuneability makes laser cooling highly attractive from an experimental point of view.
For the particular case of a Sisyphus cooling lattice \cite{dal89,coh90,cas90}, this allows temperatures of a few $\mu$K to be reached.
Despite assigning a temperature, it is important to note that atoms cooled by the Sisyphus mechanism are not in a thermal state.
Notably, their momentum probability density exhibits heavy power-law tails \cite{mar96,lut03,dou06} and their dynamics is governed by long-ranged temporal correlations and superdiffusion \cite{kat97,sag12}.
For a free (aside from the cooling lattice) particle these effects have been well understood from a semiclassical description of the atoms' dynamics \cite{dal89,coh90,cas90,lut13}, which will be discussed in more detail below.

Here, we go a step beyond the free particle by introducing an additional confining potential acting on the atoms in the lattice \cite{dec15}.
For the free particle, the only degree of freedom contributing to the energy is its momentum, so its stationary momentum probability density can trivially be related to the total energy.
Introducing confinement yields the position of the particle (with its corresponding potential energy) as an additional degree of freedom.
In this situation the relation between total energy and stationary state statistics indeed becomes nontrivial.
In order to answer the question, whether and under what conditions the energy of the system determines the stationary state phase-space probability density, we determine the latter from the semiclassical Fokker-Planck equation description of the system.
We do so analytically in three limiting cases, and show that, while the leading order results depend only on the energy, higher order corrections violate the 1-to-1 correspondence of probability and energy.

We begin with a brief review of Boltzmann-Gibbs statistics of a Brownian particle in a potential and some immediate consequences in Section \ref{sec:BG-statistics}.
Then, in Section \ref{sec:optical-lattice} we introduce the semiclassical description of Sisyphus cooling and mention some consequences for the statistics of free cooled atoms.
Adding the confining potential, we arrive at the central equation of this work in Section \ref{sec:basic-equations}, the Fokker-Plank equation for the phase-space probability density.
In Section \ref{sec:small-D}, we start analyzing the latter by showing how the Boltzmann-Gibbs density serves as a starting point for an expansion describing the center part of the density for deep lattices.
The resulting expression gives us corrections to the Boltzmann-Gibbs result that violate energy equipartition.
The next part of the analysis in Section \ref{sec:large-freq} focuses on an underdamped approximation.
We show that for strong confinement, the total energy is approximately conserved.
Thus to leading order, the resulting phase-space density depends only on energy.
However, we find heavy power-law tails for high energies, in stark contrast to the exponential energy dependence of the Boltzmann-Gibbs density.
As we show in Section \ref{sec:large-E}, these high-energy tails persist even if the confinement is weak, which is a consequence of the peculiar structure of the Sisyphus cooling force.
We obtain an explicit expression for the large-energy behavior of the phase-space density including corrections that, once again, violate energy equipartition.
In Section \ref{sec:detailed-balance} we proceed to show that our system is indeed in a non-equilibrium stationary state in that it violates detailed balance, and relate these violations to the probability current.
Finally, in Section \ref{sec:experimental}, we relate our findings to the experimental systems of Sisyphus cooling of confined atoms and discuss the relevant parameter regime and how the deviations from Boltzmann-Gibbs statistics might be observed in experiments.
Throughout the analysis, we support our findings with numerical simulations, which we then use to explore the regime where the expansions break down and show that the results are fully consistent with the behavior emerging from the limiting cases.
This paper complements our Letter \cite{dec15} with detailed derivations and a much extended discussion of the results while also providing several novel aspects.

\section{Boltzmann-Gibbs statistics \label{sec:BG-statistics}}
The paradigmatic model that leads to Boltzmann-Gibbs statistics in a natural way is a Brownian particle in a confining potential.
This situation can be cast into a Langevin equation \cite{cof04},
\begin{align}
\dot{p} &= - \gamma p - U'(x) + \sqrt{2 D_p}\xi \nonumber \\
\dot{x} &= \frac{p}{m} \label{langevin-brownian} .
\end{align}
For simplicity, we restrict our discussion to the one-dimensional case and denote by $x$ the position of the particle and by $p$ its momentum.
Equation \eqref{langevin-brownian} describes a particle of mass $m$ moving in the conservative force field $F(x) = -U'(x)$, where $U(x)$ is a confining potential.
In addition, the particle is immersed in a thermal environment, which is responsible for the Stokes friction $F_\text{fric}(p) = -\gamma p$ and the fluctuating force $F_\text{fluc} = \sqrt{2 D_p} \xi$.
These two forces describe in an effective manner the collisions between the particle and the constituent particles of the environment.
Since the environment is thermal, the damping coefficient $\gamma$ and the momentum diffusion constant $D_p$ are related via the temperature $T$ by the fluctuation-dissipation theorem, $D_p = m \gamma \kb T$.
$\xi$ is a Gaussian white noise of unit magnitude, i.~e.~ $\langle \xi(t) \xi(t') \rangle = \delta(t-t')$.
Here $\langle \ldots \rangle$ denotes an average over the ensemble of realizations of the process $\xi(t)$.
Equivalently, the system may be described via a Kramers-Fokker-Planck equation for the phase-space probability density \cite{ris86},
\begin{align}
\partial_t &P(x,p,t) \nonumber \\
& = \bigg[ - \frac{p}{m} \partial_x + \partial_p \Big[ \gamma p + U'(x) + D_p \partial_p \Big] \bigg] P(x,p,t) . \label{KFP-brownian}
\end{align}
It is then easy to show that the stationary state solution $\partial_t P(x,p,t) = 0$ is precisely given by the Boltzmann-Gibbs density \cite{ris86},
\begin{align}
P_\text{BG}(x,p) = Z^{-1} e^{-\frac{H(x,p)}{\kb T}}, \label{BG-density}
\end{align}
where the Hamiltonian gives the total energy of the particle as a function of its position and momentum, $H(x,p) = p^2/(2 m) + U(x)$, and $Z = \int \text{d}x \ \text{d}p \ e^{-H(x,p)/(\kb T)}$ is the normalizing partition function.

The Boltzmann-Gibbs density \eqref{BG-density}, which is readily generalized to higher dimensions, has a number of important properties and consequences.
It depends on the coordinate $x$ and momentum $p$ only through the Hamiltonian.
This means that equal-energy surfaces in phase-space are also surfaces of equal probability.
An immediate consequence is the equipartition theorem, which relates equilibrium expectation values of the Hamiltonian to the temperature \cite{tol18,hua87},
\begin{align}
\langle p \ \partial_p H(x,p) \rangle_\text{eq} = \langle x \ \partial_r H(x,p) \rangle_\text{eq} = \kb T. \label{equipartition-general}
\end{align}
For the canonical harmonic potential, $U(x) = m \omega^2 x^2/2$, this yields the more familiar relation
\begin{align}
\frac{\kb T}{2} = \langle E_k \rangle_\text{eq} = \langle E_p \rangle_\text{eq} , \label{equipartition-harmonic}
\end{align}
where $E_k = p^2/(2 m)$ is the kinetic and $E_p = m \omega^2 x^2/2$ the potential energy.
Thus every quadratic degree of freedom contributes $\kb T/2$ to the average energy.
Whenever the phase-space probability density is a function of $H(x,p)$ only, equality holds between the moments. 
However, the right hand side of Eq.~\eqref{equipartition-general} is only given by the temperature in the Boltzmann-Gibbs case.
Furthermore, the Boltzmann-Gibbs density is exponential in the Hamiltonian, and thus additive contributions to the system's energy factorize in terms of the phase-space density.
In particular, the kinetic and the potential energy term separate,
\begin{align}
P_\text{BG}(x,p) &= P_x(x) P_p(p) \\
\text{with} \; P_x(x) &= Z_x^{-1} e^{-\frac{U(x)}{\kb T}}, \quad P_p(p) = Z_p^{-1} e^{-\frac{p^2}{2 m \kb T}} \nonumber .
\end{align}
This means that, in equilibrium, the kinetic and potential degrees of freedom are independent of each other and can be described separately.
In particular, the equilibrium average kinetic energy is not affected by introducing or changing the potential.

\section{Semiclassical description of Sisyphus cooling \label{sec:optical-lattice}}
In the above discussion, we took into account the effect of the bath through a dissipative friction term and a fluctuating random force which causes diffusion.
As it turns out, a similar description can be employed to describe laser cooling of atoms \cite{met99}.
Here the dissipation and diffusion describe changes in the atomic momentum due to the interactions with the photons of the light field by scattering, absorption and emission.
For Sisyphus cooling \cite{dal89,coh90,cas90}, the atom interacts with a standing light wave.
As long as the atom is not localized in the resulting lattice potential, its trajectory can be described as a classical particle with momentum $p$ and position $x$.
If further the momentum of the atom is large compared to the recoil momentum by interacting with a single photon, we can treat the interactions with the light field as a continuous process rather than individual events.
Under these conditions, it can be shown that the motion of the atoms can be described by a semiclassical Langevin equation \cite{cas90},
\begin{align}
\dot{p} &= - \frac{\gamma p}{1 + \frac{p^2}{p_c^2}} - U'(x) + \sqrt{2 D_p} \odot \xi \nonumber \\
\dot{x} & = \frac{p}{m} \label{langevin-lattice} .
\end{align}
The interactions between the atom and the cooling lattice are taken into account via an effective friction and a fluctuating force.
We stress that $U(x)$ here is an additional spatial potential, which is distinct from the cooling lattice.
The Langevin equation \eqref{langevin-lattice} is quite similar to the one for a Brownian particle \eqref{langevin-brownian}, the apparent difference being the nonlinearity of the friction force
\begin{align}
F_\text{fric} = -\gamma p /(1+p^2/p_c^2) , \label{friction}
\end{align}
The latter is Stokes-like only for small momenta $|p| \ll p_c$ but decreases with momentum as $F_\text{fric} \propto - 1/p$ for large momenta $|p| \gg p_c$.
Beyond that, the diffusion coefficient now depends on the momentum, $D_p = D_0 + D_1/(1+p^2/p_c^2)$.
This implies that we need to specify the interpretation of \eqref{langevin-lattice} as a stochastic integral \cite{ris86,gar96}, here, anti-It{\=o} or endpoint interpretation, denoted by $\odot$ is the appropriate one, leading to the correct Kramers-Fokker-Planck equation \eqref{KFP-lattice}.

For a detailed explanation of why the nonlinear friction and the momentum-dependent diffusion coefficient appear, the reader is referred to \cite{coh90,cas90}.
We note that the four parameters $p_c$, $\gamma$, $D_0$ and $D_1$ can be expressed in terms of the experimental control parameters like the intensity and detuning of the cooling lasers, see Section \ref{sec:experimental}.
Very roughly, the cooling lattice consists of two superimposed standing waves with orthogonal polarization.
The splitting of the atomic Zeemann-sublevels and the transition rates between them thus depend on the atoms' position in the lattice.
Conversely, the potential seen by the atom depends on which state it is in.
If the frequency of the lattice beams is chosen in the right way, the atom will climb a potential barrier, transition to a different state for which the same position is a potential minimum and then has to climb yet another barrier.
Repeating this process, the atom dissipates kinetic energy via the emission of photons.
It is intuitively clear that how well this process works depends crucially on the speed of the atoms: atoms at rest remain in the lowest energy state corresponding to their position, while very fast atoms cannot distinguish between going uphill and downhill in the optical lattice.
This explains why the friction force is weak if the atom is either very slow or very fast.
The momentum-dependent part of the diffusion coefficient $D_1/(1+p^2/p_c^2)$ reflects the fact that the transitions between different atomic states are probabilistic and thus the atom will occasionally transition at the wrong time.
This part of the diffusion coefficient satisfies a fluctuation-dissipation relation with the friction force.
The first, momentum-independent part of the diffusion coefficient $D_0$, however, is due to spontaneous emission events and has no counteracting friction term, thereby driving the system out of equilibrium.
We remark that deriving the semiclassical equations of motion \eqref{langevin-lattice} relies on averaging the dynamics over one wavelength of the cooling lattice \cite{cas90} and thus the former do not depend on the spatial modulation of the laser field.
It can be shown that this approximation does not change the qualitative dynamics \cite{hol15}, however, as discussed in Section \ref{sec:experimental}, it requires that the scale of the confining potential should be much larger than the lattice wavelength.

In the absence of the confining potential $U(x)=0$, the resulting dynamics have been discussed in detail in Refs.~\cite{mar96,lut03,kes10,kes12,dec12}. 
The fact that the friction force is weak for fast atoms means that those atoms dissipate little energy and thus can stay fast for a very long time.
This induces a broad stationary state momentum distribution \cite{lut03},
\begin{align}
P_p(p) &= Z^{-1} \bigg(1 + \frac{D_0}{D_0+D_1} \frac{p^2}{p_c^2} \bigg)^{-\frac{1}{2 D}} \label{momentum-PDF-free},
\end{align}
long-ranged temporal correlations \cite{mar96} in the atomic momentum, which lead to superdiffusion for shallow lattices \cite{kes12,dec12}.
Here we stress that the temporal correlations \cite{kat97}, the asymptotic power-law tails of the momentum density \cite{dou06} and the superdiffusive motion \cite{sag12} have all been observed in experiments.
In general, the dynamics are controlled by the dimensionless parameter $D \equiv D_0/(\gamma p_c^2)$.
This parameter is related to the depth $U_0$ of the cooling lattice by $D = c E_r/U_0$ \cite{kes12}, where $E_r$ is the photon recoil energy and $c \sim 20$ is a constant whose precise value depends on the details of the experiment.
Briefly summarizing the qualitative results for the free case, for $D < 1/5$ the momentum correlations decay sufficiently fast in time that the diffusion is normal. For $D > 1/5$, superdiffusion sets in and for $D>1$ there is no longer a stationary momentum probability density \cite{kes10}.
Since without the confining potential, the only degree of freedom entering the Hamiltonian is the momentum, $H(p) = p^2/(2 m)$, we can trivially write the stationary density Eq.~\eqref{momentum-PDF-free} as a function of the Hamiltonian,
\begin{align}
P_p(p) = Z^{-1} \bigg(1 + \frac{H(p)}{E_c} \bigg)^{-\frac{1}{2D}},
\end{align}
with $E_c = p_c^2(1+D_1/D_0)/(2 m)$ and $H(p) = p^2/(2 m)$.
The above form of the probability density is formally equivalent to a Tsallis distribution \cite{tsa88,tsa95,lut03}. 
As shown below, this equivalence no longer holds for the confined system \cite{dec15}.
Equivalently, we find the probability distribution of energy,
\begin{align}
P_E(E) = \tilde{Z}^{-1} \sqrt{\frac{E_c}{E}} \bigg(1 + \frac{E}{E_c} \bigg)^{-\frac{1}{2D}}. \label{energy-PDF-free}
\end{align}
We note that the parameter $D$ depends only on $D_0$. 
By contrast, $D_1$ appears in the asymptotic power-law behavior of the momentum density \eqref{momentum-PDF-free}, $P_p(p) \propto p^{-1/D}$ only as a prefactor \cite{mar96}.
Given this, for the sake of simplicity, we will consider the case $D_1 = 0$ going forward, which does not change the qualitative results.
We will briefly discuss the effect of nonzero $D_1$ in Section \ref{sec:experimental}.
Formally, Eq.~\eqref{langevin-lattice} with $p$ interpreted as the position is related to the diffusion of a particle in a logarithmic potential \cite{hir11,dec11,hir12}, which can be used to model a range of physical systems \cite{man69,bra00,fog07}.
In these cases, however, it is not clear what the physical equivalent of the potential $U(x)$ is.

Taking into account the confining potential with the Boltzmann-Gibbs result in mind immediately raises the question whether the phase-space density may be written as a function of the Hamiltonian.
In this case, the average kinetic energy would be independent of the confinement, energy equipartition would hold and some effective temperature could be assigned to the system.
In the following, we will address this question and investigate the similarities between the thermal Boltzmann-Gibbs case and Sisyphus cooling.

\section{Confinement and Sisyphus cooling \label{sec:basic-equations}}
As long as the semiclassical picture is valid, the motion of the atoms can be described by Eq.~\eqref{langevin-lattice}, where we consider a pure harmonic potential, $U(x) = m \omega^2 x^2/2$.
In the following, we will focus only on the steady state, $\partial_t P(x,p,t) = 0$.
The stationary phase-space density $P(x,p)$ is then given in terms of a Kramers-Fokker-Planck equation \cite{ris86},
\begin{align}
\Bigg[ \Omega \bigg( - p \partial_{x} + x \partial_{p} \bigg) +  \partial_{p} \bigg(\frac{{p}}{1+{p}^2} + D \partial_{p} \bigg) \Bigg] P(x,p) = 0 . \label{KFP-lattice}
\end{align}
In order to simplify the notation, we have changed to dimensionless position and momentum variables $x =  m \omega \tilde{x}/p_c$, $p = \tilde{p}/p_c$.
Since going forward, we will mostly use the dimensionless variables, we from now on refer to the dimensionful variables as $\tilde{x}$ and $\tilde{p}$.
We have further re-introduced the dimensionless parameter $D \equiv D_0/(\gamma p_c^2)$ and defined $\Omega \equiv \omega/\gamma$.
In terms of this dimensionless description, it is clear that the properties of the system are governed by the two parameters $D$ and $\Omega$.
Equation \eqref{KFP-lattice} constitutes the main object of investigation of this work.
The first term on the left-hand side of \eqref{KFP-lattice} describes the Hamiltonian part of the evolution, the oscillation of a particle in an harmonic well.
The second term contains the friction and the noise, which are effects of the \enquote{bath}.

At this point, we also introduce two equivalent ways of writing Eq.~\eqref{KFP-lattice}, which will turn out to be convenient for the following discussion.
The first alternate representations follows from a simple rescaling of position and momentum, $z = x/\sqrt{D}$ and $u = p/\sqrt{D}$,
\begin{align}
\Bigg[ \Omega \bigg( - u \partial_z + z \partial_u \bigg) +  \partial_u \bigg(\frac{u}{1+D u^2} + \partial_u \bigg) \Bigg] P_D(z,u) = 0, \label{KFP-lattice-D}
\end{align}
with $P(x,p) = P_D[x/\sqrt{D}, p/\sqrt{D}]/D$.
Clearly \eqref{KFP-lattice-D} reduces to the equation for the Boltzmann-Gibbs case \eqref{KFP-brownian} in the limit $D \rightarrow 0$.
We will use this in the next section to perform a systematic expansion around this limit.
Secondly, we change to a polar representation of the phase space, by introducing the energy $\varepsilon = (x^2+p^2)/2$ and the phase-space angle $\alpha = \arctan(p/x)$, $0 \leq \alpha < 2 \pi$.
In terms of these coordinates, Eq.~\eqref{KFP-lattice} reads
\begin{align}
\Bigg[ &\underbrace{\Omega \partial_\alpha}_{\begin{array}{c}  \text{ \scriptsize Hamiltonian} \\[-1 ex]\text{ \scriptsize evolution} \end{array}} + \underbrace{\mathcal{L}_{\varepsilon,\alpha}}_{\begin{array}{c}\text{ \scriptsize friction} \\[-1 ex]  \text{\scriptsize and noise} \end{array}} \Bigg] P_P(\varepsilon,\alpha) = 0 \quad \text{with} \label{KFP-lattice-E} \\
\mathcal{L}_{\varepsilon,\alpha} &= \partial_\alpha \frac{\sin(\alpha)\cos(\alpha)}{1+2 \varepsilon \sin^2(\alpha)} + \partial_\varepsilon \frac{2 \varepsilon \sin^2(\alpha)}{1+2 \varepsilon \sin^2(\alpha)}  \nonumber \\
& + D \bigg[ 2 \partial_\alpha \sin(\alpha) \cos(\alpha) \partial_\varepsilon  + \frac{1}{2 \varepsilon} \partial_\alpha \cos^2(\alpha) \partial_\alpha \nonumber \\
& + (\sin^2(\alpha) - \cos^2(\alpha)) \partial_\varepsilon + 2 \sin^2(\alpha) \partial_\varepsilon \varepsilon \partial_\varepsilon \bigg] \nonumber ,
\end{align}
with $P(x,p) = P_P[(x^2+p^2)/2,\arctan(p/x)]$, since the Jacobian of the transformation is unity.
The advantage of this representation is that the Hamiltonian part of the evolution consists of just a derivative with respect to $\alpha$, whereas the friction and noise terms are now more complicated.
We will use Eq.~\eqref{KFP-lattice-E} extensively in Sections \ref{sec:large-freq} and \ref{sec:large-E} to find the behavior at large frequencies and large energies, respectively.
Note that the dimensionless energy $\varepsilon$ is related to the physical energy $E$ via $E = \varepsilon p_c^2/m$.
In the following Sections \ref{sec:small-D}, \ref{sec:large-freq} and \ref{sec:large-E}, we will discuss the mathematical properties of the solution of Eqs.~\eqref{KFP-lattice}, \eqref{KFP-lattice-D} and \eqref{KFP-lattice-E} and the physical implications.
This solution can be obtained analytically in terms of expansions in certain limits.
The general features elucidated in these expansions, however, persist even beyond the validity of the expansions themselves, as we will establish via numerical simulations.
In Section \ref{sec:experimental}, we will come back to the actual experimental system of Sisyphus cooling with added confinement and discuss the relevant parameter regime, as well as other physical effects that need to be taken into account.

\section{Deep lattices: Deviations from Boltzmann-Gibbs \label{sec:small-D}}
The dimensionless parameter $D$ is inversely proportional to the depth of the cooling lattice (see Section \ref{sec:optical-lattice}).
For deep lattices, $D$ is thus small.
As noted before, Eq.~\eqref{KFP-lattice-D} reduces to a Boltzmann-Gibbs-like equation in the limit of $D \rightarrow 0$.
In particular, the normalized solution for $D = 0$ is
\begin{align}
P_D^{(0)}(z,u) = \frac{1}{2 \pi} e^{-\frac{z^2+u^2}{2}} , \label{small-d-0order}
\end{align}
which is precisely the Boltzmann-Gibbs density \eqref{BG-density}.
Starting from this, we define an auxiliary function $g(z,u)$ via
\begin{align}
P_D(z,u) = P_D^{(0)}(z,u) g(z,u) .
\end{align}
Plugging this into Eq.~\eqref{KFP-lattice-D}, we obtain an equation for $g(z,u)$,
\begin{align}
\Big[\mathcal{L}_0 &+ D \mathcal{L}_1 + D^2 \mathcal{L}_2\Big] g(z,u) = 0 \label{small-d-auxilary} \\
\mathcal{L}_0 &= \Omega(z \partial_{u} - u \partial_{z}) - u \partial_{u} + \partial_{u}^2 \nonumber \\
\mathcal{L}_1 &= 2 \Omega (z u^2 \partial_{u} - u^3 \partial_{z}) - 3 u^2 + u^4 - 3 u^3 \partial_{u} + 2 u^2 \partial_{u}^2 \nonumber \\
\mathcal{L}_2 &= \Omega (z u^4 \partial_{u} - u^5 \partial_{z}) - u^4 + u^6 - 2 u^5 \partial_{u} + u^4 \partial_{u}^2 \nonumber .
\end{align}
We see that the equation for the function $g(z,u)$ contains three terms multiplied by different powers of $D$.
For $D = 0$, the resulting equation $\mathcal{L}_0 g(z,u) = 0$, obviously has the solution $g(z,u) = 1$, which yields the Boltzmann-Gibbs result \eqref{small-d-0order}.
To proceed, we assume that we can expand $g(z,u)$ for small $D$,
\begin{align}
g(z,u) = 1 + D g^{(1)}(z,u) + \mathcal{O}(D^2) . \label{small-d-auxilary-expansion}
\end{align}
Plugging this into Eq.~\eqref{small-d-auxilary}, and equating orders in $D$ we obtain an equation for $g^{(1)}(z,u)$,
\begin{align}
\mathcal{L}_0 g^{(1)}(z,u) - 3 u^2 + u^4 = 0 . \label{small-d-auxilary-1order}
\end{align}
Examining the operator $\mathcal{L}_0$, we can already guess what the solution to this equation might be.
If $g^{(1)}(z,u)$ is a polynomial in $z$ and $u$ of total order $N$, then $\mathcal{L}_0$ leaves the total order unchanged.
The inhomogeneous terms in \eqref{small-d-auxilary-1order}, on the other hand, are of order $4$, so if there exists a polynomial solution, we necessarily have $N=4$,
\begin{align}
g^{(1)}(z,u) = \sum_{k=0}^{4} \sum_{l=0}^{4-k} a^{(1)}_{k l} z^{k} u^{l} \label{small-d-polynomial} .
\end{align}
Plugging this into Eq.~\eqref{small-d-auxilary-1order} then gives us a set of linear equations for the coefficients $a^{(1)}_{k l}$.
For the first order expansion, we explicitly give the coefficients in Appendix \ref{app:small-d}.
We can continue the expansion in terms of $D$, Eq.~\eqref{small-d-auxilary-expansion}, to higher orders.
Before we proceed to do so, however, let us remark on the validity of the expansion.
We implicitly assumed that $D g^{(1)}(z,u)$ is small.
Since we know that $g^{(1)}(z,u)$ is a polynomial of total order $4$ in $z$ and $u$, we need not only for $D$ to be small but also terms of the form $D z^k u^l$ with $k + l \leq 4$.
This means that our expansion is valid for small $D$ in the center part of the phase-space density.
Up to first order in $D$, we find for the normalized phase-space probability density
\vspace*{-2em}
\begin{widetext}
\begin{align}
P^{(1)}_D(z,u) = \frac{e^{-\frac{z^2+u^2}{2}}}{2 \pi} \bigg[ 1 + \frac{D}{4 (3+4\Omega^2)} \Big[ 3 u^4 + 18 z^2 - 27 + \Big( 4 u^3 z - 12 u z \Big) \Omega + \Big(3(u^2+z^2)^2 - 24 \Big) \Omega^2 \Big] \bigg] . \label{small-d-1order}
\end{align}
\end{widetext}
Comparing this first order result to the Boltzmann-Gibbs one, we note that, in contrast to the latter, Eq.~\eqref{small-d-1order} does not depend on $u$ and $z$ as a function of only the Hamiltonian $H(u,z) = (z^2 + u^2)/2$.
Consequently, it has less symmetry than the zero-order Boltzmann-Gibbs approximation: 
It is not symmetric with respect to interchanging $z$ and $u$, or reversing $z \rightarrow -z$ or $u \rightarrow -u$ individually.
However, we note that all these symmetries are restored in the limit $\Omega \gg 1$, i.e.~ strong confinement, where to leading order in $\Omega$, we find
\begin{align}
P^{(1)}_D(z,u) \simeq \frac{e^{-H(z,u)}}{2 \pi} \bigg[ 1 + \frac{D}{4} \Big[ \Big(3 H^2(z,u) - 8 \Big) \Big] \bigg]. \label{small-d-1order-omega}
\end{align}
A similar behavior is found for any $D$ and large $\Omega$ in Section \ref{sec:large-freq}.

The equation for the second-order correction $g^{(2)}(z,u)$ is easy to derive,
\begin{align}
\mathcal{L}_0 g^{(2)}(z,u) + \mathcal{L}_1 g^{(1)}(z,u) - u^4 + u^6 = 0.
\end{align}
Since the term $\mathcal{L}_1 g^{(1)}(z,u)$ contains contributions proportional to $p^8$, $g^{(2)}(z,u)$ is a polynomial of total degree $N=8$.
Generally, we can write the expansion of $g(z,u)$ up to order $D^M$ as
\begin{align}
g(z,u) = 1 + \sum_{n = 1}^{M} D^{n} \sum_{k=0}^{4 n} \sum_{l=0}^{4 n - k} a^{(n)}_{k,l} z^k u^l ,
\end{align}
thus $g^{(M)}(z,u)$ is a polynomial of order $N = 4 M$.
For the following analysis, we perform this expansion up to third order $M=3$, where we use Mathematica to handle the cumbersome algebra.
\begin{figure}[ht!]
\includegraphics[width=0.47\textwidth, clip, trim=0mm 0mm 20mm 10mm]{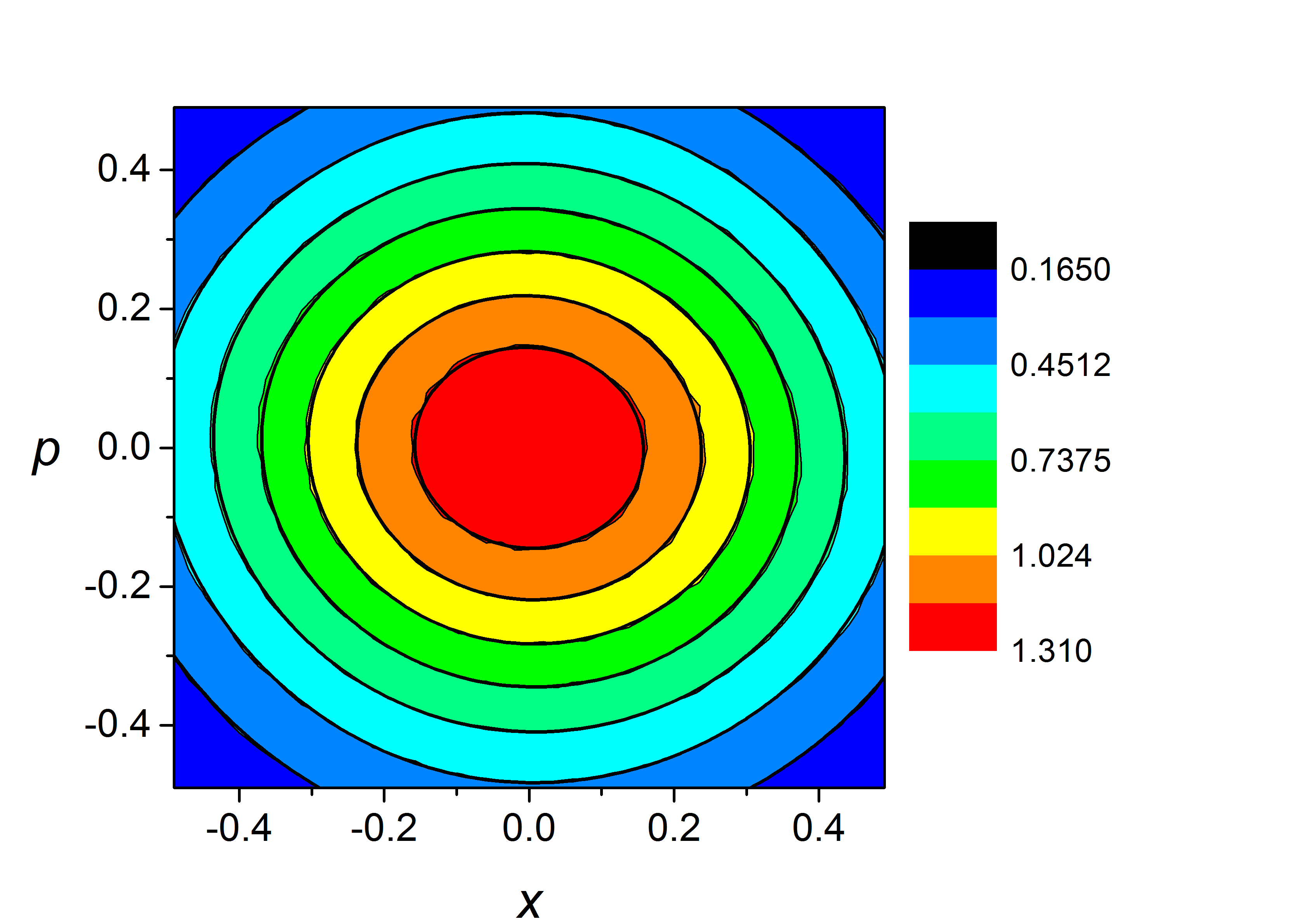} \\
\includegraphics[width=0.47\textwidth, clip, trim=0mm 0mm 20mm 10mm]{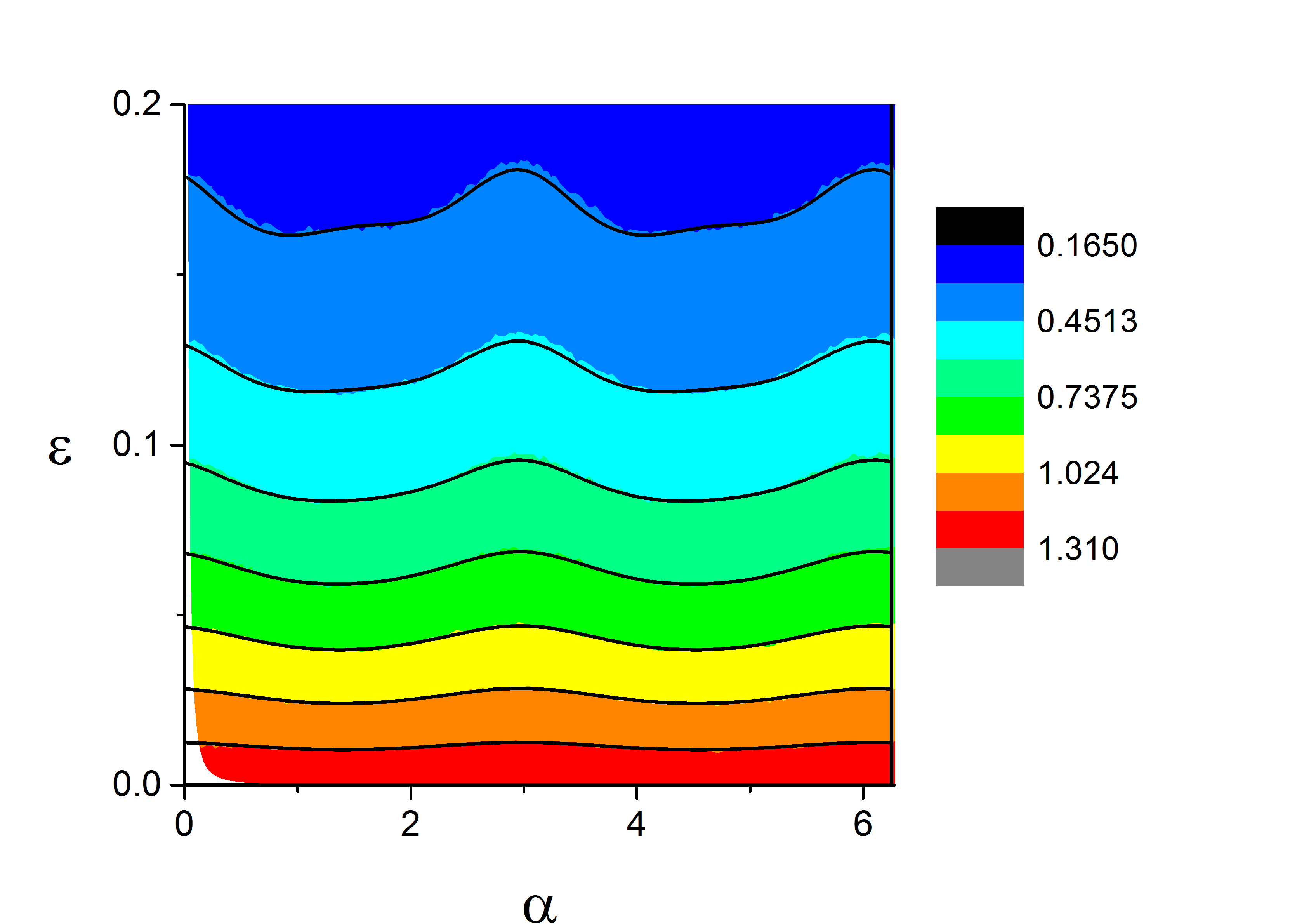}
\caption{Phase-space probability density for $\Omega = 0.5$ and $D = 0.1$ plotted using the position and momentum $(x,p)$ (top) respectively energy and phase-space angle $(E,\alpha)$ (bottom). The colored surfaces are the results of numerical Langevin simulations, the black lines represent the third-order small-$D$ expansion. For the Boltzmann-Gibbs case $D \rightarrow 0$, these plots would display perfect cirlces (top) respectively straight lines (bottom). The slightly tilted elliptical shape (top) respectively wavy lines (bottom) are due to the deviations from Boltzmann-Gibbs for finite $D$. \label{fig:jdist}}
\end{figure}
The resulting phase-space density is shown in Fig.~\ref{fig:jdist}.
Plotting the latter using the energy-angle coordinates introduced in Eq.~\eqref{KFP-lattice-E} clearly exhibits the deviations from the Boltzmann-Gibbs density, which is independent of the angle $\alpha$.
This density has a number of features which distinguish it from the Boltzmann-Gibbs density, which we discuss in the following.
In Section \ref{sec:BG-statistics}, we saw that a phase-space density that only depends on the Hamiltonian leads to energy equipartition, which in our rescaled variables corresponds to $\langle z^2 \rangle = \langle u^2 \rangle = 1$ for $D=0$.
In order to quantify the deviations induced by finite $D$, we define the equipartition ratio
\begin{align}
\chi = \langle z^2 \rangle/\langle u^2 \rangle = \langle x^2 \rangle/\langle p^2 \rangle = \langle \varepsilon_p \rangle/\langle \varepsilon_k \rangle, \label{equipart}
\end{align} 
where $\varepsilon_p = x^2/2$ and $\varepsilon_k = p^2/2$ are the dimensionless potential and kinetic energy.
Obviously $\chi = 1$ corresponds to energy equipartition.
From the third order expansion, we find
\begin{align}
\chi^{(3)} = 1 + \frac{6}{3+4\Omega^2} D^2 - \frac{6 (38 \Omega^2 - 21)}{(3+4\Omega^2)^2} D^3 . \label{equipart-order3}
\end{align}
We see that there is no linear term in this expansion, meaning that the first order result \eqref{small-d-1order} does induce any violation of energy equipartition, even though it is not a function of the Hamiltonian only.
Further, for large $\Omega$, the deviations from equipartition are also of order $\Omega^{-2}$, which agrees with the finding that the phase-space density becomes a function of the Hamiltonian in this limit.
\begin{figure}[ht!]
\includegraphics[width=0.47\textwidth, clip, trim=0mm 0mm 20mm 10mm]{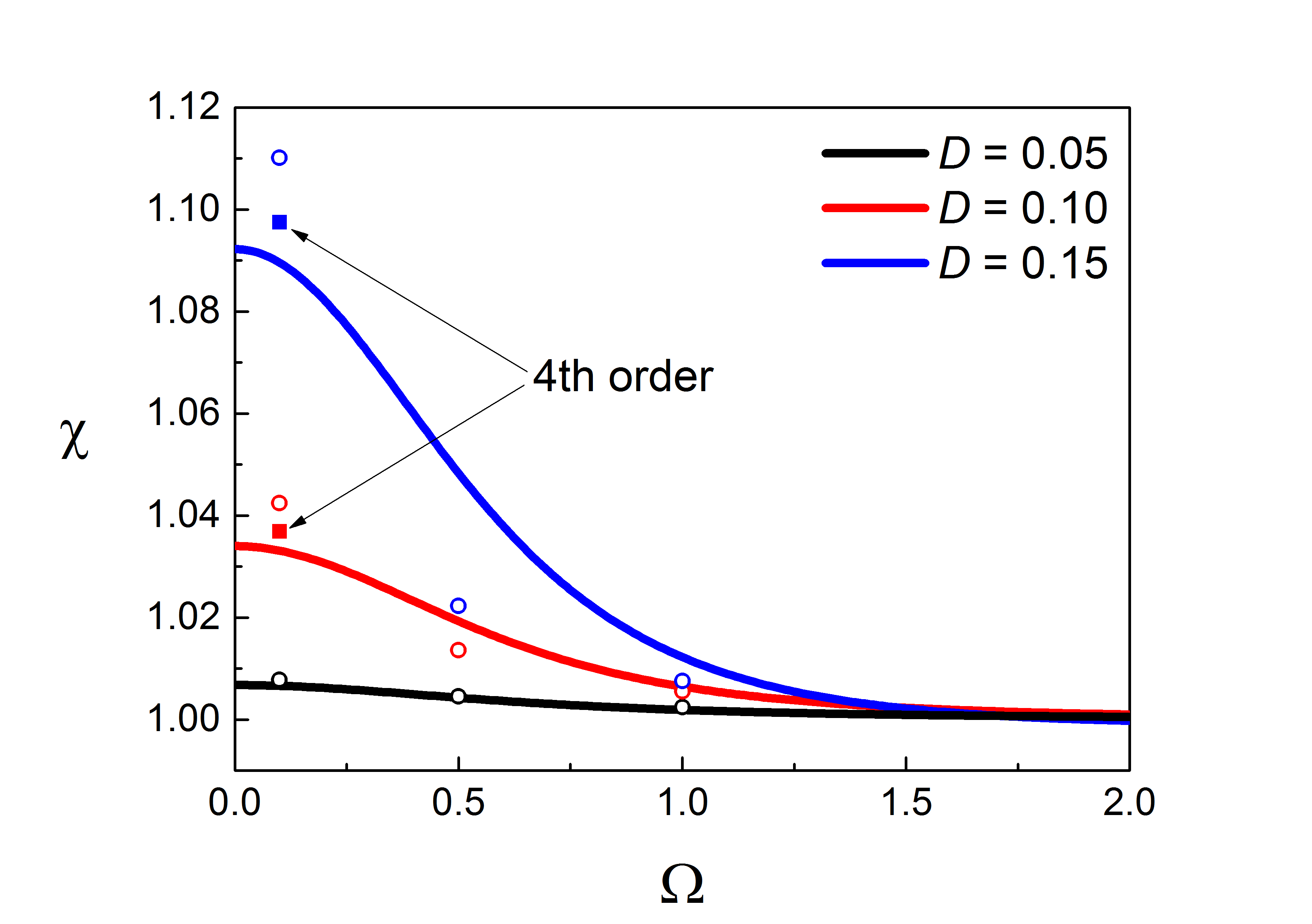}
\caption{Equipartition ratio $\chi$ as a function of the trap strength $\Omega$ for different values of $D$. The solid lines are the third-order result Eq.~\eqref{equipart-order3}, the circles are the result of numerical Langevin simulations. For comparison, we also show the results of a fourth-order expansion in $D$ at $\Omega = 0.1$, which are closer to the numerical results. \label{fig:equipart}}
\end{figure}
\begin{figure}[ht!]
\includegraphics[width=0.47\textwidth, clip, trim=0mm 0mm 20mm 10mm]{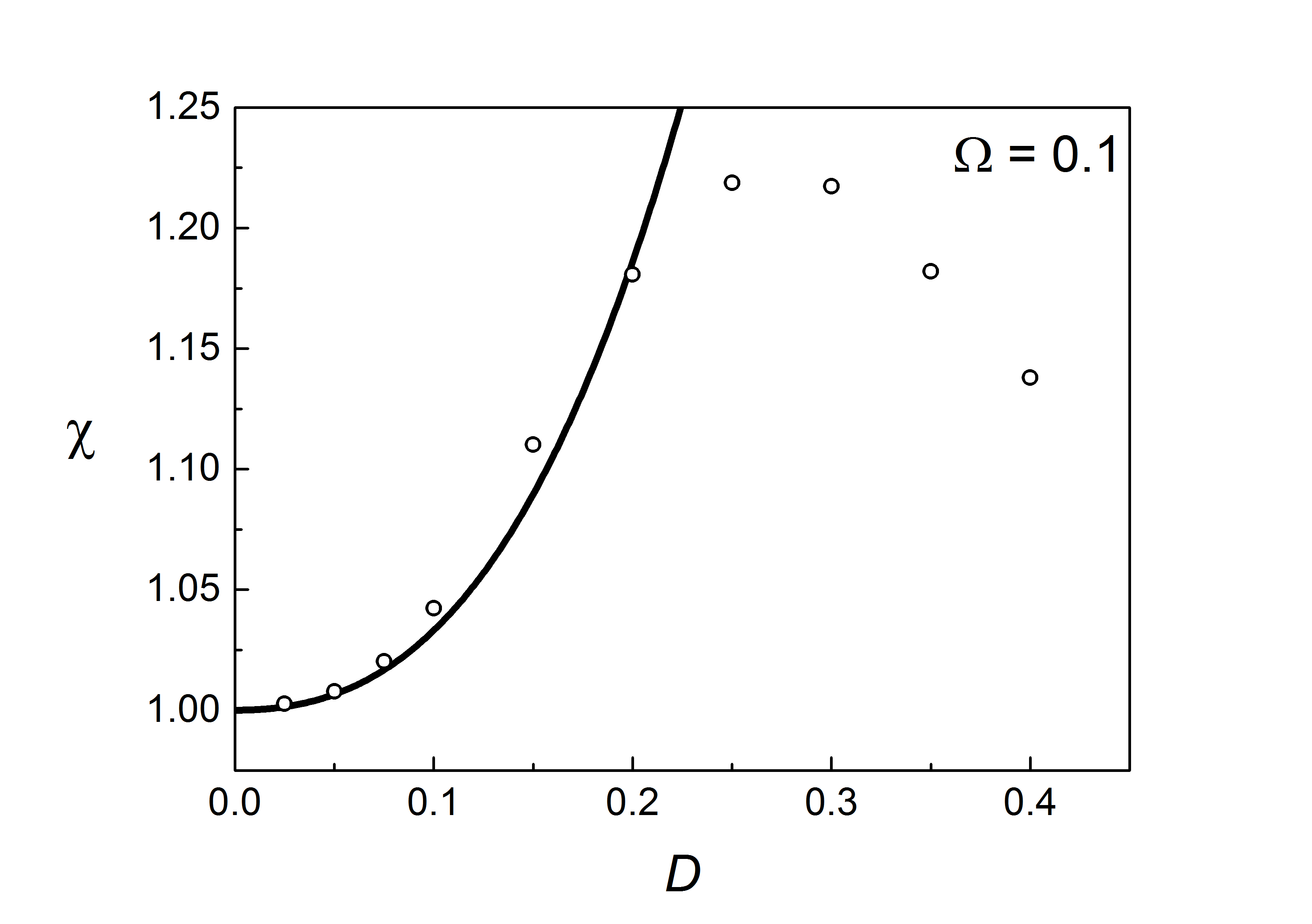}
\caption{Equipartition ratio $\chi$ as a function of the parameter $D$ for $\Omega = 0.1$. The solid line is the third-order result Eq.~\eqref{equipart-order3}, the circles are the result of numerical Langevin simulations. For small $D$, the agreement between numerics and theory is good, at larger $D$, the two results deviate. In particular, the numerically obtained equipartion ratio exhibits a turnover and decreases for larger $D$. \label{fig:equipart2}}
\end{figure}
\begin{figure}[ht!]
\includegraphics[width=0.47\textwidth, clip, trim=0mm 0mm 20mm 10mm]{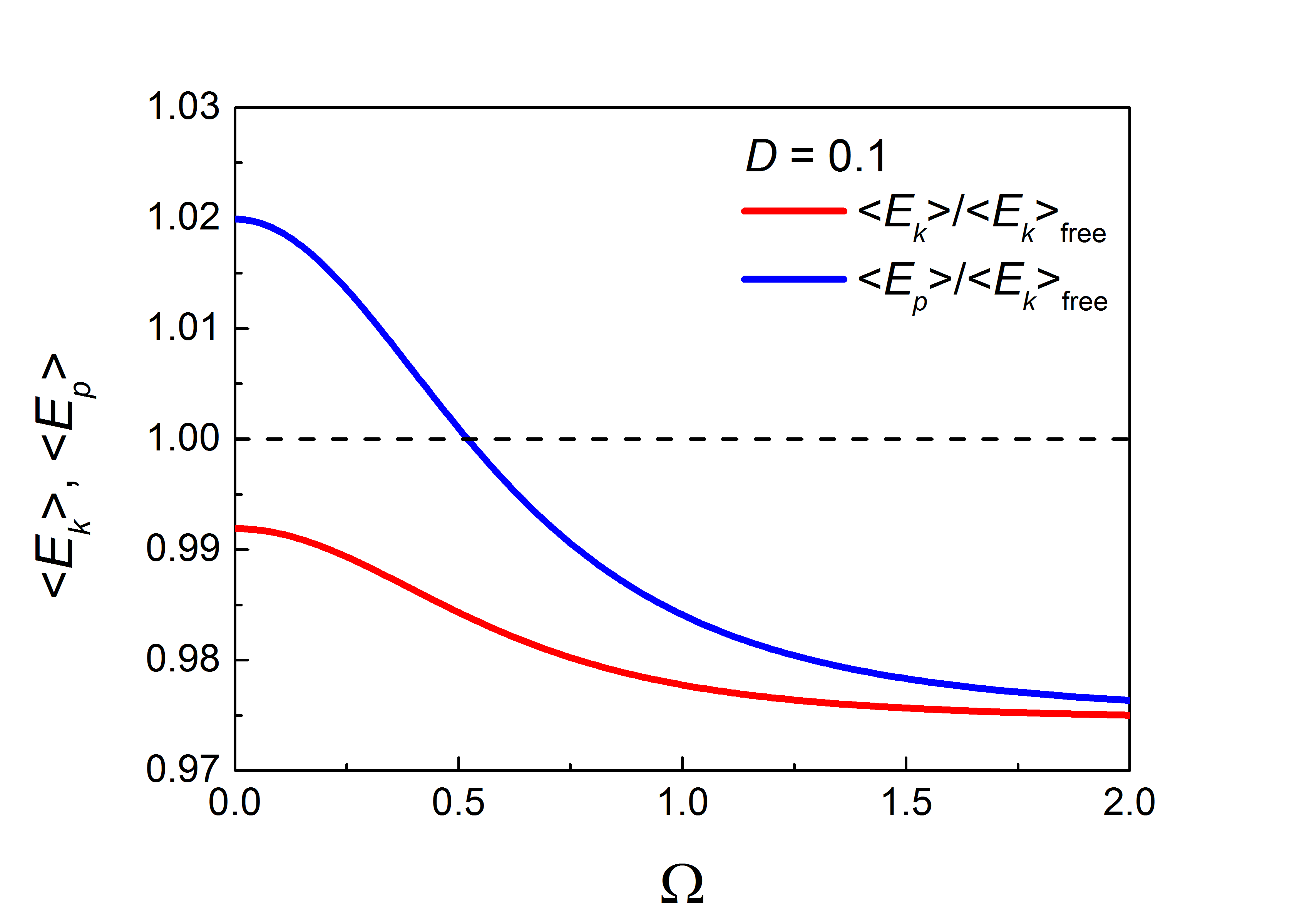}
\caption{Average potential and kinetic energy from the third-order small-$D$ expansion as a function of $\Omega$ for $D = 0.1$. Both have been divided by the average kinetic energy of the free particle. Note that the kinetic energy is reduced compared to the latter. \label{fig:equipart3}}
\end{figure}
The average potential energy is always larger than the average kinetic energy.
This is reflected in the slightly elongated shape of the phase-space density along the $x$-axis, see Fig.~\ref{fig:jdist}.
These deviations from energy equipartition are strongest for small trapping strengths $\Omega$ and increase with increasing $D$, see Fig.~\ref{fig:equipart}.
Not surprisingly, the results of the small-$D$ expansion deviate from the numerical ones for larger values of $D$, where the expansion breaks down.
Nevertheless, the trends with respect to $\Omega$ and $D$ are captured correctly.
When plotted as a function of $D$ for $\Omega = 0.1$, we see that the deviations from equipartition are maximal for $D \approx 0.3$ where the potential energy is about $20 \%$ larger than the kinetic one see Fig.~\ref{fig:equipart2}.
The reason for the decrease at larger $D$ is due to the fact that the average energy in the stationary state diverges for $D > 1/2$.
This is discussed in more detail in Section \ref{sec:large-E}.

Instead of looking at the equipartition ratio, it is worthwhile to also discuss the behavior of the potential and kinetic energy individually.
The results from the third-order expansion are shown in Fig.~\ref{fig:equipart3}.
Here, we normalized the energy to the kinetic energy of a free particle at the same value of $D$.
Even for large frequencies, where the phase-space density can be expressed as a function of the Hamiltonian, both the kinetic and potential energy are less than the kinetic energy of a free particle.
This underlines the nontrivial interplay between the confining potential and the nonlinear friction force.
Since the friction force is strongest for particles of moderate momentum, the slowing down near the turning points of the oscillatory motion in the potential increases the dissipation.
This stays true in the limit of small $\Omega$, where the average kinetic energy is still reduced compared to the free case, even though the potential energy is now larger.
Due to the perturbative nature of the above results, the overall effect shown here is small.
However, it can be quite substantial at realistic parameter values beyond the validity of the perturbation expansion, where we numerically find an average potential energy that is several times larger than the kinetic one, see Section \ref{sec:experimental}.
At first glance it might seem disconcerting that even as $\Omega \rightarrow 0$, we do not recover the free particle result.
However, here we are only discussing the stationary behavior of the system.
For very small $\Omega$, it will take longer and longer to reach this stationary state, as the oscillation period in the potential grows as $1/\Omega$ until at $\Omega = 0$ there is no stationary state at all.
The stationary state limit $t \rightarrow \infty$ and the limit $\Omega \rightarrow 0$ thus do not commute.

The expansion for the phase-space density \eqref{small-d-1order} for $D \neq 0$ also differs from the Boltzmann-Gibbs case in that it is no longer symmetric with respect to $u \rightarrow -u$ or $z \rightarrow -z$; only the combination of both operations is a symmetry of the problem: $P_D(z,u) = P_D(-z,-u)$.
This can be conveniently visualized by defining the antisymmetric and symmetric part of the probability density,
\begin{align}
P_D^{\text{sy}}(z,u) &= \frac{1}{2} \Big( P_D(z,u) + P_D(-z,u) \Big) \label{sym-asym} \\
\Rightarrow &P_D^{\text{sy}}(z,u) = P_D^{\text{sy}}(-z,u) = P_D^{\text{sy}}(z,-u) \nonumber \\
P_D^{\text{as}}(z,u) &= \frac{1}{2} \Big( P_D(z,u) - P_D(-z,u) \Big) \nonumber \\
\Rightarrow &P_D^{\text{as}}(z,u) = -P_D^{\text{as}}(-z,u) = -P_D^{\text{as}}(z,-u) . \nonumber
\end{align}
\begin{figure}[ht!]
\includegraphics[width=0.47\textwidth, clip, trim=0mm 0mm 20mm 10mm]{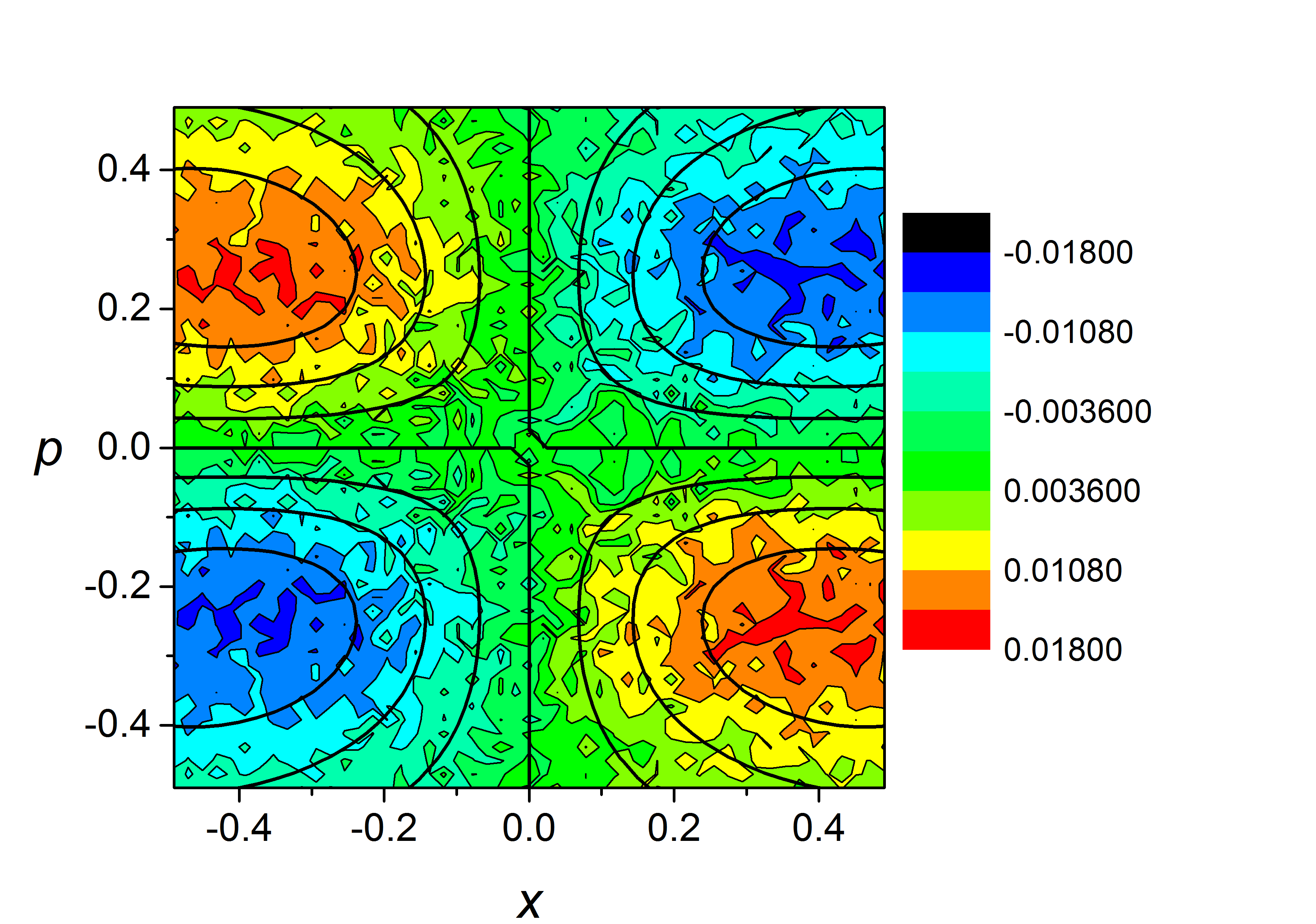}
\caption{Antisymmetric part Eq.~\eqref{sym-asym} of the phase-space probability density for $\Omega = 0.5$ and $D = 0.1$. The colored surfaces are the results of numerical Langevin simulations, the black lines represent the third-order small-$D$ expansion. \label{fig:asdist}}
\end{figure}
For the zeroth-order Boltzmann-Gibbs result, the antisymmetic part of the phase-space density vanishes.
For non-zero $D$, however, this antisymmetric part highlights a feature that is hard to discern in the total density:
the probability of position and momentum having opposite signs is increased with respect to equal signs; a particle to the right of the center of the potential is more likely to be moving left, and vice versa, see Fig.~\ref{fig:asdist}.
This kind of behavior can be encoded in a single quantity by defining
\begin{align}
\eta &= \frac{\int_{-\infty}^{0} \text{d}z \int_{0}^{\infty}\text{d}u P_D(z,u) + \int_{0}^{\infty} \text{d}z \int_{-\infty}^{0}\text{d}u P_D(z,u)}{\int_{-\infty}^{0} \text{d}z \int_{-\infty}^{0}\text{d}u P_D(z,u) + \int_{0}^{\infty} \text{d}z \int_{0}^{\infty}\text{d}u P_D(z,u)} \nonumber \\
&= 1-2 \frac{\int_{0}^{\infty} \text{d}z \int_{0}^{\infty} \text{d}u P_D^{\text{as}}(z,u)}{\int_{0}^{\infty} \text{d}z \int_{0}^{\infty} \text{d}u \Big[P_D^\text{sy}(z,u)+P_D^{\text{as}}(z,u)\Big]}, \label{asym-ratio}
\end{align}
where we split the phase-space density into symmetric and antisymmetric part.
\begin{figure}[ht!]
\includegraphics[width=0.47\textwidth, clip, trim=0mm 0mm 20mm 10mm]{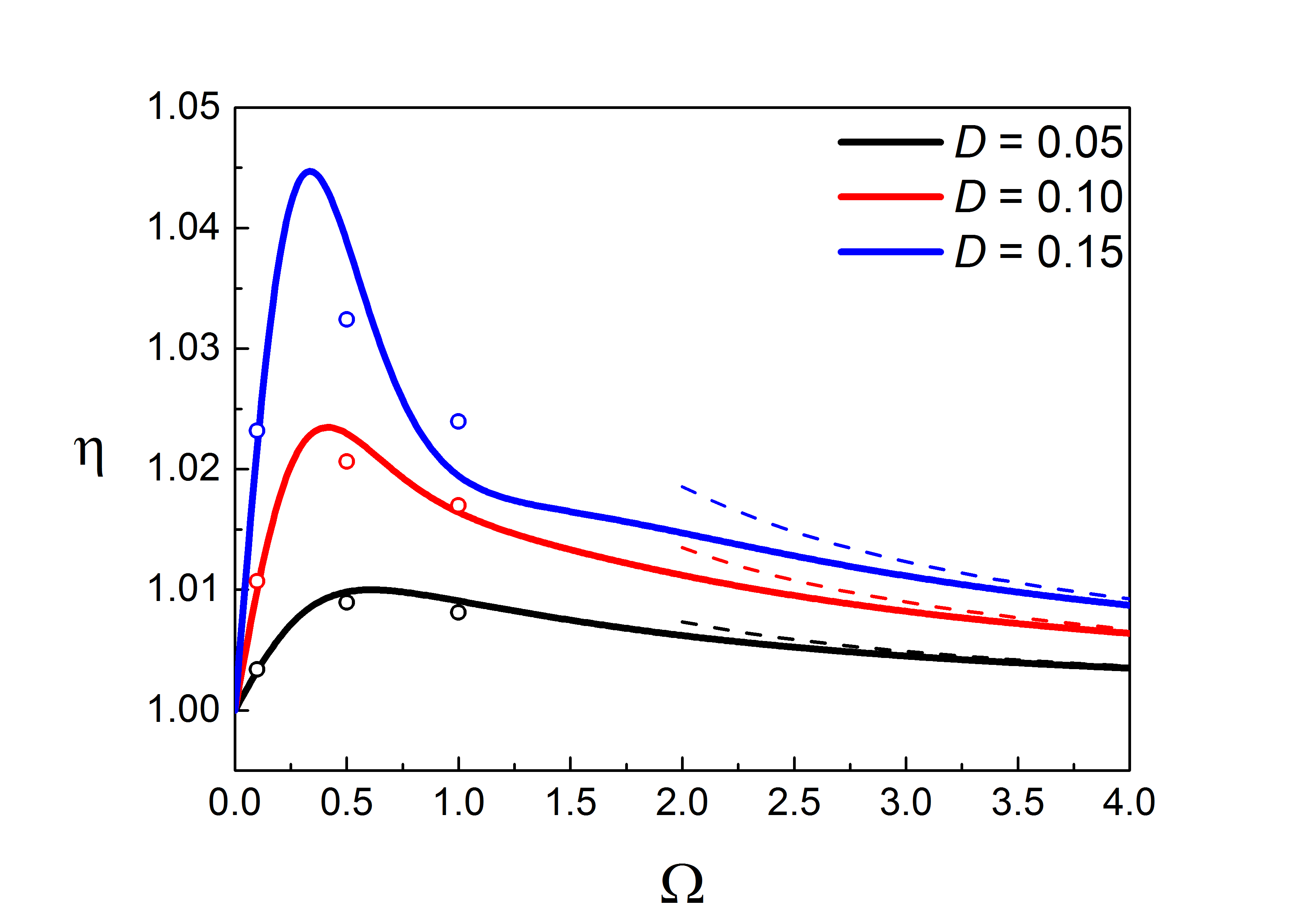}
\caption{Antisymmetry ratio $\eta$ Eq.~\eqref{asym-ratio} as a function of $\Omega$ for different values of $D$. The solid lines are the third-order result Eq.~\eqref{equipart-order3}, the circles are the result of numerical Langevin simulations and the dashed lines are the large-frequency expansion discussed in Section \ref{sec:large-freq}. \label{fig:asmeas}}
\end{figure}
The parameter $\eta$ is precisely ratio of the probabilities of having position and momentum in opposite and same directions.
Contrary to the equipartition ratio, $\eta$ has a non-monotonous behavior as a function of frequency, see Fig.~\ref{fig:asmeas}.
While for both small and large $\Omega$, having position and momentum in the same and opposite directions is equally likely, for intermediate values of $\Omega$, opposite directions are more likely.
For large $\Omega$ this agrees with the observation that the phase-space density is a function of the Hamiltonian only in this limit and thus has rotational symmetry in phase-space.
In this limit the motion of the particle can be described as oscillations in the potential with a slowly changing energy due to dissipation and diffusion, see Section \ref{sec:large-freq} for a more detailed discussion.
For small $\Omega$, on the other hand, the kinetic energy of the particle typically varies on much shorter time scales than the potential one.
Thus position and momentum are less correlated and $\eta$, which essentially measures these correlations, is small.

\section{Strong confinement: Underdamped approximation \label{sec:large-freq}}
In the previous section, we noticed that the center part of the phase-space probability density can be written as a function of the Hamiltonian for $\Omega \gg 1$, i.e.~a strong confining potential.
We now want to see whether this observation carries over to the entire distribution as well.
Our starting point is Eq.~\eqref{KFP-lattice-E}, the parameterization of the phase-space density in terms of the energy $E$ and the phase-space angle $\alpha$.
Dividing by $\Omega$, Eq.~\eqref{KFP-lattice-E} can be written as
\begin{align}
\partial_\alpha P_P(\varepsilon,\alpha) = -\frac{1}{\Omega} \mathcal{L}_{\varepsilon,\alpha} P_P(\varepsilon,\alpha) \label{KFP-lattice-E2}.
\end{align}
For $\Omega \gg 1$, we then have to leading order $\partial_\alpha P_P(\varepsilon,\alpha) = 0$.
Thus the phase-space density is to leading order independent of the angle $\alpha$.
Physically $\Omega \gg 1$ means that the oscillation frequency in the trap is much larger than the damping rate.
A particle would thus complete many oscillations before its energy changes substantially due to dissipation and its motion can be approximately described by the Hamiltonian dynamics.
In this sense, the expansion in large $\Omega$ is a weak damping limit, also referred to as the underdamped approximation \cite{str63}.
This can be made more rigorous by comparing the change in energy due to damping $\Delta E_\text{diss}$ and diffusion $\Delta E_\text{diff}$ per period of the unperturbed oscillation,
\begin{align}
\Delta E_\text{diss} = \oint_\mathcal{T} \text{d}\tilde{x} \ F_\text{fric}(\tilde{p}(\tilde{x})) .
\end{align}
Here we use the dimensionful units and denote by $\mathcal{T}$ the closed circular Hamiltonian orbit.
Employing the form for the friction force \eqref{friction} and changing to dimensionless units, we find
\begin{align}
\Delta \varepsilon_\text{diss} = -\Omega^{-1} \oint_\mathcal{T} \text{d}x \frac{p(x)}{1+p^2(x)} .
\end{align}
Since the energy $\varepsilon = (x^2+p^2)/2$ is conserved along Hamiltonian orbits, we may use this to express $p$ as a function of $x$ and obtain
\begin{align}
\Delta \varepsilon_\text{diss} &= 4 \Omega^{-1} \int_{0}^{\sqrt{2 \varepsilon}} \text{d}x \frac{\sqrt{2 \varepsilon - x^2}}{1 + 2 \varepsilon - x^2} \nonumber \\
&= 2 \pi \Omega^{-1} \bigg(1 - \frac{1}{\sqrt{1+2\varepsilon}}\bigg). \label{dissipation}
\end{align}
The energy change due to diffusion is given by the momentum diffusion coefficient,
\begin{align}
\Delta E_\text{diff} = \frac{D_0 T}{m},
\end{align}
where $T = 2 \pi/\omega$ is the period of the oscillation.
In dimensionless units, this reads
\begin{align}
\Delta \varepsilon_\text{diff} = 2 \pi D \Omega^{-1} . \label{diffusion}
\end{align}
From Eqs.~\eqref{dissipation} and \eqref{diffusion}, we conclude that the energy change due to dissipation and diffusion is proportional to $\Omega^{-1}$, which justifies the approach of using the Hamiltonian evolution with a slowly changing energy.
Comparing the relative change in energy per oscillation, we have
\begin{align}
\frac{\Delta \varepsilon_\text{diss}}{\varepsilon} &=  \frac{2 \pi \big(1 - \frac{1}{\sqrt{1+2\varepsilon}}\big)}{\Omega \ \varepsilon} \nonumber \\
\frac{\Delta \varepsilon_\text{diff}}{\varepsilon} &= \frac{2 \pi D }{\Omega \ \varepsilon} \label{dissipation-diffusion}
\end{align}
We notice that both quantities are small not only for large frequency $\Omega$, but also for large energy $\varepsilon$.
For the diffusive contribution this would be true even for linear friction.
For the dissipative term, on the other hand, for linear friction, $\Delta \varepsilon_\text{diss}/\varepsilon$ is found to be independent of $\varepsilon$ and thus small only for large frequency.
Thus for the nonlinear friction force discussed above, the system becomes underdamped both for large frequency and for large energy.
The former limit will be discussed in the following, whereas we will exploit the latter property in Section \ref{sec:large-E}.

Having assured ourselves that the underdamped approximation works for our system, we proceed to discuss what results we can obtain from it.
From the leading order behavior $\partial_\alpha P_P(\varepsilon,\alpha) = 0$, we conclude $P_P(\varepsilon,\alpha) \simeq P_\varepsilon(\varepsilon)/(2 \pi) + \mathcal{O}(\Omega^{-1})$.
Plugging this into Eq.~\eqref{KFP-lattice-E} and integrating over $\alpha$, we obtain an equation for the leading order energy density $P_\varepsilon(\varepsilon)$,
\begin{align}
\partial_\varepsilon \bigg[ 1 - \frac{1}{\sqrt{1+2\varepsilon}} + D \varepsilon \partial_\varepsilon \bigg] P_\varepsilon(\varepsilon) = 0 . \label{energy-density-FP}
\end{align}
Solving this energy diffusion equation, we find the main result of this section,
\begin{align}
P_\varepsilon(\varepsilon) &= Z_\varepsilon^{-1} \big(1 + \sqrt{1+2\varepsilon}\big)^{-\frac{2}{D}} \label{energy-density-stationary} \\
\text{with} \quad Z_\varepsilon &= \frac{2^{-\frac{2}{D}} D}{(1-D)(2-D)} \nonumber .
\end{align}
Clearly this density is very different both from the Boltzmann-Gibbs density Eq.~\eqref{BG-density} and the Tsallis-like form Eq.~\eqref{energy-PDF-free} obtained without the confining potential.
\begin{figure}[ht!]
\includegraphics[width=0.47\textwidth, clip, trim=10mm 0mm 10mm 10mm]{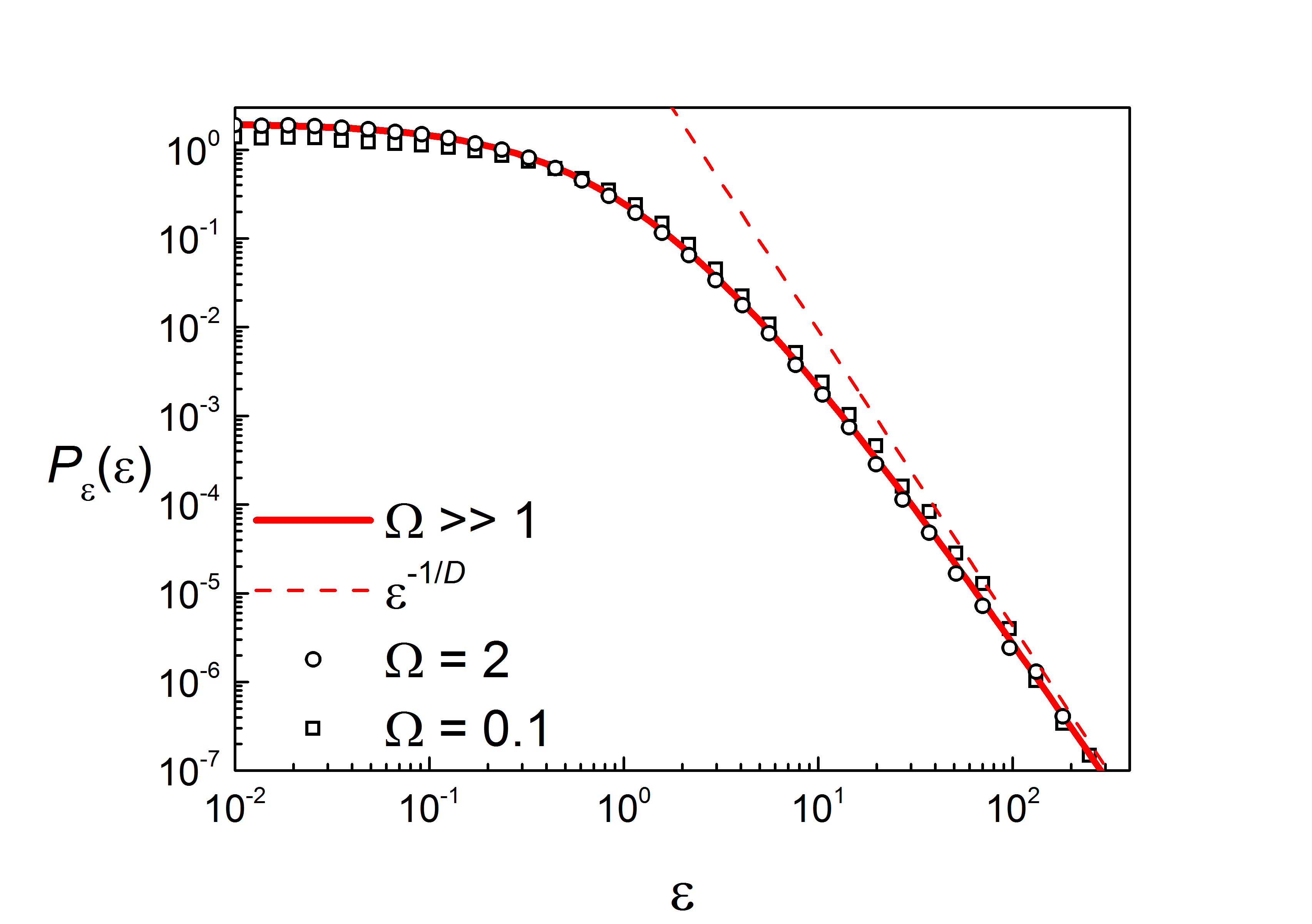}
\caption{Energy density as a function of energy for $D=0.3$ and different values of $\Omega$. The solid red line is Eq.~\eqref{energy-density-stationary}, the dashed red line the corresponding asymptotic $\varepsilon^{-1/D}$-behavior. The circles are numerical simulations for $\Omega = 2$, the squares are for $\Omega = 0.1$. \label{fig:edist}}
\end{figure}
\begin{figure}[ht!]
\includegraphics[width=0.47\textwidth, clip, trim=10mm 0mm 10mm 10mm]{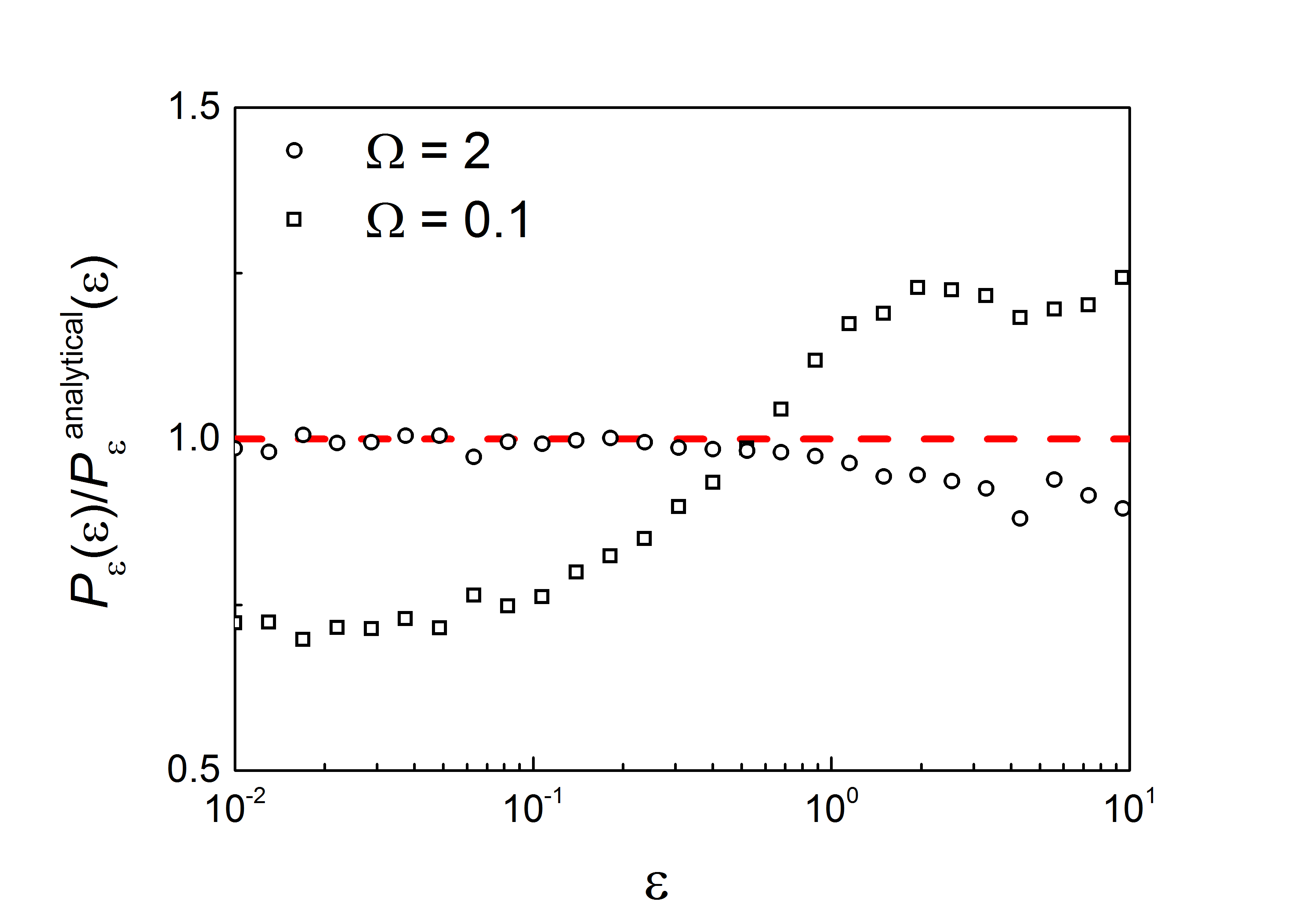}
\caption{Ratio of the numerically obtained energy density and the analytical large-frequency result Eq.~\eqref{energy-density-stationary}. Circles are for $\Omega = 2$, squares for $\Omega = 0.1$. The red dashed line indicating the value of $1$ has been added for clarity. \label{fig:edist-rel}}
\end{figure}
We compare Eq.~\eqref{energy-density-stationary} to numerical simulations in Figs.~\ref{fig:edist} and \ref{fig:edist-rel}.
Good agreement is already obtained for moderate values of $\Omega \approx 2$, suggesting that the corrections to the leading-order behavior are generally small.
This is substantiated by Fig.~\ref{fig:edist-rel}, where we show the ratio between the numerically obtained energy density and Eq.~\eqref{energy-density-stationary}.
For $\Omega = 2$ this ratio is close to $1$, whereas for $\Omega = 0.1$, we see that the likelihood of large energies is increased, in agreement with the observation from Section \ref{sec:small-D} that the total energy increases towards small $\Omega$.
As for the case without the confining potential, Eq.~\eqref{energy-PDF-free}, this stationary density has power-law tails in the energy, $P_\varepsilon(\varepsilon) \propto \varepsilon^{-1/D}$.
Importantly, the exponent of the tails is universal for large energies, even when $\Omega$ is not large, see Fig.~\ref{fig:edist}.
Compared to the free case $P_\varepsilon(\varepsilon) \propto \varepsilon^{-1/(2D)-1/2}$, however, the exponent of the tails is different, and the energy density for the confined system decays more rapidly at large energies.
Note that in both cases $D=1$ marks a transition where the stationary density is no longer normalizable and thus the solution becomes time-dependent for $D>1$.
A time-dependent solution is also needed to describe the moments of the density beyond certain values of $D$.
For example, the average energy is given by
\begin{align}
\langle \varepsilon \rangle = \frac{D(2-D)}{(2-3D)(1-2D)} . \label{average-energy}
\end{align}
This expression diverges at $D = 1/2$, and the average energy increases as a function of time for $D > 1/2$.
This behavior is akin to the divergence of the kinetic energy at $D = 1/3$ for the free case \cite{kat97,lut03,kes10}.
While we do not investigate the time-dependent solution at this point, we note that the divergence of the average energy occurs at larger values of $D$ when we include the confining potential as compared to the free case.
Since $D$ is a measure of the ratio of heating due to spontaneous emission and cooling due to the lattice, this implies that the potential actually improves the dissipation mechanism.
This is in agreement with the observation from the small-$D$ expansion in Section \ref{sec:small-D}, where we found that in the presence of the confining potential, the kinetic energy is reduced.
The divergence of the average energy at $D = 1/2$ also explains the results for the equipartition ratio depicted in Fig.~\ref{fig:equipart2}.
For $D$ slightly below $1/2$, the average of the energy is dominated by the large-energy tails of the phase-space density.
Since, as we saw before, the underdamped limit is also valid for large energies, the average energy is well-described by a phase-space density that depends only on the Hamiltonian and consequently, we have equipartition of energy in this regime.
For large $\Omega$ this holds for any $D$ and the kinetic and potential energy are equal to leading order in $\Omega$, $\langle \varepsilon_k \rangle = \langle \varepsilon_p \rangle = \langle \varepsilon \rangle/2$.

\begin{figure}[ht!]
\includegraphics[width=0.47\textwidth, clip, trim=0mm 0mm 20mm 10mm]{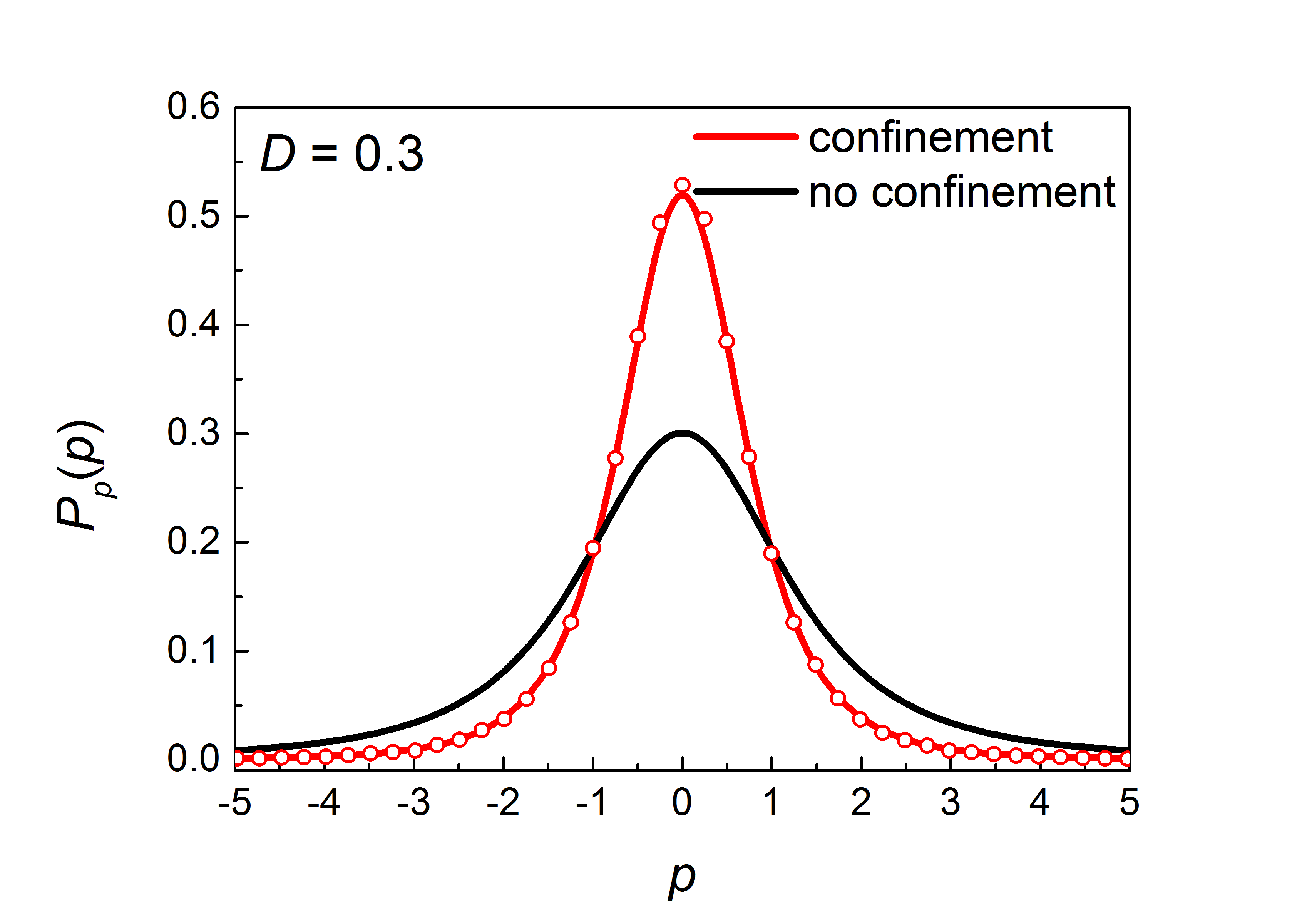}
\caption{Momentum distribution for $D = 0.3$. The black line is the result \eqref{momentum-PDF-free} without the confining potential, the red line is obtained by numerically integrating the large-$\Omega$ expansion of the phase-space density \eqref{phase-space-large-omega} over $x$. The circles are the result of Langevin simulations for $\Omega = 4$. Clearly, the confinement leads to a narrower momentum distribution, which suggests improved cooling (see Sec.~\ref{sec:experimental} for a more detailed discussion). \label{fig:pdist}}
\end{figure}
From Eq.~\eqref{energy-density-stationary}, we can immediately deduce the phase-space density for large $\Omega$,
\begin{align}
P(x,p) &= P_\varepsilon(H(x,p))/(2 \pi) \nonumber \\
& = (2 \pi Z_\varepsilon)^{-1} \big(1 + \sqrt{1+x^2+p^2}\big)^{-\frac{2}{D}} . \label{phase-space-large-omega}
\end{align}
By integrating over $p$ or $x$, respectively, we can in principle obtain the marginal densities $P_x(x)$ and $P_p(p)$.
These have no closed-form representation, however it is easy to see that their asymptotic tails behave as
\begin{align}
P_p(p) \simeq \frac{\sqrt{\pi} \Gamma\big(\frac{2-D}{2 D}\big)}{2 \pi Z_\varepsilon \Gamma\big(\frac{1}{D}\big)} |p|^{-\frac{2}{D}+1}, \label{momentum-PDF-confined}
\end{align}
and likewise for $P_x(x)$.
As with the energy density, the tails $P_p(p) \propto |p|^{-2/D+1}$ are markedly different from the free case $P_p(p) \propto |p|^{-1/D}$ Eq.~\eqref{momentum-PDF-free}.
The resulting momentum distribution is shown in Fig.~\ref{fig:pdist}.

So far, we have focused on the leading order term, which depends only on energy.
We expand the full phase space density with respect to $\Omega$, we obtain
\begin{align}
P_P(\varepsilon,\alpha) = \frac{P_\varepsilon(\varepsilon)}{2 \pi} \Bigg[1 + \frac{f_1(\varepsilon,\alpha)}{\Omega} + \mathcal{O}(\Omega^{-2}) \Bigg]. \label{large-omega-expansion}
\end{align}
The function $f_1(\varepsilon,\alpha)$ represents the (generally angle-dependent) first order correction to the angle-independent leading order result. 
Plugging this expansion into the Kramers-Fokker-Planck-equation \eqref{KFP-lattice-E2}, we find an equation for $f_1(\varepsilon,\alpha)$ by equating the coefficients of the terms of order $\Omega^{-1}$,
\begin{align}
\partial_\alpha f_1(\varepsilon,\alpha) &= - \frac{\mathcal{L}_{\varepsilon,\alpha} P_\varepsilon(\varepsilon)}{P_\varepsilon(\varepsilon)} \label{f_1-equation} \\
\Rightarrow f_1(\varepsilon,\alpha) &= - \frac{1}{P_\varepsilon(\varepsilon)} \int_{0}^{\alpha} \text{d}\alpha' \mathcal{L}_{\varepsilon,\alpha'} P_\varepsilon(\varepsilon) + \tilde{f}_1(\varepsilon) \nonumber  .
\end{align}
Here $\tilde{f}_1(\varepsilon)$ is some function that depends only on $\varepsilon$.
Performing the integral over $\alpha$, we get
\begin{widetext}
\begin{align}
f_1(\varepsilon,\alpha) &= -\frac{\sin(\alpha)\cos(\alpha)}{1+2\varepsilon \sin^2(\alpha)} - \frac{\phi(\varepsilon,\alpha)}{(1+2\varepsilon)^{\frac{3}{2}}} - \frac{\sin(2\alpha)}{2(1+2\varepsilon)(\varepsilon(\cos(2\alpha)-1)-1)} - \Bigg[\alpha - \frac{\phi(\varepsilon,\alpha)}{\sqrt{1+2\varepsilon}} + D \cos(\alpha) \sin(\alpha) \Bigg] \frac{\partial_\varepsilon P_\varepsilon(\varepsilon)}{P_\varepsilon(\varepsilon)} \nonumber \\
&\qquad \qquad - D \Bigg[\alpha-\cos(\alpha)\sin(\alpha)\Bigg] \frac{\partial_\varepsilon \varepsilon \partial_\varepsilon P_\varepsilon(\varepsilon)}{P_\varepsilon(\varepsilon)} + \tilde{f}_1(\varepsilon). \label{anglefunc1}
\end{align}
\end{widetext}
Here, we introduced the function
\begin{align}
\phi(\varepsilon,\alpha) = \arctan\Big(\sqrt{1+2\varepsilon}\tan(\alpha)\Big) + \pi \Big\lfloor \frac{\alpha}{\pi} + \frac{1}{2} \Big\rfloor ,
\end{align}
where $\lfloor x \rfloor$ denotes the floor function, i.e.~the largest integer smaller than $x$.
This function removes the discontinuities in $\arctan(\tan(\alpha))$ introduced by the divergence of $\tan(\alpha)$ at $\alpha = (2k+1)\pi/2$.
Equation \eqref{anglefunc1} also places a condition on $P_\varepsilon(\varepsilon)$.
In order to be consistent, we require $f_1(\varepsilon,\alpha) = f_1(\varepsilon,\alpha+2\pi)$.
Using the above definition of the function $\phi(\varepsilon,\alpha)$, we find from Eq.~\eqref{anglefunc1}
\begin{align}
0 &= f_1(\varepsilon,\alpha)-f_1(\varepsilon,\alpha + 2 \pi) \nonumber \\
& = \frac{2\pi}{P_\varepsilon(\varepsilon)}\bigg( \partial_\varepsilon \Big(1-\frac{1}{\sqrt{1+2\varepsilon}}\Big) P_\varepsilon(\varepsilon) + D \partial_\varepsilon \varepsilon \partial_\varepsilon P_\varepsilon(\varepsilon) \bigg). \label{pe-consistency}
\end{align}
Taking the derivative outside the parentheses, we recover precisely Eq.~\eqref{energy-density-FP} and thus the solution \eqref{energy-density-stationary} is consistent.
Plugging in Eq.~\eqref{energy-density-stationary}, we find for the corresponding terms in Eq.~\eqref{anglefunc1}
\begin{align}
\frac{\partial_\varepsilon P_\varepsilon(\varepsilon)}{P_\varepsilon(\varepsilon)} &= -\frac{2}{D(1+2\varepsilon + \sqrt{1+2\varepsilon})} \nonumber \\
\frac{\partial_\varepsilon \varepsilon \partial_\varepsilon P_\varepsilon(\varepsilon)}{P_\varepsilon(\varepsilon)} &=  \frac{4\varepsilon \sqrt{1+2\varepsilon}-2D(1+\varepsilon+\sqrt{1+2\varepsilon})}{D^2(1+2\varepsilon)^{\frac{3}{2}}(1+\sqrt{1+2\varepsilon})^2}
\end{align}
Before we proceed to discuss this result, let us determine the as of yet unknown function $\tilde{f}_1(\varepsilon)$.
In order to find it, we require a consistency condition like \eqref{pe-consistency}, which we obtain from the second order of the large-$\Omega$ expansion.
For the latter, we have the equation
\begin{align}
\partial_\alpha f_2(\varepsilon,\alpha) = - \frac{1}{P_\varepsilon(\varepsilon)} \mathcal{L}_{\varepsilon,\alpha} \Big[ f_1(\varepsilon,\alpha) P_\varepsilon(\varepsilon) \Big] \label{f2}.
\end{align}
In particular, we want $f_2(\varepsilon,\alpha)$ to be periodic in $\alpha$, so we necessarily have
\begin{align}
\int_{0}^{2\pi} \text{d}\alpha' \mathcal{L}_{\varepsilon,\alpha'} \Big[f_1(\varepsilon,\alpha') P_\varepsilon(\varepsilon) \Big] = 0.
\end{align}
Since the entire problem is $\pi$-periodic (corresponding to $(x,p) \rightarrow (-x,-p)$), we only need to consider the integral from $0$ to $\pi$.
We split the integral at $\alpha' = \pi/2$,
\begin{align}
0 &= \int_{0}^{\pi/2} \text{d}\beta \ \mathcal{L}'_{\varepsilon,\beta} \Big[f_1\Big(\varepsilon,\frac{\pi}{2}-\beta\Big) P_\varepsilon(\varepsilon) \Big] \nonumber \\
& \quad + \int_{0}^{\pi/2} \text{d}\gamma \ \mathcal{L}''_{\varepsilon,\gamma} \Big[f_1\Big(\varepsilon,\frac{\pi}{2}+\gamma\Big) P_\varepsilon(\varepsilon) \Big],
\end{align}
where we introduced the new variables $\beta = \pi/2-\alpha$ and $\gamma = \alpha-\pi/2$.
Here $\mathcal{L}'_{\varepsilon,\beta}$ and $\mathcal{L}''_{\varepsilon,\gamma}$ are the correspondingly transformed operators.
From the symmetry of $\mathcal{L}_{\varepsilon,\alpha}$ (see Eq.~\eqref{KFP-lattice-E}), it can easily be seen that the operator transforms in the same manner for $\beta$ and $\gamma$, so that $\mathcal{L}'_{\varepsilon,\alpha} = \mathcal{L}''_{\varepsilon,\alpha}$ and we can thus write
\begin{align}
0 &= \int_{0}^{\pi/2} \text{d}\beta \mathcal{L}'_{\varepsilon,\beta} \Big[\Big(f_1\Big(\varepsilon,\frac{\pi}{2}-\beta\Big)+f_1\Big(\varepsilon,\frac{\pi}{2}+\beta\Big)\Big) P_\varepsilon(\varepsilon) \Big].
\end{align}
Now, the angle-dependent part of Eq.~\eqref{anglefunc1} is antisymmetric around $\alpha = \pi/2$.
While this is not immediately apparent, it can easily be shown after a little algebra.
Thus this angle dependent part drops out of the integral and we have
\begin{align}
0 &= \int_{0}^{\pi/2} \text{d}\alpha \ \mathcal{L}_{\varepsilon,\alpha} \Big[\tilde{f}_1(\varepsilon) P_\varepsilon(\varepsilon) \Big],
\end{align}
where we reverted to the original angle $\alpha$.
This gives us
\begin{align}
\partial_\varepsilon \bigg[ 1 - \frac{1}{\sqrt{1+2\varepsilon}} + D \varepsilon \partial_\varepsilon \bigg]\tilde{f}_1(\varepsilon) P_{\varepsilon}(\varepsilon) = 0, \label{f1-energy}
\end{align}
which is exactly Eq.~\eqref{energy-density-FP} but now for $\tilde{f}_1(\varepsilon) P_{\varepsilon}(\varepsilon)$.
An obvious solution to this equation is $\tilde{f}_1(\varepsilon) = C$ where $C$ is an arbitrary constant.
Demanding that the first order phase-space density is normalized then immediately yields $C=0$.
In principle, there exists a second solution to Eq.~\eqref{f1-energy}, however, this can be shown to be non-normalizable.
Thus the angle-independent part $\tilde{f}_1(\varepsilon)$ in Eq.~\eqref{anglefunc1} is indeed zero.

Armed with this knowledge, we can proceed to examine the solution \eqref{anglefunc1} in more detail.
\begin{figure}[ht!]
\includegraphics[width=0.47\textwidth, clip, trim=0mm 0mm 20mm 10mm]{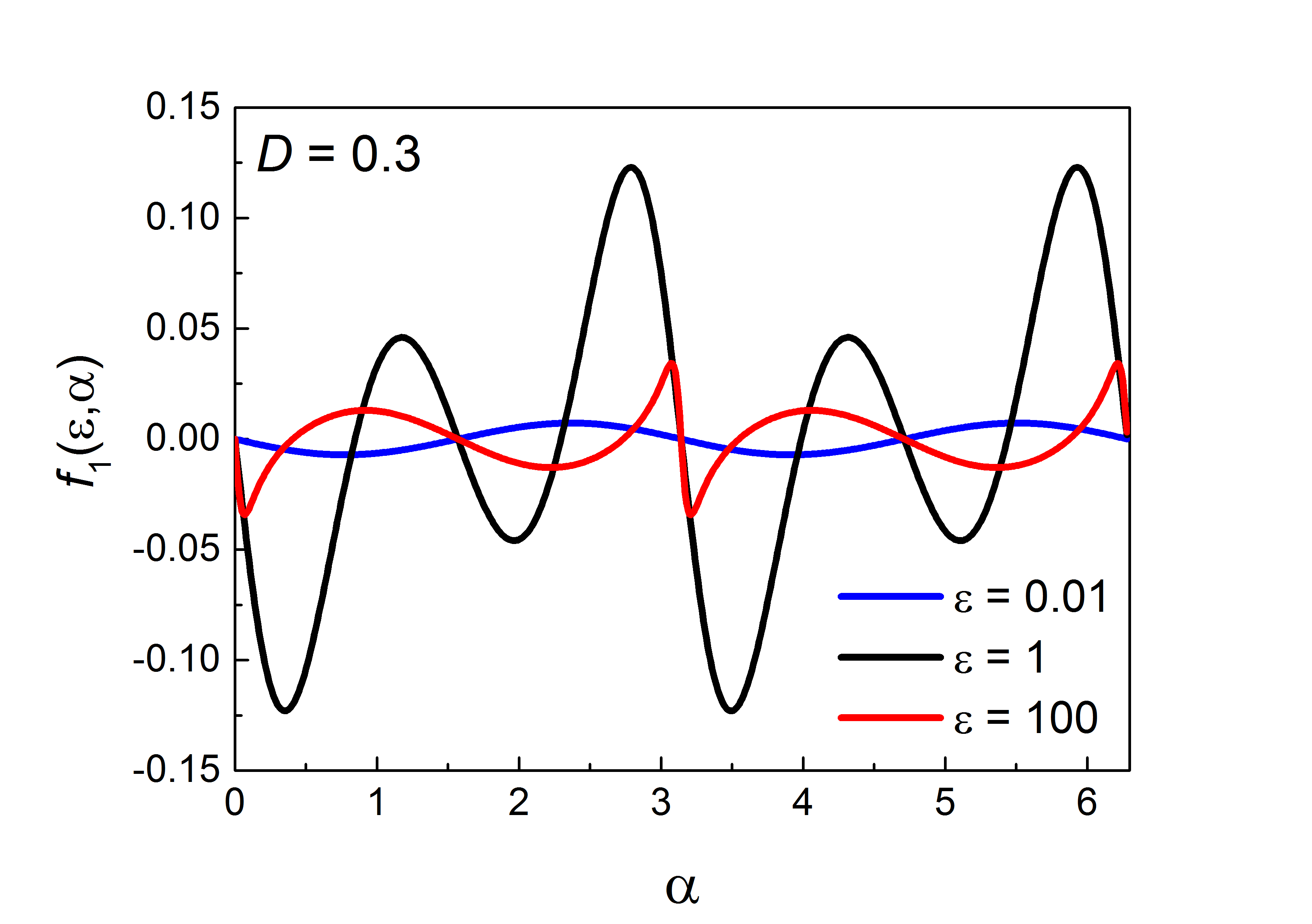}
\caption{First order angle-dependent correction to the phase-space density as a function of the angle for different energies and $D=0.3$. \label{fig:anglefunc}}
\end{figure}
The function $f_1(\varepsilon,\alpha)$ is shown in Fig.~\ref{fig:anglefunc}.
First of all, we note that the correction is small both for large and for small energies and thus the dependence on the angle $\alpha$ is strongest for intermediate energies.
For small energies, the friction force is almost linear and we recover the Boltzmann-Gibbs result with small corrections, see Section \ref{sec:small-D}.
%behavior of the correction mirrors what we already saw from the small-$D$ expansion: The probability to have momentum and position in opposite directions (corresponding to $\pi/2 < \alpha < \pi$ and $3\pi/2 < \alpha < 2 \pi$) is enhanced compared to the probability of finding them in the same direction.
%For larger energies the phase-space structure is more intricate and the probability density is very sensitive to the precise angle.
For large energies, we can expand the expression \eqref{anglefunc1} and find
\begin{align}
f_1(\varepsilon,\alpha) \simeq \frac{-D \cot(\alpha)+(1+D)\sin(2\alpha)}{2 D \varepsilon} + \mathcal{O}(\varepsilon^{-\frac{3}{2}}). \label{anglefunc-large-E}
\end{align}
We see that indeed the function is of order $\varepsilon^{-1}$ for large energies.
However, the resulting expression diverges at $\alpha = 0$ and all $\alpha = \pi$, which would lead to (nonintegrable) divergences in the probability density.
To resolve this issue, let us expand around $\alpha = 0$ instead,
\begin{align}
f_1(\varepsilon,\alpha) \simeq -\frac{1+2\varepsilon+2\sqrt{1+2\varepsilon}}{(1+2\varepsilon)(1+\sqrt{1+2\varepsilon})^2} 2 \varepsilon \alpha + \mathcal{O}(\alpha^2). \label{anglefunc-small-alpha}
\end{align}
This expression is of course regular at $\alpha = 0$, but tends to $-\alpha$ for large $\varepsilon$, in contrast to the behavior observed in Fig.~\ref{fig:anglefunc}.
So, which of the two expansions is correct?
The answer is of course, both, depending on the relative size of $\alpha$ and $\varepsilon$, as the limits $\varepsilon \rightarrow \infty$ and $\alpha \rightarrow 0$ do not commute.
As long as $\alpha$ is farther away than $1/\sqrt{2 \varepsilon}$ from $0$ or $\pi$, the large-energy expansion $\eqref{anglefunc-large-E}$ yields the correct result.
However, as we approach one of the points $\alpha = 0$ or $\alpha = \pi$, we have to use the small-angle expansion $\eqref{anglefunc-small-alpha}$.
These non-commuting limits can be glimpsed from Fig.~\ref{fig:anglefunc}, where we see sharp features emerging at $\alpha = 0$ and $\pi$ for large energies.
Physically, these features stem from the nonlinearity of the friction force.
As a high-energy particle oscillates in the confining potential, it is fast during most of its orbit, and the dissipation is weak, leading to an overall decrease of the effects with energy.
Close to the turning points $\alpha = 0, \pi$, however, the particle slows down and the friction increases, up to the point where it becomes Stokes-like.
Thus, even at high energies, there always remains a small section of phase space where the friction is relevant.

\begin{figure}[ht!]
\includegraphics[width=0.47\textwidth, clip, trim=0mm 0mm 45.5mm 20mm]{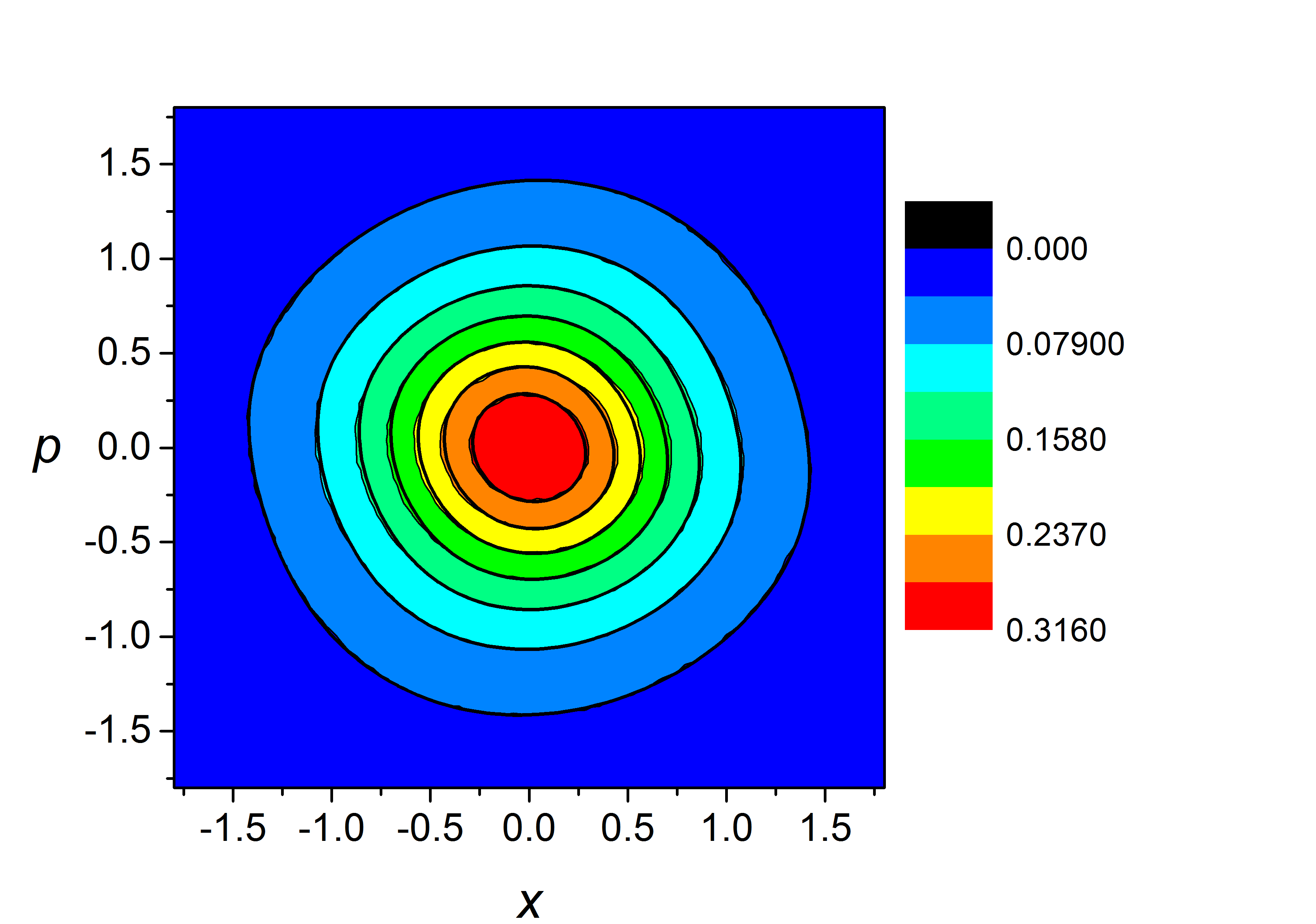}\\
\includegraphics[width=0.47\textwidth, clip, trim=0mm 0mm 35mm 20mm]{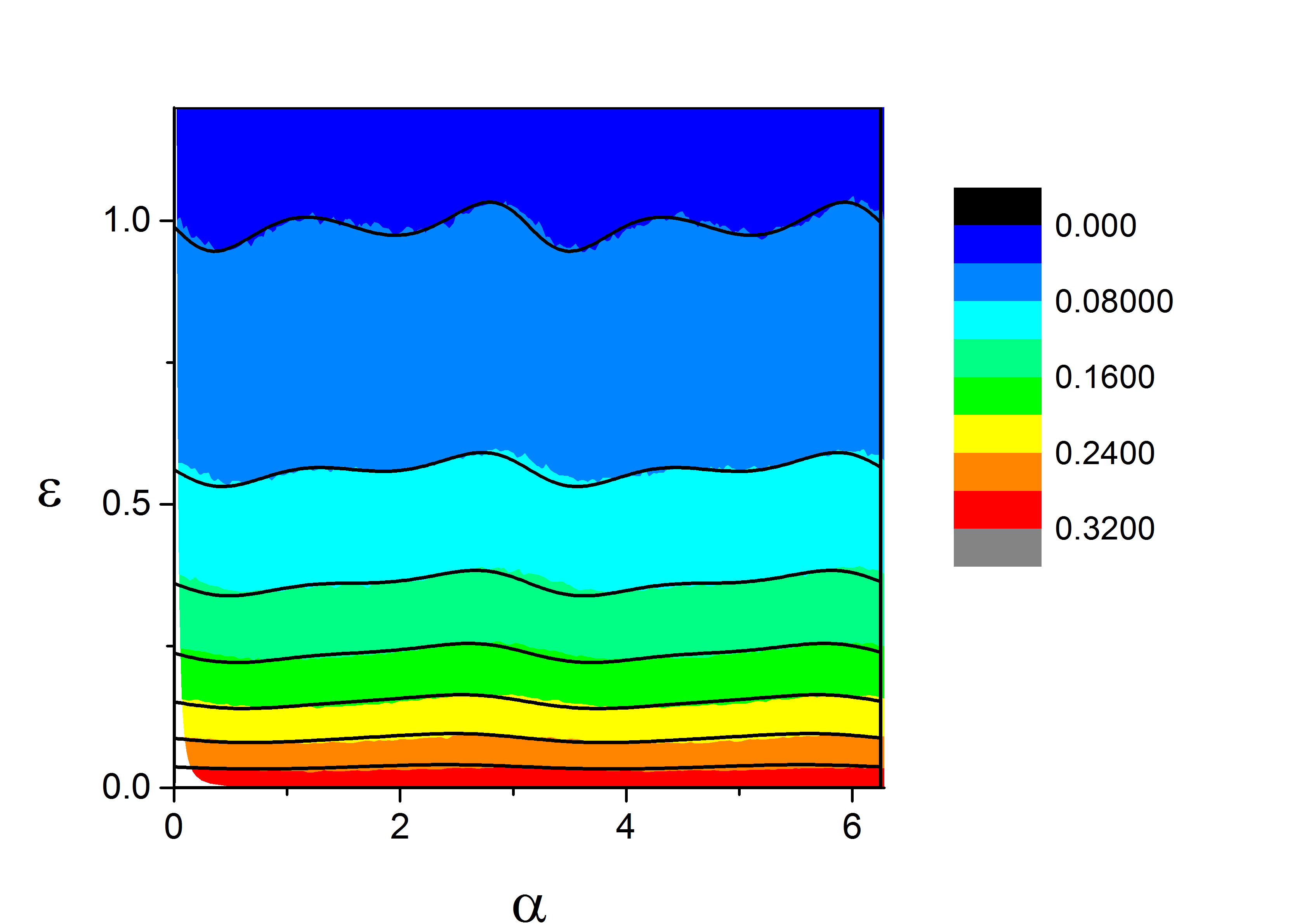}
\caption{Phase-space probability density for $D = 0.3$ and $\Omega = 2$, plotted using the position and momentum $(x,p)$ (top) respectively energy and phase-space angle $(E,\alpha)$ (bottom). The colored surfaces are the results of numerical Langevin simulations, the black lines correspond to the first-order large-$\Omega$ expansion Eq.~\eqref{large-omega-expansion}. Both results agree well and exhibt an obvious asymmetry in phase-space.  \label{fig:phasedist}}
\end{figure}
The phase-space density up to first order in $\Omega$ is shown in Fig.~\ref{fig:phasedist}.
Even for moderately large values of $\Omega$, this first-order result captures the probability density very well, since the correction term $f_1(\varepsilon,\alpha)$ by itself is small.
\begin{figure}[ht!]
\includegraphics[width=0.47\textwidth, clip, trim=0mm 0mm 35mm 20mm]{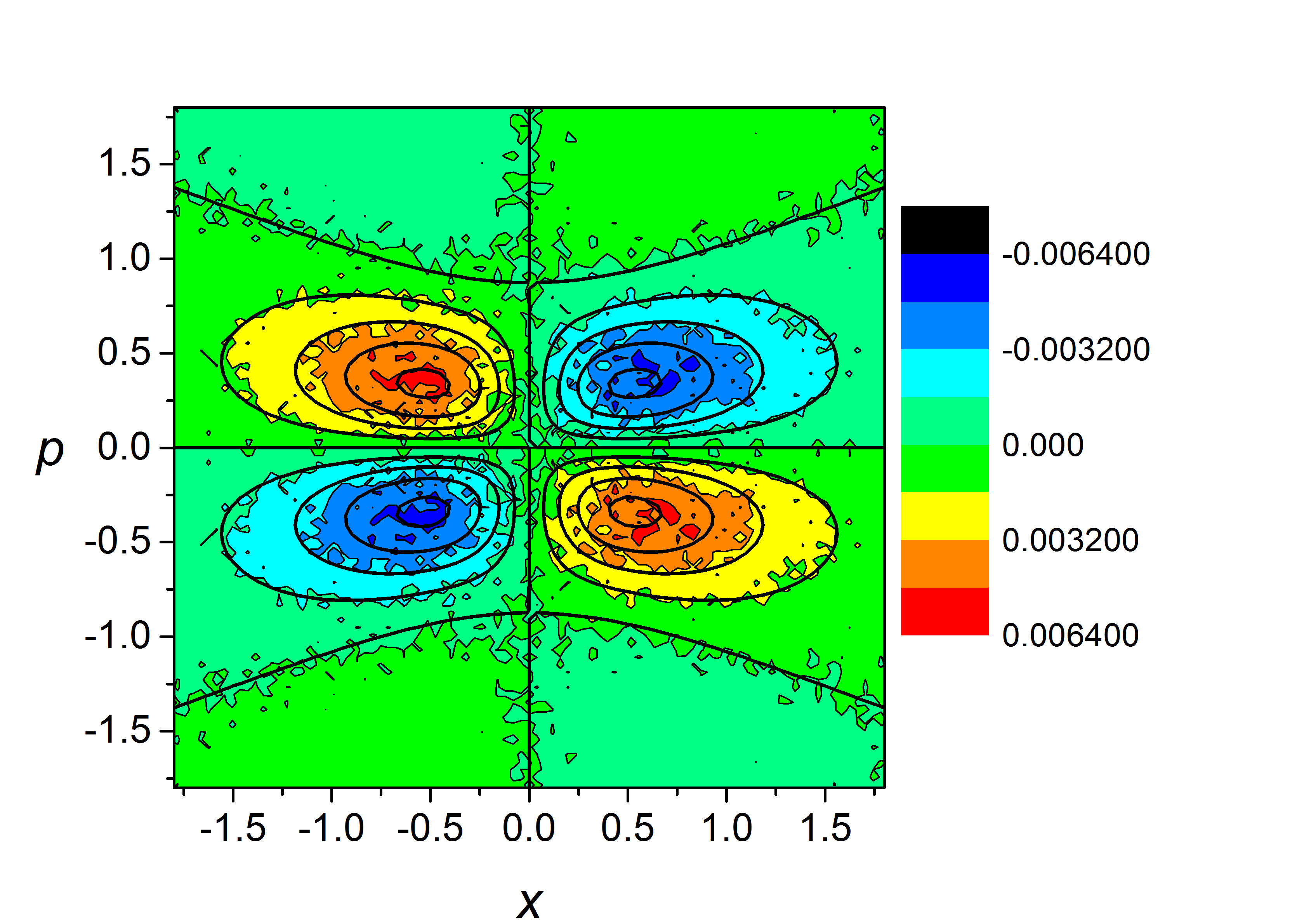}
\caption{Asymmetric part of the phase-space probability density for $D = 0.3$ and $\Omega = 2$. The colored surfaces are the results of numerical Langevin simulations, the black lines correspond to the first-order large-$\Omega$ expansion Eq.~\eqref{large-omega-expansion}.  \label{fig:phasedistas}}
\end{figure}
In Fig.~\ref{fig:phasedistas}, we show the corresponding antisymmetric part, as defined in Eq.~\eqref{sym-asym}.
The general features are very similar to the results from the small-$D$ expansion, which is no longer valid at these moderate values of $D$ (here $D=0.3$).
This hints at the generality of the asymmetry in phase-space and the deviations from equipartition, even beyond the range of validity of the respective approximations.
Note that, similarly to the small-$D$-expansion, the violation of energy equipartition is a second-order effect that is not included in Eq.~\eqref{anglefunc1}.
While it is in principle also possible to obtain higher order corrections, the calculations for obtaining them are much more involved.
\begin{figure}[ht!]
\includegraphics[width=0.47\textwidth, clip, trim=0mm 0mm 20mm 10mm]{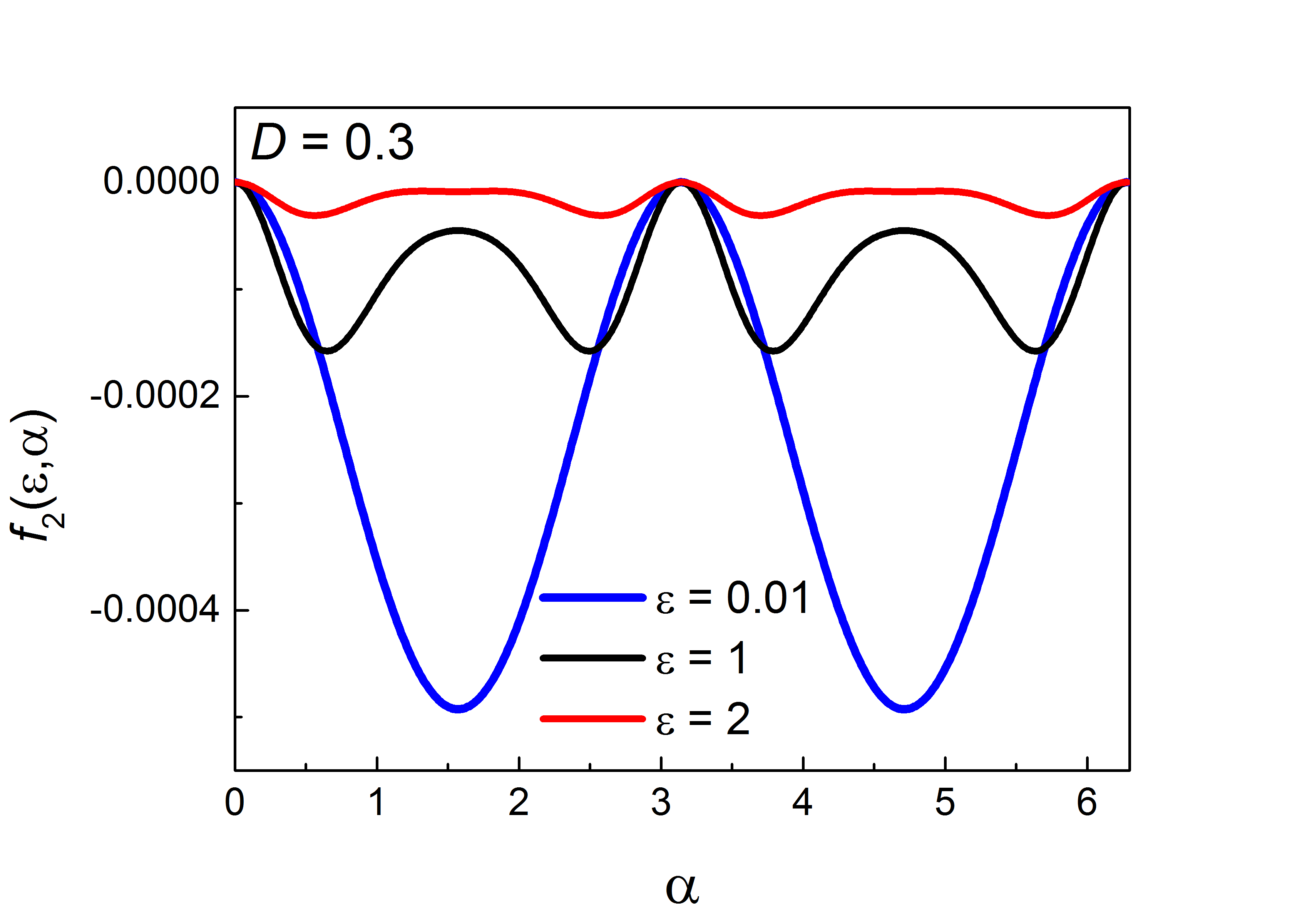}
\caption{Second order angle-dependent correction to the phase-space density as a function of the angle for different energies and $D=0.3$. \label{fig:anglefunc2}}
\end{figure}
In Fig.~\ref{fig:anglefunc2}, we show the result for the second order correction obtained by numerically solving Eq.~\eqref{f2}.
This does indeed exhibit the expected feature, an increased probability along the $x$-axis ($\alpha = 0$ and $\alpha = \pi$), which is responsible for the enhancement of the potential energy compared to the kinetic one.
Besides the fact that violations of equipartition are of order $\Omega^{-2}$, we note that the coefficient $f_2(\varepsilon,\alpha)$ of the second order term is numerically much smaller than the first order one.
This explains why we see good agreement with first order result even at moderate $\Omega$, see Figs.~\ref{fig:phasedist} and \ref{fig:phasedistas}.

\section{Large energies: Asymptotic power laws \label{sec:large-E}} 
In the previous Section, we discussed that the dynamics becomes underdamped not only for large frequencies (strong trapping) but also generically for large energies, see Eq.~\eqref{dissipation-diffusion} and discussion following it.
In this Section we want to formalize this and show that many of the features of the large-$\Omega$ expansion in fact carry over to the large-energy behavior at arbitrary $\Omega$.
We adopt a similar formalism as before, this time, however, expanding in terms of energy,
\begin{align}
P_P(\varepsilon,\alpha) \simeq N \varepsilon^{\beta} \Bigg[1 + \frac{g_{1/2}(\alpha)}{\varepsilon^{\frac{1}{2}}} + \frac{g_1(\alpha)}{\varepsilon} + \mathcal{O}(\varepsilon^{\frac{3}{2}}) \Bigg] , \label{large-energy-expansion-outside}
\end{align}
with some normalization constant $N$.
This expansion warrants some explanation.
Since we know that the system becomes underdamped for large energies, we anticipate that, similar to the result \eqref{energy-density-stationary}, the phase-space density is to leading order independent of the angle $\alpha$.
Another way to see this is that the operator $\mathcal{L}_{\varepsilon,\alpha}$ in Eq.~\eqref{KFP-lattice-E} is apparently of order $E^{-1}$ for large energies, so that we have $\Omega \partial_\alpha P(\varepsilon,\alpha) = \mathcal{O}(E^{-1})$.
The power-law behavior of the leading order term and half-integer order terms in the expansion are motivated by the large-energy expansion of the stationary energy density \eqref{energy-density-stationary},
\begin{align}
P_\varepsilon(\varepsilon) \simeq Z_\varepsilon^{-1} (2 \varepsilon)^{-\frac{1}{D}} \bigg( 1 - \frac{\sqrt{2}}{D \varepsilon^{\frac{1}{2}}} + \frac{1}{D^2 \varepsilon} + \mathcal{O}(\varepsilon^{-\frac{3}{2}}) \bigg) \label{large-omega-large-e},
\end{align}
where similar terms occur.
We plug the expansion \eqref{large-energy-expansion-outside} into the Kramers-Fokker-Planck equation \eqref{KFP-lattice-E} and expand for large energies, resulting in an equation containing powers of $\varepsilon$.
Equating orders of $\varepsilon$, we find conditions on the functions $g_{1/2}(\alpha)$, $g_1(\alpha)$ and so on.
Up to order $\varepsilon^{-1}$, these read
\begin{align}
\partial_\alpha &g_{1/2}(\alpha) = 0 \nonumber \\
\partial_\alpha &g_{1}(\alpha) = \frac{1}{2 \Omega} \Big( - (1+2\beta)(2 D \beta \sin^2(\alpha) - 1) \nonumber \\
&\qquad + D \beta \cos^2(\alpha) + \cot^2(\alpha)  \Big). 
\end{align}
The first condition obviously demands that $g_{1/2}(\alpha) = \tilde{g}_{1/2}$ is a constant, while the second one gives us after integration
\begin{align}
g_1(\alpha) &= -\frac{1}{2 \Omega} \Big( \cot(\alpha) + \beta(2 \alpha(D \beta-1) \nonumber \\
&\qquad - D(1+\beta) \sin(2\alpha) ) \Big) + \tilde{g}_1 . \label{large-E-order1}
\end{align}
Demanding that $g_1(\alpha)$ should be $2\pi$-periodic, we get a condition on $\beta$ that yields $\beta = 0$ or $\beta = -1/D$.
Since the first solution is non-normalizable at large energies, $\beta = -1/D$ is the relevant solution.
This shows that the power-law tails $P_\varepsilon(\varepsilon) \propto \varepsilon^{-1/D}$ are in fact the universal large energy behavior.
Apart from the constant $\tilde{g}_1$, the above is then completely equivalent to the large-energy expansion of the large-$\Omega$ solution, Eq.~\eqref{anglefunc-large-E}, which is reassuring.
The constants $\tilde{g}_{1/2}$ and $\tilde{g}_1$ have to be determined from next order in the expansion.

However, Eq.~\eqref{large-E-order1} has the same issue that we already observed with Eq.~\eqref{anglefunc-large-E}:
It diverges at $\alpha = 0, \pi$.
The reason for this is that we expanded $\mathcal{L}_{\varepsilon,\alpha}$ for large $\varepsilon$ without paying attention to $\alpha$. 
But as we saw before, the limits $\varepsilon \rightarrow \infty$ and $\alpha \rightarrow 0$ do not commute.
Because of this, we need to perform a similar expansion, but in the region where $\alpha$ is close to $0$ or $\pi$, i.e.~where close to the turning points, where the energy is large but the momentum is not.
In this region, however, the position $x$ is of order $\sqrt{\varepsilon}$, so we can expand Eq.~\eqref{KFP-lattice} for large $x$.
Recalling equation \eqref{KFP-lattice},
\begin{align}
\Bigg[ \Omega \bigg( - p \partial_x + x \partial_p \bigg) +  \partial_p \bigg(\frac{p}{1+p^2} + D \partial_p \bigg) \Bigg] P_S(x,p) = 0 , \label{KFP-lattice-p}
\end{align}
where we explicitly denote the solution inside the strip by $P_S$.
we see that for $x$ large and $p$ of order $1$, the very first term is of order $1/x$, the second one of order $x$ and the final two terms are of order $1$.
Note that $P_P(\varepsilon,\alpha) = P_S(\sqrt{2 \varepsilon} \cos(\alpha),\sqrt{2 \varepsilon} \sin(\alpha))$, i.~e.~the Jacobian of the variable transformation is unity.
We write down an expansion for the phase-space density similar to Eq.~\eqref{large-energy-expansion-outside},
\begin{align}
P_S(x,p) \simeq M |x|^{\gamma} \Bigg[1 + \frac{h_{1/2}(p)}{x} + \frac{h_1(p)}{x^2} + \mathcal{O}(x^{-3}) \Bigg] , \label{large-energy-expansion-inside}
\end{align}
where $M$ is a normalization constant.
We plug this into Eq.~\eqref{KFP-lattice-p}, expand for large $x$ and evaluate the coefficients,
\begin{align}
\Omega \partial_p &h_{1/2}(p) = -\partial_p \frac{p}{1+p^2} \nonumber \\
\Omega \partial_p &h_{1}(p) = -\partial_p \bigg(\frac{p}{1+p^2} + D \partial_p \bigg) h_{1/2}(p).
\end{align}
Solving for $h_{1/2}(p)$ and $h_1(p)$, we have
\begin{align}
&h_{1/2}(p) = -\frac{1}{\Omega} \frac{p}{1+p^2} + \tilde{h}_{1/2} \nonumber \\
&h_1(p) = \frac{D + p^2(1-D)}{\Omega^2 (1+p^2)^2} - \frac{\tilde{h}_{1/2} \ p}{\Omega(1+p^2)} + \tilde{h}_1 .
\end{align}
So far, the two expansions Eqs.~\eqref{large-energy-expansion-outside} and \eqref{large-energy-expansion-inside} are independent of each other.
Whereas Eq.~\eqref{large-energy-expansion-outside} is valid at most points in phase space, where $|p| \gg 1$ (or equivalently $|\alpha| \gg 1/\sqrt{2\varepsilon}$), Eq.~\eqref{large-energy-expansion-inside} describes the strip where $|p| \lesssim 1$ (or equivalently $|\alpha| \lesssim \sqrt{2\varepsilon}$).
However, they describe an expansion of the same function in these different areas of phase-space, and consequently should be related.
In particular taking the $|p| \rightarrow \infty$ limit of Eq.~\eqref{large-energy-expansion-inside} inside the strip should match onto the $\alpha \rightarrow 0$ limit of Eq.~\eqref{large-energy-expansion-outside} outside the strip.
From the zeroth-order term this immediately gives $\gamma = 2 \beta = -2/D$ and $M = 2^{1/D} N$.
Further expanding $g_{1/2}$ and $g_1(\alpha)$ around $\alpha = 0$ gives
\begin{align}
&g_{1/2}(\alpha) \varepsilon^{-\frac{1}{2}} = \tilde{g}_{1/2} \varepsilon^{-\frac{1}{2}} \nonumber \\
&g_1(\alpha) \varepsilon^{-1} \simeq \Big[\tilde{g}_1 -\frac{1}{2 \Omega \alpha} +  \mathcal{O}(\alpha) \Big] \varepsilon^{-1}. \label{outside-inside}
\end{align}
We need to compare this to the large-$p$ expansion of $h_{1/2}(p)$ and $h_{1}(p)$,
\begin{align}
&h_{1/2}(p) x^{-1} \simeq \Big[\tilde{h}_{1/2} - \frac{1}{\Omega} \Big( \frac{1}{p} - \frac{1}{p^3} + \mathcal{O}(p^{-5}) \Big)\Big] x^{-1} \nonumber \\
&h_1(p) x^{-2} \simeq \Big[\tilde{h}_1 + \frac{1}{\Omega^2} \frac{1-D}{p^2} - \frac{\tilde{h}_{1/2}}{\Omega} \Big(\frac{1}{p} - \frac{1}{p^3} \Big) \nonumber \\
& \qquad \qquad \qquad \qquad + \mathcal{O}(p^{-4}) \Big] x^{-2}.
\end{align}
Close to $\alpha = 0$, we further have $p \simeq \sqrt{2 \varepsilon} (\alpha + \mathcal{O}(\alpha^3))$ and $x \simeq \sqrt{2 \varepsilon} (1 + \mathcal{O}(\alpha^2))$  and thus
\begin{align}
&h_{1/2}(p) (2 \varepsilon)^{-\frac{1}{2}} \simeq \tilde{h}_{1/2} (2 \varepsilon)^{-\frac{1}{2}} - \frac{1}{2\Omega} \frac{1}{\alpha \varepsilon} + \mathcal{O}(\varepsilon^{-3/2}) \nonumber \\
&h_1(p) (2\varepsilon)^{-1} \simeq \tilde{h}_1 (2\varepsilon)^{-1} + \mathcal{O}(\varepsilon^{-\frac{3}{2}}). \label{inside-outside}
\end{align}
Matching the coefficients of different orders in $\varepsilon$ between Eqs.~\eqref{outside-inside} and \eqref{inside-outside} then connects the integration constants of the two expansions: $\tilde{h}_{1/2} = \tilde{g}_{1/2}/\sqrt{2}$ and $\tilde{h}_1 = \tilde{g}_1/2$.
As remarked before, we need to go to higher orders in $\varepsilon$ to find the integration constants $\tilde{g}_{1/2}$ and $\tilde{g}_1$.
This is similar to the function $\tilde{f}_1(\varepsilon)$ we found for the large-$\Omega$ expansion in Eq.~\eqref{anglefunc1}.
This procedure is carried out up to order $\varepsilon^{-2}$ in Appendix \ref{app:large-e}.
The result, plotted in Fig.~\ref{fig:phasedist-large-e}, for the expansion up to first order, in terms of $\varepsilon$ and $\alpha$ is
\begin{widetext}
\begin{align}
P_{P}(\varepsilon,\alpha) \simeq N \varepsilon^{-\frac{1}{D}} \left\lbrace \begin{array}{ll} 
1 - \frac{\sqrt{2}}{D}  \varepsilon^{-1/2} + \Big[ \frac{1}{D^2} + \frac{1}{2 \Omega} \Big( \big(1 + \frac{1}{D}\big) \sin(2\alpha) - \cot(\alpha) \Big) \Big] \varepsilon^{-1},& \text{for} \; \sqrt{2 \varepsilon} \alpha \gg 1 \\[2 ex]
1 - \Big[\frac{\sqrt{2}}{D} + \frac{\sqrt{\varepsilon} \alpha}{ \Omega (1+2 \varepsilon \alpha^2)} \Big] \varepsilon^{-1/2} + \Big[\frac{1}{D^2} + \frac{1}{2 \Omega^2} \Big[ \frac{2 D -1}{(1+2 \varepsilon \alpha^2)^2} + \frac{D-D^2 + 2 \sqrt{2 \varepsilon} \alpha \Omega}{D(1+2 \varepsilon \alpha^2)} \Big] \Big] \varepsilon^{-1},  & \text{for} \; \sqrt{2 \varepsilon} \alpha \lesssim 1 . \label{large-e-final}
\end{array} \right.
\end{align}
\end{widetext}

\begin{figure}[ht!]
\includegraphics[width=0.47\textwidth, clip, trim=0mm 0mm 10mm 10mm]{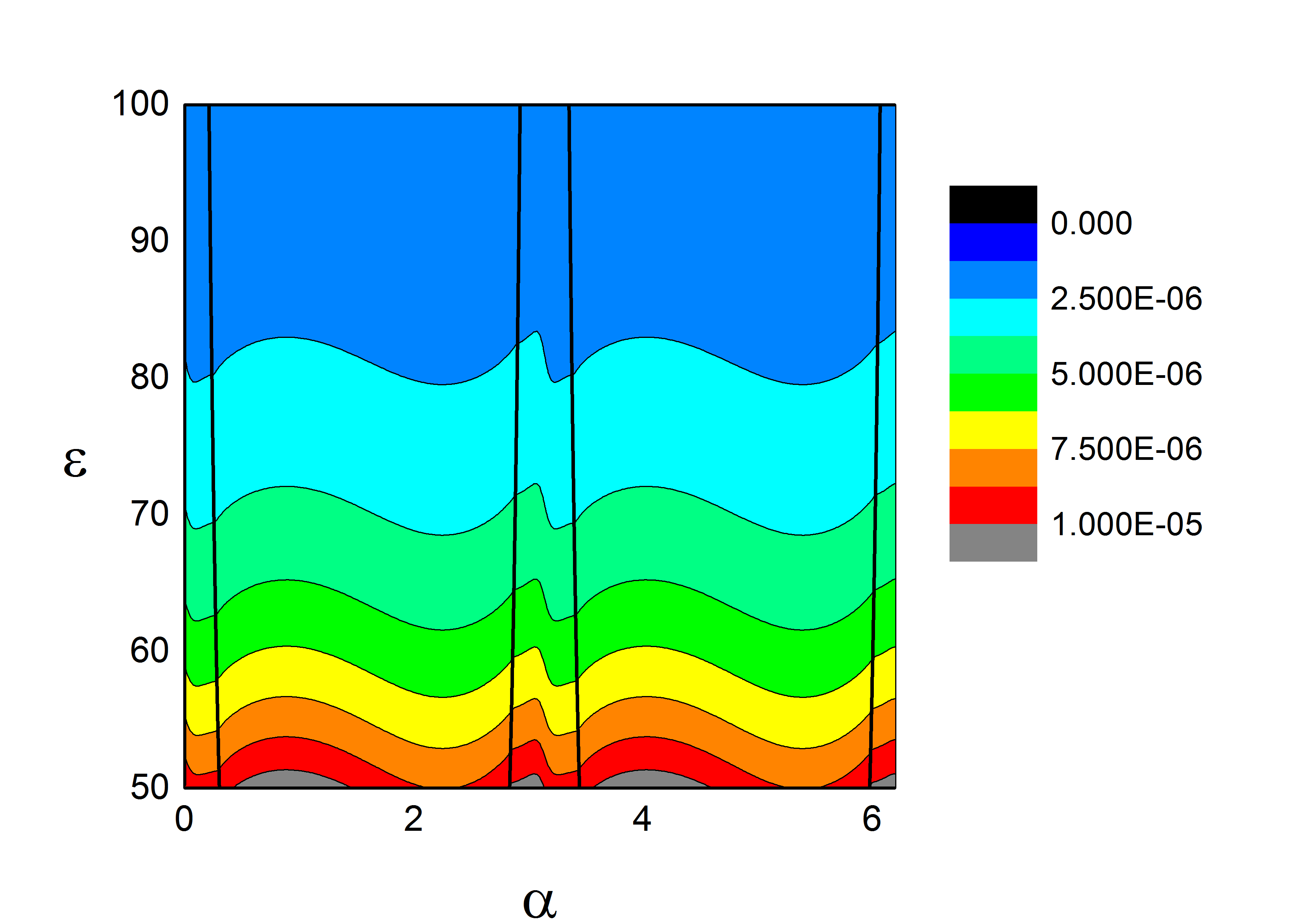}
\caption{Phase-space probability density for $D = 0.3$ and $\Omega = 0.5$ from the large-energy expansion Eq.~\eqref{large-e-final}, plotted using energy and phase-space angle $(E,\alpha)$. The thick black lines denote the approximate boundary between the strip and outside area. Note that for large energies, the matching at the boundary works increasingly well.  \label{fig:phasedist-large-e}}
\end{figure}

The above discussion shows that the power-law tail found in the large-$\Omega$ result \eqref{energy-density-stationary} is in fact generic for large energies, where the system always is underdamped, independent of $\Omega$.
To leading order the stationary state is thus characterized by the energy.
Just like for small $D$ and large $\Omega$, the corrections lead to a complex structure of the phase-space density.
We note that while the large-energy expansion agrees with the first order large-$\Omega$ expansion in the appropriate limits, the contribution from the strip actually also contains terms of order $\Omega^{-2}$. 
In that sense, just as the first order large-$\Omega$ expansion contains information that is not contained in the first order large-energy one, the converse is also true.

\section{Probability currents and detailed balance \label{sec:detailed-balance}}
In the previous sections we discussed the unusual stationary state behavior of confined atoms in Sisyphus cooling.
We already saw that the stationary state is non-thermal in that there exists no well-defined temperature and that energy equipartition does not hold.
In that sense, we may refer to this stationary state as a non-equilibrium stationary state.
In order to differentiate this further, we want to see whether or not detailed balance holds.
Detailed balance, which is often taken as the defining property of an equilibrium system, means that for every possible transition in the system, the forward and backward process are equally likely \cite{ris86,gar96}.
In terms of the transition probabilities, this can be expressed as \cite{gar96}
\begin{align}
P(x',p',t+\tau ; x,p,t) = P(x,-p,t+\tau ; x',-p',t), \label{detailed-balance}
\end{align}
where $P(x',p',t' ; x, p, t)$ is the joint probability density for finding a particle at $(x,p)$ at time $t$ and at $(x',p')$ at time $t'$.
Here the minus sign in front of the momentum is a consequence of the momentum being odd under time reversal.
For the stationary state in a Markovian system, we can express this in terms of the conditional probability density $P(x',p',\tau \vert x, p, 0)$,
\begin{align}
P(x',p',\tau &\vert x,p,0) P(x,p) \nonumber \\
& = P(x,-p,\tau \vert x',-p',0) P(x',-p').
\end{align}
For $\tau = 0$, the conditional probability density reduces to a delta-function, $P(x',p',0 \vert x,p,0) = \delta(x-x') \delta(p-p')$.
Since this function is even under $(p,p') \rightarrow (-p,-p')$, detailed balance implies that $P(x,p) = P(x,-p)$, i.e.~the stationary probability density has to be even under momentum reversal \cite{gar96}.
As we saw in Sections \ref{sec:small-D} and \ref{sec:large-freq}, the stationary state of our system of confined cold atoms does not satisfy this property and therefore does not respect detailed balance.

For a Kramers-Fokker-Planck equation describing one-dimensional underdamped motion in a potential,
\begin{align}
\Bigg[ - p \partial_x + \partial_p \bigg[ U'(x) - F_\text{fric}(p) + D_p \partial_p \bigg] \Bigg] P(x,p) = 0, \label{KFP-general}
\end{align}
detailed balance is equivalent to the conditions \cite{gra71,ris86}
\begin{subequations}
\begin{align}
D_{p} - D_{-p} &= 0 \label{db-conditions-1} \\
\partial_x J_x^\text{rev} + \partial_p J_p^\text{rev} &= 0 \label{db-conditions-2}\\
J_x^\text{ir} = J_p^\text{ir} &= 0 \label{db-conditions-3}
\end{align}\label{detailed-balance-conditions}%
\end{subequations}
Note that we use dimensionless variables here.
For a confined particle, the reversible currents $J_i^\text{rev}$ correspond to the oscillatory motion of the particle in the potential.
The irreversible currents, on the other hand are induced by the dissipative and fluctuating forces due to the bath.
Equation~\eqref{db-conditions-1} implies that the momentum diffusion coefficient $D_p$ should be an even function of $p$.
This condition holds for the optical lattice system.
For Eq.~\eqref{KFP-general} the reversible and irreversible probability currents are given by
\begin{align}
J_x^\text{rev} = p P(x,p), \qquad &J_p^\text{rev} = -U'(x) P(x,p), \label{prob-current} \\
J_x^\text{ir} = 0, \qquad &J_p^\text{ir} = \Big( F_\text{fric}(p) - D_p \partial_p \Big) P(x,p) . \nonumber
\end{align}
Plugging this into Eq.~\eqref{detailed-balance-conditions}, we see that detailed balance is equivalent to stationary solution $P(x,p)$ solving both the Hamiltonian and \enquote{bath} part of the Kramers-Fokker-Planck equation individually,
\begin{align}
\bigg[ - p \partial_x + U'(x) \partial_p  \bigg] P(x,p) &= 0 \label{db-hamilton} \\
\bigg[ - F_\text{fric}(p) + D_p \partial_p \bigg]  P(x,p) &= 0 \label{db-bath} .
\end{align}
Equation \eqref{db-hamilton} implies that $P(x,p) = f(p^2/2 + U(x))$.
From Eq.~\eqref{db-bath}, we further have
\begin{align}
P(x,p) = g(x) \exp \bigg[{\int_{0}^{p} \text{d}p' \ \frac{F_\text{fric}(p')}{D_p'}} \bigg] .
\end{align}
The two expressions for $P(x,p)$ are compatible only if a fluctuation-dissipation relation holds,
\begin{align}
\frac{F_\text{fric}(p)}{D_p} = - \tilde{\beta} p \label{db-flucdiss},
\end{align}
and we have the Boltzmann-Gibbs density 
\begin{align}
P(x,p) = N \exp \bigg[ -\tilde{\beta} \Big( \frac{p^2}{2} + U(x) \Big) \bigg],
\end{align}
where $\tilde{\beta} > 0$ plays the role of an inverse effective temperature.
Thus for the general class of systems described by Eq.~\eqref{KFP-general}, an (effective) Boltzmann-Gibbs ensemble is the only possible solution obeying detailed balance.
For Sisyphus cooling, we have from Eq.~\eqref{langevin-lattice} in terms of dimensionless variables $F_\text{fric} = -p/(1+p^2)$, $D_p = D(1 + \mathfrak{D} (1+p^2))$ and thus
\begin{align}
\frac{F_\text{fric}(p)}{D_p} = \frac{-\frac{p}{1+p^2}}{D + \frac{D \mathfrak{D}}{1+p^2}},
\end{align}
with $\mathfrak{D} = D_1/D_0$, see Eq.~\eqref{friction}. 
This satisfies the condition \eqref{db-flucdiss} only for $D \rightarrow 0$, where we obtain the Boltzmann-Gibbs density, see Section \ref{sec:small-D} and Appendix \ref{app:d1}.
In this limit $D_1/\gamma p_c^2$ corresponds to the above defined effective temperature $\tilde{\beta}$.
For $D_1 = 0$, as assumed in the previous sections, the effective temperature in the Boltzmann-Gibbs limit $D = D_0/\gamma p_c^2 \rightarrow 0$ vanishes.
The violation of detailed balance and thus the non-equilibrium nature are due to the absence of a fluctuation-dissipation relation like Eq.~\eqref{db-flucdiss} between the noise and the friction force.
While the friction force and the momentum-dependent part $D \mathfrak{D}/(1+p^2)$ of the diffusion coefficient (see Eq.~\eqref{langevin-lattice} and the following discussion) are both due to the motion of the atoms in the optical lattice and thus obey a fluctuation-dissipation relation, the momentum-independent part $D$ of the diffusion coefficient, which represents spontaneous emission of photons, has no dissipative counterpart.

\begin{figure}[ht!]
\includegraphics[width=0.47\textwidth, clip, trim=0mm 0mm 0mm 0mm]{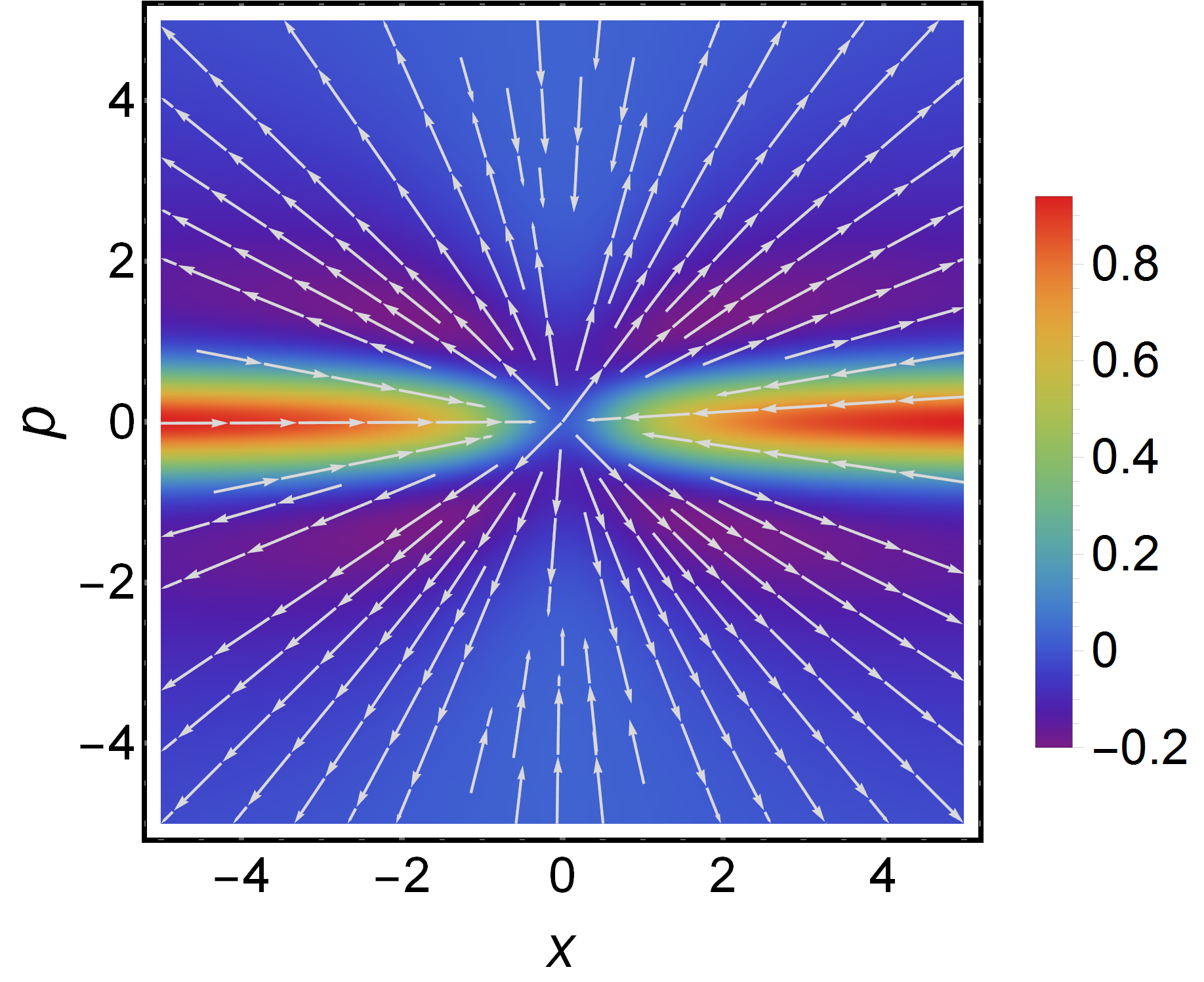}
\caption{The scalar product between the mean velocity and the gradient of the phase-space density $\phi = \vec{v}_P \cdot \vec{\nabla} P /P$, Eq.~\eqref{phi-result}. The color corresponds to the value of $\phi$, the arrows denote the direction of the current parallel to the gradient; inward for positive $\phi$, outward for negative $\phi$. \label{fig:current-angle}}
\end{figure}

We now want to quantify the violation of the detailed balance conditions Eq.~\eqref{detailed-balance-conditions}.
We focus on the case where the diffusion coefficient is even in the momentum so that Eq.~\eqref{db-conditions-1} is satisfied.
The remaining two conditions are then equivalent to Eqs.~\eqref{db-hamilton} and \eqref{db-bath}.
These two equations imply a geometric property for the total probability current $\vec{J} = (J_x, J_p)$, where $J_i = J_i^\text{rev} + J_i^\text{ir}$.
As is easily verified by direct calculation, as long as detailed balance holds, the probability current is perpendicular to the gradient of the density $\vec{\nabla}P = (\partial_x P, \partial_p P)$, i.~e.~$\vec{J} \cdot \vec{\nabla} P = 0$ \cite{gra71b}.
This means that for a system with detailed balance, the probability current always flows along equi-probability lines.
For the optical lattice system with $D \neq 0$, on the other hand, detailed balance is broken and thus $\vec{J} \cdot \vec{\nabla} P \neq 0$. 
The current is related to the local mean phase-space velocity via $\vec{v}_P = \vec{J}/P$, where by definition
\begin{align}
\partial_t \left( \begin{array}{ll} \langle x \rangle \\ \langle p \rangle \end{array} \right) = \langle \vec{v}_P \rangle .
\end{align}
In the steady state, the left hand side vanishes and thus the average phase-space velocity $\langle \vec{v}_P \rangle$ is zero, however, it is generally still non-zero locally.
We use the scalar product between the phase-space velocity and the normalized gradient of the density 
\begin{align}
\phi = \vec{v}_P \cdot \frac{1}{P}\vec{\nabla} P = \frac{\vec{J} \cdot \vec{\nabla} P}{P^2} \label{phidef}
\end{align}
as a measure of the misalignment between current and equi-probability lines and hence detailed balance violation.
In the large-$\Omega$ limit and for constant $D_p \equiv D_0$, we can write the probability density as
\begin{align}
P(x,p) \simeq P_0(x,p) \Big(1 + \frac{f(x,p)}{\Omega} \Big) + \mathcal{O}(\Omega^{-2})
\end{align}
with $P_0(x,p) = P_\varepsilon[(x^2+p^2)/2]/(2 \pi)$ and $f(x,p) = f_1[(x^2+p^2)/2,\arctan(p/x)]$, see Eq.~\eqref{large-omega-expansion}. 
From the definition of the probability current Eq.~\eqref{prob-current}, we then find
\begin{align}
\phi &= \frac{1}{P_0(x,p)} \Bigg[\bigg[ P_0(x,p) \Big[ p \partial_x - x \partial_p \Big] f(x,p) \bigg] \\
& \qquad - \bigg[ \frac{p}{1+p^2} \partial_p + D \big\lbrace \partial_p \ln(P_0(x,p)) \big\rbrace \partial_p \bigg] P_0(x,p) \Bigg] \nonumber \\
& \qquad + \mathcal{O}(\Omega^{-1}). \nonumber
\end{align}
Note that $\phi$ depends on the first order correction $f(x,p)$ even in the limit $\Omega \rightarrow \infty$.
Further recognizing that $[p \partial_x - x \partial_p] f(x,p) = -\partial_\alpha f_1(\varepsilon,\alpha)$ and using Eqs.~\eqref{energy-density-stationary} and \eqref{f_1-equation}, we finally find,
\begin{align}
\phi &= \partial_p \frac{p}{1+p^2} + D \partial_p^2 \ln(P_0(x,p)) \label{phi-result} \\
	&= \partial_p \frac{p}{1+p^2} - 2 \partial_p^2 \ln(1+\sqrt{1+p^2 + x^2}) + \mathcal{O}(\Omega^{-1}) \nonumber.
\end{align}
Intriguingly, $\phi$ is to leading order independent of $D$.
This seems counter-intuitive, since the detailed balance violations should vanish for $D \rightarrow 0$.
As it turns out, this is an artifact of setting $D_1 = 0$, since then, as discussed above, the effective temperature for $D \rightarrow 0$ is zero and there is thus no well-defined detailed-balance preserving state in this limit.
Repeating the calculation for a non-zero $D_1$ yields $\lim_{D \rightarrow 0} \phi = 0$ as it should, see Appendix \ref{app:d1-omega}.
The average $\langle \phi \rangle$ over all phase-space is zero, reflecting the fact that there is no global probability current in the steady state.
We thus interpret $\phi$ as a measure of local flow due to the detailed balance violation of the system.
The quantity $\phi$ for $\Omega \rightarrow \infty$ is shown in Fig.~\ref{fig:current-angle}, where we see that the local phase-space velocity can be both parallel and antiparallel to the density gradient.
The local mean phase-space velocity $\vec{v}_P = (\dot{x},\dot{p})$ represents the average change in position and momentum of particles located at a phase space point $(x,p)$.
In areas where $\phi$ is positive, particles on average move from a low-density to a high-density region.
This is mostly observed for small momenta, where the friction is effective and thus causes a flow towards the central, low energy region.
In most areas of phase-space, the flow is directed towards larger energies due to the weak friction, balancing out the inward flow close to the $(p = 0)$-axis.

We stipulate that probability currents that do not flow along equi-probability lines are generally connected to a non-zero entropy production as a measurable consequence of the non-equilibrium nature of the system.
The breaking of detailed balance necessarily implies the presence of irreversible probability currents in the system \cite{cho11}.
As was shown in Ref.~\cite{esp10}, a distinct contribution to the entropy production also arises from the breaking of detailed balance.
Indeed, the entropy production can be directly expressed via the irreversible currents \cite{tom10}, which in the case of momentum-dependent forces leads to an anomalous entropy production \cite{kwo16}.
We suggest that the geometric properties of the probability current discussed above may yield a more detailed understanding of the precise way the system deviates from equilibrium.
We leave this investigation to future work.

\section{Physical considerations for the cold atom system \label{sec:experimental}}
So far, we treated the solution of our main equation \eqref{KFP-lattice} from a mathematical point of view, with only occasional reference to the actual system of Sisyphus cooling of confined atoms.
In the following, we first delineate the experimentally relevant parameter regime in terms of the dimensionless parameters $D$ and $\Omega$.
We also discuss the effects of the momentum dependent diffusion coefficient and Stark shifts induced by introducing the confining potential.
Finally, we present numerical simulations for two example sets of parameters to estimate the magnitude of the observed effects.
In the following we will refer to the dimensionful coordinates as $\tilde{x}$ and $\tilde{p}$, while $x$ and $p$ denote their dimensionless counterparts, see Eq.~\eqref{KFP-lattice}.

\vspace*{0.2cm}

\textit{Magnitude of the confinement.} As the semiclassical description of Sisyphus cooling relies on spatially averaging the motion of the atoms over a wavelength of the cooling lattice \cite{cas90}, it will work reliably only when the atoms are able to move over several lattice periods on time scale of interest. 
This puts a natural constraint on the magnitude of the confining potential; the latter has to be weak enough so as to not localize the atoms on the length scale of the cooling lattice.
We can estimate the magnitude of the confining field by demanding that the typical spread of the atomic cloud should be much larger than the lattice period, $\langle \tilde{x}^2 \rangle \gg (2 \pi/k)^2$, where $k$ is the wave vector of the cooling laser.
As long as this holds and our theory is valid, we have $\langle \tilde{x}^2 \rangle = p_c^2/(m \omega)^2 \langle x^2 \rangle$, where $\langle x^2 \rangle$ is of order $1$.
Thus we should have $\omega \ll p_c k/(2 \pi m)$, or $\Omega \ll p_c k/(2 \pi m \gamma)$.
In terms of the lattice parameters, the parameters $p_c$, $\gamma$, $D_0$ and $D_1$ can be expressed as \cite{cas90}
\begin{align}
p_c &= \frac{m \Gamma s_0}{9 k}, \qquad \gamma = \frac{3 \hbar k^2 |\delta|}{m \Gamma}, \\
D_0 &= \frac{11 \hbar^2 k^2 \Gamma s_0}{18}, \qquad D_1 = \frac{\hbar^2 k^2 \delta^2 s_0}{\Gamma}, \nonumber
\end{align}
where $\Gamma$ is the natural linewidth of the atomic transition, $\delta$ is the laser detuning and $s_0 = I/I_s/(1+4\delta^2/\Gamma^2)$ is the saturation parameter with the laser intensity $I$ and the saturation intensity $I_s$.
Consequently, we get the condition
\begin{align}
\Omega \ll \frac{1}{108 \pi} \frac{\hbar \Gamma}{E_r} \frac{\Gamma s_0 }{\delta} ,
\end{align}
where $E_r = \hbar^2 k^2/(2 m)$ is the photon recoil energy.
For Cesium ${}^{133}\text{Cs}$, which is commonly used in Sisyphus cooling experiments $E_r = 5.4 \cdot 10^{-29} \text{J}$ and $\Gamma = 7.4 \cdot 10^{6} \text{s}^{-1}$ \cite{ste03}, and thus 
\begin{align}
\Omega \ll 1.19 \frac{\Gamma  s_0}{\delta} .
\end{align}
In the following, we consider two exemplary sets of parameters, $\delta = 10 \Gamma$ and $I = 20 I_s$ (large detuning) and $\delta = 1.5 \Gamma$ and $I = 3.3 I_s$ (moderate detuning).
Both choices lead to a lattice depth $U_0 \approx 130 E_r$, which is the point where the minimal kinetic energy for free atoms is reached according to the semiclassical treatment \cite{cas90,per00}.
Since $D$ is related to $U_0$ by $D = 22 E_r/U_0$, this leads to a value of $D \approx 0.17$ \footnote{The precise relation between $D$ and $U_0$ depends on the details of the atomic transition and the parametrization, however this only causes a slight change in the proportionality constant \cite{cas90,mar96}}.
For these kinds of lattice parameters, we are thus at moderate values of $D$, which are far enough from the point where the average energy diverges ($D = 1/2$, see Eq.~\eqref{average-energy}) to be able to characterize the system by the stationary solution, yet still large enough for the deviations from Boltzmann-Gibbs to become important, see Section \ref{sec:small-D}.
In the large detuning limit the bound on $\Omega$ turns out to be $\Omega \ll 5.9 \cdot 10^{-3}$, whereas for moderate detuning we find $\Omega \ll 0.26$. 
In both cases, $\Omega$ has to be very small, which means that we cannot employ the underdamped description of Section \ref{sec:large-freq}.
However, we saw in Eq.~\eqref{equipart-order3} that the deviations from energy equipartition are most pronounced at small $\Omega$ and thus these effects are rather important in this regime.
For moderate detuning, and not too small $\Omega$, also the asymmetry of the phase-space density, quantified in Eq.~\eqref{asym-ratio}, might be observed.
For the simulations, we take $\Omega$ to be one tenth of the above limit, i.~e.~$\Omega = 5.9 \cdot 10^{-4}$ for large detuning and $\Omega = 0.026$ for moderate detuning.

\vspace*{0.2cm}

\textit{Momentum-dependent diffusion coefficient.} In Section \ref{sec:optical-lattice}, we saw that the diffusion coefficient $D_p$ actually depends on momentum, but we so far ignored this since it does not change the qualitative results of our analysis.
The momentum-dependent diffusion coefficient is given by
\begin{align}
D_{\tilde{p}} = D_0 + \frac{D_1}{1+\frac{\tilde{p}^2}{p_c^2}} ,
\end{align}
which reduces to the momentum-independent case for $D_1 = 0$.
The momentum-dependent part is small for large momenta $|\tilde{p}| \ll p_c$, where the nonlinearity of the friction force becomes important.
Precisely for this reason, the qualitative results, which hinge on this nonlinearity, are not changed by the introduction of a finite $D_1$, in particular the parameter $D = D_0/(\gamma p_c^2)$, which controls the behavior is unchanged.
As long as the ratio $\mathfrak{D} = D_1/D_0$ is not too large, we can repeat the same procedure as in Section \ref{sec:small-D} for the small-$D$ expansion including the momentum-dependent diffusion coefficient, see Appendix \ref{app:d1}.
As a result, we see that a nonzero $D_1$ essentially amplifies the deviations from the Boltzmann-Gibbs behavior.
This is intuitively reasonable, as an enhanced diffusion coefficient at small momenta will push the particles to higher momenta, where they feel the effect of the nonlinear friction more strongly.
In terms of the lattice parameters, the ratio $D_1/D_0$ can be expressed as \cite{cas90}
\begin{align}
\mathfrak{D} = \frac{D_1}{D_0} = \frac{18}{11} \frac{\delta^2}{\Gamma^2} .
\end{align}
For moderate detuning, $D_1$ is thus comparable to $D_0$, $\mathfrak{D} \approx 3.7$; for large detuning, $D_1$ is much bigger than $D_0$, $\mathfrak{D} \approx 160$. 
In both cases the modified small-$D$ expansion presented in Appendix \ref{app:d1} is not applicable and we have to rely on numerical simulations.
We note, however, that the trend of increasing deviations from Boltzmann-Gibbs behavior with increasing $\mathfrak{D}$ persists.

\vspace*{0.2cm}

\textit{Stark shifts.} In the previous discussion, we treated the harmonic confinement as an additional force that acts on the atoms but does not otherwise affect the friction or diffusion terms in the Kramers equation \eqref{KFP-lattice}.
Experimentally, the confinement could be realized by a static, or low-frequency, electric field.
Then, the atomic states experience a position-dependent light shift due to the interaction with the trapping field.
This light shift affects the detuning $\delta$ and thus the cooling mechanism.
For a two-level atom, the DC Stark shift to the ground state is can be obtained from second order perturbation theory \cite{gri00},
\begin{align}
\delta \epsilon_{g} = \frac{|\langle e | H_{i} | g \rangle|^2}{\epsilon_{e}-\epsilon_{g}},
\end{align}
where $g$ and $e$ denote the ground and excited state, $H_{I}$ is the interaction Hamiltonian and $(\delta)\epsilon$ the respective energy (shift).
In a static electric field of amplitude $\mathcal{E}$, the interaction Hamiltonian reads
\begin{align}
H_i = - \vec{\mathcal{E}} \vec{\mu} ,
\end{align}
with the atomic dipole operator $\mu$.
This gives us
\begin{align}
\Delta \epsilon_{g} = - \frac{1}{2} \alpha_0 \mathcal{E}^2 \qquad \text{with} \qquad \alpha_0 = -2 \frac{|\langle e | \mu | g \rangle|^2}{\epsilon_{e}-\epsilon_{g}},
\end{align}
$\alpha_0$ being the static polarizability of the atom.
This quasistatic approximation is valid when the frequency of the electric field that constitutes the confinement is much lower than the frequency of the atomic transition $\omega_0$ \cite{gri00}.
Introducing the intensity $\mathcal{I} = 2 c \varepsilon_0 \mathcal{E}^2$, this reads
\begin{align}
\Delta \epsilon_{g} = - \frac{\alpha_0}{4 c \epsilon_0} \mathcal{I},
\end{align}
or in terms of the potential $U = -\alpha_0/(2 \epsilon_0 c) \mathcal{I}$,
\begin{align}
\Delta \epsilon_{g}(\tilde{x}) = \frac{1}{2} U(\tilde{x}).
\end{align}
The excited state has the opposite shift $\Delta \epsilon_e = -\Delta \epsilon_g$.
In our case, the trapping potential is harmonic, and we get the estimate estimate for the total light shift $\Delta \epsilon = \Delta \epsilon_g - \Delta \epsilon_e$ relative to the detuning,
\begin{align}
\frac{\Delta \epsilon}{\hbar \delta} = \frac{1}{\hbar \delta} U(\tilde{x}) = \frac{m \omega^2 }{2 \hbar \delta} \tilde{x}^2 = \frac{m v_c^2}{2 \hbar \delta} x^2,
\end{align}
where in the last step, we replaced the physical position $\tilde{x}$ by the our dimensionless variable $x$.
Using that $\langle x^2 \rangle$ is of order one, we find for the average relative light shift
\begin{align}
\frac{\Delta \epsilon}{\hbar \delta} \approx \frac{1}{324} \frac{\hbar \Gamma}{E_r} \frac{\Gamma s_0^2}{\delta} \approx 1.2 \frac{\Gamma s_0^2}{\delta} ,
\end{align}
where the rightmost expression is again for Cesium.
For large detuning, we find that the relative light shifts are approximately $3.1\cdot 10^{-4}$.
In the large-detuning regime, we can thus treat the confining potential as a classical force without worrying about the additional induced light shifts.
For moderate detuning we find relative shifts of around $0.09$, which are small but might still lead to a position-dependent detuning and thus cooling rate.
From this point of view, it is thus advantageous to work in the large-detuning regime.

\vspace*{0.2cm}

\begin{figure}[ht!]
\includegraphics[width=0.47\textwidth, clip, trim=0mm 0mm 20mm 10mm]{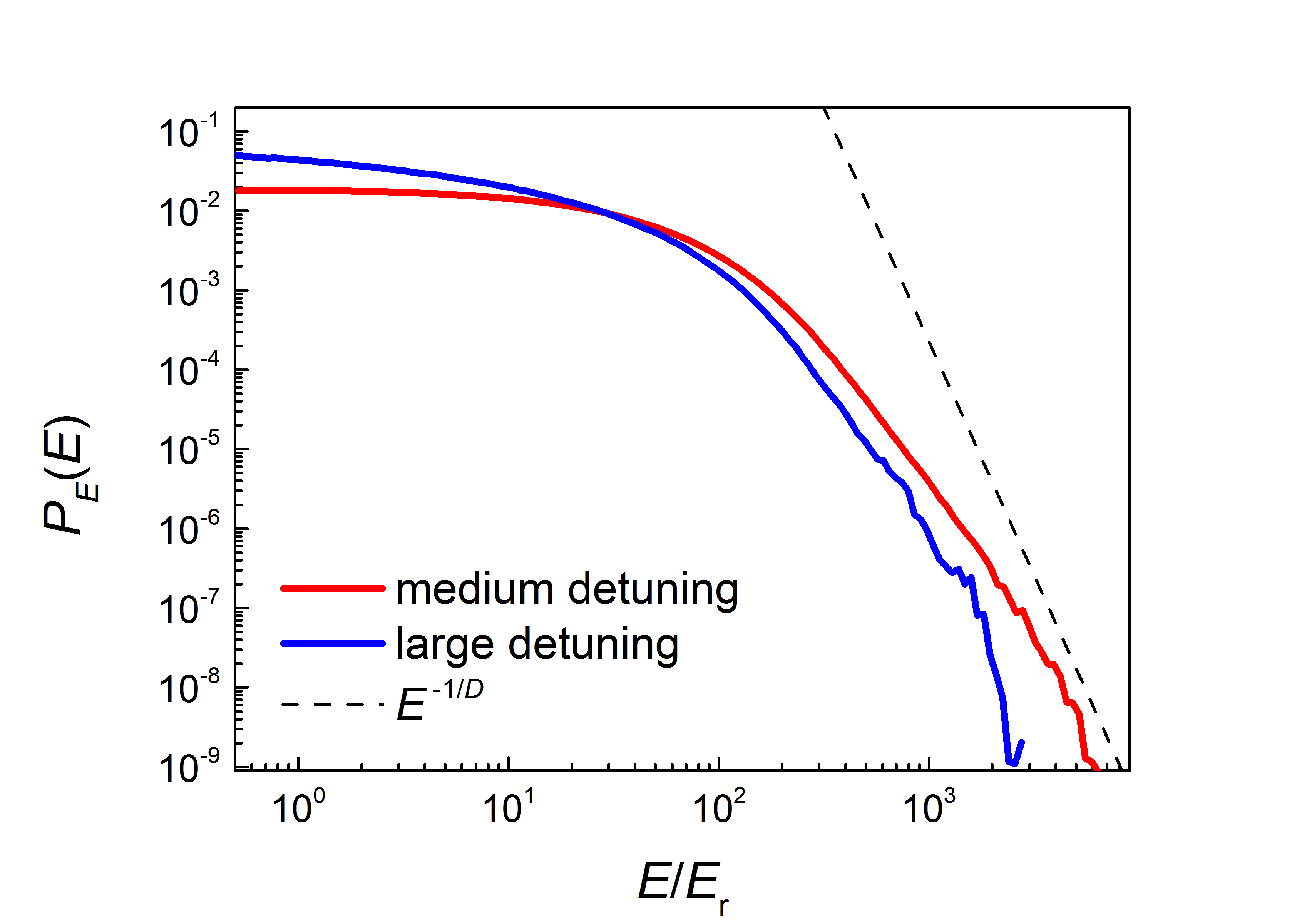}
\caption{Energy probability density for model Cesium as a function of energy in units of the recoil energy. The red line corresponds to medium detuning ($\Delta = 1.5 \Gamma$, $I = 3.3 I_s$), the blue line to large detuning ($\Delta = 10 \Gamma$, $I = 20 I_s$). The dashed line is the expected asymptotic power law $P_E(E) \propto E^{-1/D}$ for large energies. \label{fig:exper-edist}}
\end{figure}

\begin{figure}[ht!]
\includegraphics[width=0.47\textwidth, clip, trim=0mm 0mm 20mm 10mm]{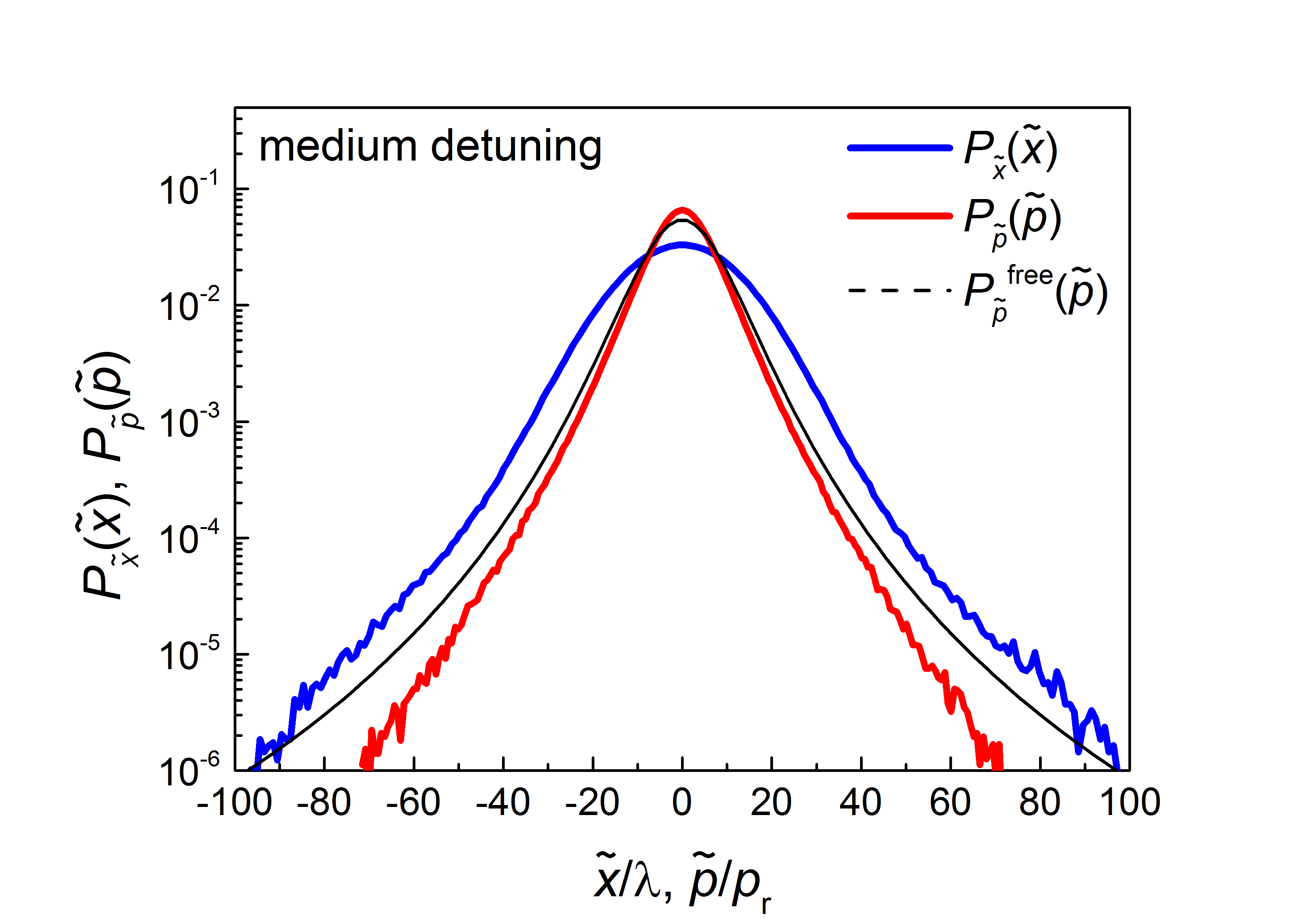}
\caption{Momentum (red) and position (blue) probability density versus positon/momentum in units of the recoil momentum $p_r = \hbar k/m$ and the lattice wavelength $\lambda = 2 \pi/k$ for medium detuning. The black line is the momentum probability density for the same parameters without the confining potential. \label{fig:exper-xpdist1}}
\end{figure}

\begin{figure}[ht!]
\includegraphics[width=0.47\textwidth, clip, trim=0mm 0mm 20mm 10mm]{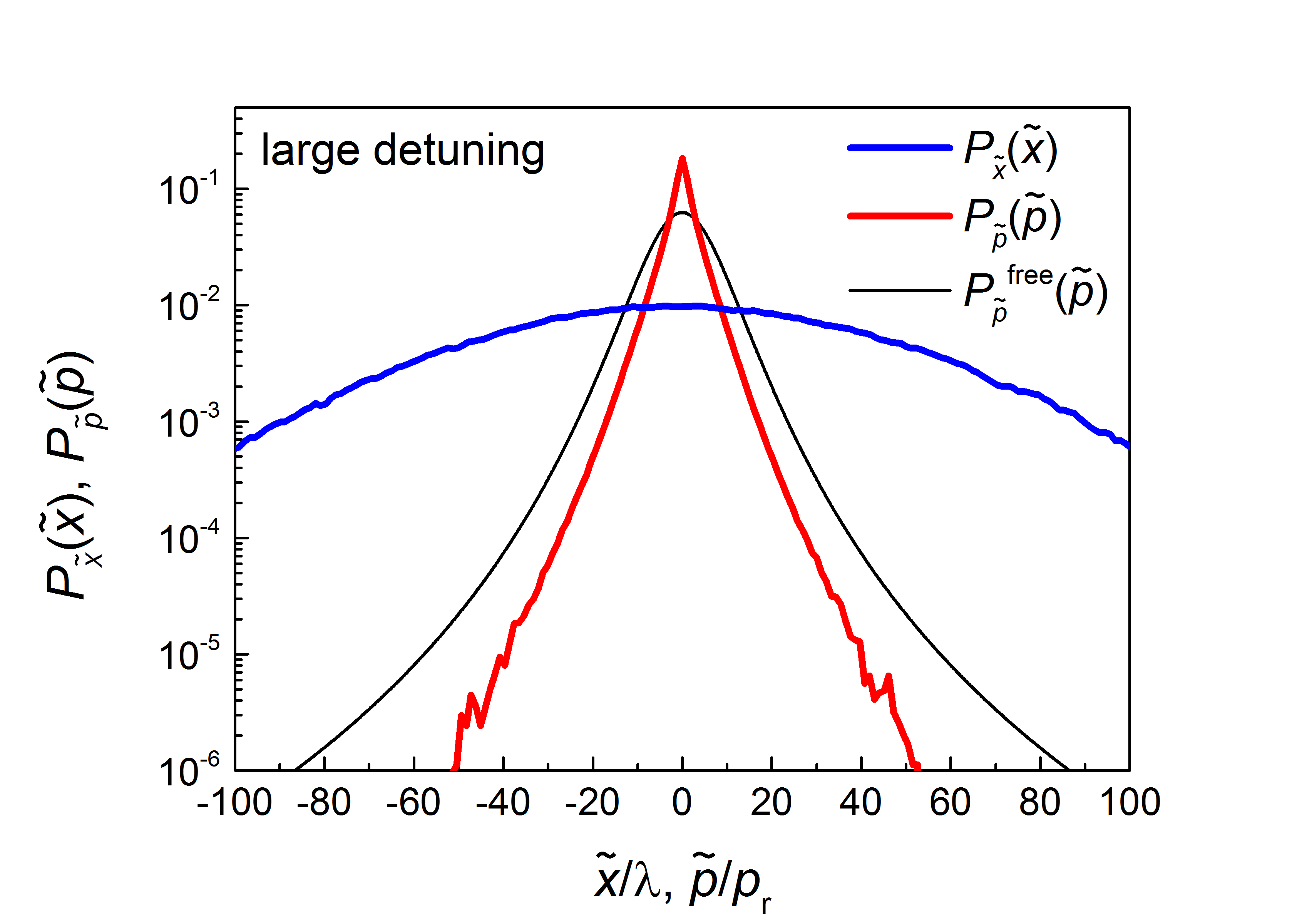}
\caption{Momentum (red) and position (blue) probability density versus positon/momentum in units of the recoil momentum $p_r = \hbar k/m$ and the lattice wavelength $\lambda = 2 \pi/k$ for large detuning. The black line is the momentum probability density for the same parameters without the confining potential. \label{fig:exper-xpdist2}}
\end{figure}

\textit{Numerical results.} Since our limiting expansions are not valid for the parameters stated above, let us discuss some results from numerical Langevin simulations with the above parameters.
%While theses results do not offer a quantitative prediction for experiments due to the simplifications and approximations entering the derivation of the semiclassical description Eq.~\eqref{langevin-lattice}, they nevertheless offer a good estimate for the order of magnitude of the effects and thus whether the latter will be observable in experiments. 
Let us first discuss the case of moderate detuning. 
Here, we find a stationary average potential energy $\langle E_p \rangle \approx 90 E_r$ and average kinetic energy $\langle E_k \rangle \approx 60 E_r$, which yields an equipartition ratio of $1.5$.
The potential energy is thus significantly enhanced with respect to the kinetic one.
This is mirrored in a discernible difference between the position and momentum distributions, see Fig.~\ref{fig:exper-xpdist1}.
Interestingly, the average kinetic energy is smaller than the value for the same parameters without the confining potential, $\langle E_k \rangle_\text{free} \approx 83 E_r$, which was obtained in Refs.~\cite{hod95,per00}.
We further find a small but discernible asymmetry in the phase-space density, resulting in an asymmetry parameter of $\eta \approx 1.02$.
The power law tails of the energy distribution can only be observed at very large energies $E \gtrsim 10^3 E_r$, see Fig.~\ref{fig:exper-edist}.
For large detuning, this asymmetry vanishes within the accuracy of the simulations, however, both the potential and kinetic energy are reduced considerably, to $\langle E_p \rangle \approx 68 E_r$ and $\langle E_k \rangle \approx 21 E_r$.
The imbalance between potential and kinetic energy becomes even larger at an equipartition ratio of $3.2$ and the position and momentum distribution differ significantly, see Fig.~\ref{fig:exper-xpdist2}.
Most strikingly, the average kinetic energy is well below the minimum value for Sisyphus cooling without confinement, $\langle E_k \rangle_\text{free} \approx 66 E_r$, within the semiclassical picture of course \cite{cas90,per00}.
At large detuning, the power law tails of the energy distribution cannot be observed for reasonable values of the total energy, see Fig.~\ref{fig:exper-edist}.
We note that a larger potential energy compared to the kinetic one was found in terms of a diffusion approximation in Ref.~\cite{per00}.
However, this approximation predicts the divergence of potential and kinetic energy at different values of $D$, which is in contradiction to the result Eq.~\eqref{average-energy} that is confirmed by the numerical simulations.
Further, within the diffusion approximation, the kinetic energy is unaffected by the confinement and always corresponds to the result for unconfined atoms, again in contradiction to our findings.

\vspace*{0.2cm}

\textit{Measurement protocol.} A possible protocol to measure both the position and momentum density of the atoms independently may be the following:
Taking into account the above considerations, the atoms could trapped within a confining field, provided by a laser beam that is very far red-detuned from the atomic transitions, or by an electrostatic potential.
For simplicity one could trap the atoms in all spatial directions but only have the optical lattice along one direction, similar to the setup employed in Ref.~\cite{sag12} to observe the diffusion of the atoms.
Then we let the atoms relax into a stationary state, the characteristic time scale for this is estimated from our simulations being around 10 ms for $\Delta = 1.5 \Gamma$ and around 100 ms for $\Delta = 10 \Gamma$. 
In the stationary state, we could then determine the position density of the cloud, by illuminating and imaging it.
In order to obtain the momentum density, we turn off both the confining field and the optical lattice rapidly and take snapshots of the ensuing ballistic expansion of the cloud.
Once the size of the cloud is significantly bigger than its stationary extension in the trapped state, the position of the atoms will essentially be proportional to their momentum at the time of release times the flight time and we can thus extract the momentum density in the trapped state.
We note that also this kind of time-of-flight measurement was applied before in \cite{dou06} to determine the steady momentum density of Sisyphus cooled atoms without confinement.
From the position- and momentum- density one can then obtain the average potential and kinetic energy by integration.

\section{Conclusion \label{sec:conclusion}}

In the preceding sections, we have discussed the nonequilibrium stationary state that results from confining cold atoms that are subject to Sisyphus cooling.
We obtained analytical results in terms of expansions in three limiting regimes.
The common theme of all these expansions is that the leading order phase-space density is a function of the Hamiltonian only.
The advantage of this description is obvious: Instead of position and momentum, we only need the total energy to characterize the stationary state of the system.
For small values of the parameter $D$, which corresponds to deep optical lattices in the physical system, we indeed recover a Boltzmann-Gibbs-like density describing the center, small-energy part of the phase-space density.
However, this is only correct to leading order, as the correction terms for any finite $D$ break the equivalence between energy and probability and we need to describe the system in terms of position and momentum.
A measurable consequence is that the average potential energy of the atoms will be larger than their average kinetic energy.
This expansion does not, however, make any predictions about the tails of the probability density, which are always exponential within the approximation, while in reality, they are power laws.

A complementary viewpoint is the underdamped limit, which is attained either when the confining potential is strong relative to the damping coefficient, or for very large energies.
In this underdamped limit, the atoms perform oscillations in the confining potential, with an energy that changes slowly in time.
A natural consequence is that the total energy describes the statistics to leading order.
In this limit, we ascertain that the phase-space density and its marginals are indeed heavy tailed power-law densities.
We stress that the exponent of the power-law is more negative than for the case without the confining potential, implying that confining the atoms does alter their statistics in a qualitative manner.
Once we go beyond the leading order, we again find that the correction terms break the equivalence of energy and probability.

Our analytical results are based on perturbation theory. 
In the non-perturbative regime, where we expect larger effects, we performed numerical simulations that confirm that the relevant features persist even beyond the validity of the expansions.
In particular, the imbalance between potential and kinetic energy can potentially be very large, so that measuring the average potential energy in the trapped state does not allow one to infer their kinetic energy.
Rephrasing this in terms of temperature, this means that the temperature extracted from a measurement of the potential energy will be larger than the actual kinetic temperature of the atoms, which often is the quantity of interest.
The fact that the minimum kinetic temperature in the trapped state can actually be lower than without the confining potential is surprising and would not have been obtained from naively treating the system within the thermal Boltzmann-Gibbs approximation.

Our results are thus interesting both from the viewpoint of statistical mechanics and cold atom physics.
From a statistical mechanics perspective we have here a system that closely approximates the equivalence of energy and probability, that is a central hypothesis of equilibrium statistical mechanics, in certain limits, which, however, can also show strong deviations from this paradigm in other regimes.
Predicting these intricate statistics for a well-controlled and existing experimental systems means that their experimental confirmation is within reach.
As for the more practical application, by combining trapping and Sisyphus cooling, a continuously cooled atomic cloud with a well-defined stationary state can be realized.
Our results suggest that in this situation, the kinetic temperature of the cloud can be even lower than without the trapping, as the confinement improves the efficiency of the cooling mechanism.

\begin{acknowledgments}
\textbf{Acknowledgments.} This work was supported by the Israel Science Foundation. A.~D.~was partly employed as an International Research Fellow of the Japan Society for the Promotion of Science.
\end{acknowledgments}

\appendix

\begin{widetext}

\section{Coefficients for the small-$D$-expansion \label{app:small-d}}

The coefficients in the expansion \eqref{small-d-polynomial} can be obtained by plugging the expansion into Eq.~\eqref{small-d-auxilary-1order},
\begin{align}
\sum_{k=0}^{4} \sum_{l=0}^{4-k} a^{(1)}_{k l} &\mathcal{L}_0 z^{k} u^{l} - 3 u^2 + u^4 = 0 \\
\text{with} \qquad &\mathcal{L}_0 = \Omega(z \partial_{u} - u \partial_{z}) - u \partial_{u} + \partial_{u}^2 . \nonumber
\end{align}
Acting with $\mathcal{L}_0$,
\begin{align}
\mathcal{L}_0 z^k u^l = \Omega (l z^{k+1} u^{l-1} - k z^{k-1} u^{l+1} ) - l u^l + l (l-1) u^{l-2},
\end{align}
and demanding that the resulting equation should be valid independent of $u$ and $z$, we can equate the coefficient of any specific combination of powers $z^k u^l$ to zero.
The resulting equations can be solved for the coefficients $a^{(1)}_{k l}$.
We can drastically reduce the number of equations by noting the symmetry of our problem.
The original equation for $g(z,u)$, is, just like the Fokker-Planck equation \eqref{KFP-lattice-D}, invariant under an inversion $(z,u) \rightarrow (-z,-u)$.
As the solution should obey the same symmetry, this immediately implies that all coefficients where $k+l$ is odd should be zero.
The remaining equations then determine all of the remaining coefficients save $a^{(1)}_{0 0}$,
\begin{align}
a^{(1)}_{k l} &= 0 \qquad \text{for} \ k + l \ \text{odd}, \ \text{or} \ k + l > 4, \nonumber \\
a^{(1)}_{0 2} &= a^{(1)}_{3 1} = 0, \quad a^{(1)}_{0 4} = \frac{3 (1 + \Omega^2)}{4 (3 + 4 \Omega^2)}, \nonumber \\
a^{(1)}_{1 1} &= - \frac{3 \Omega}{4 (3 + 4 \Omega^2)}, \quad a^{(1)}_{1 3} = \frac{\Omega}{4 (3 + 4 \Omega^2)} \nonumber \\
a^{(1)}_{2 0} &= \frac{9}{2 (3 + 4 \Omega^2)}, \quad a^{(1)}_{2 2} = \frac{3 \Omega^2}{2 (3 + 4 \Omega^2)} \nonumber \\
a^{(1)}_{4 0} &= - \frac{3 \Omega)}{4 (3 + 4 \Omega^2)} .
\end{align}
The final coefficient $a^{(1)}_{0 0}$ is fixed by demanding that the resulting probability density should be normalized,
\begin{align}
a^{(1)}_{0 0} = -\frac{3 (9 + 8\Omega^2)}{4 (3 + 4 \Omega^2)} .
\end{align}

\section{Asymptotic matching for the large-energy expansion \label{app:large-e}}

From the discussion in Section \ref{sec:large-E}, we know that the expansion \eqref{large-energy-expansion-outside} outside the strip has singularities at $\alpha = 0, \pi$.
This means that the expansion coefficients $g_k(\alpha)$ can have different values in the upper ($p>0$) and lower ($p < 0$) half-plane of phase space. 
To account for this, we modify Eq.~\eqref{large-energy-expansion-outside},
\begin{align}
P^{\pm}_P(\varepsilon,\alpha) \simeq N \varepsilon^{\beta} \Bigg[1 + \sum_{k = 1}^{K} \frac{g^{\pm}_{k/2}(\alpha)}{\varepsilon^{\frac{k}{2}}} + \mathcal{O}(\varepsilon^{-\frac{K+1}{2}}) \Bigg], \label{appB-expansion-outside}
\end{align}
where \enquote{+} denotes $p > 0$ and \enquote{--} denotes $p < 0$.
Similarly, the approximate solutions to Eq.~\eqref{KFP-lattice-p} inside the strip may be different depending on whether we are in the right $x > 0$ or left $x < 0$ half-plane of phase space.
Instead of Eq.~\eqref{large-energy-expansion-inside}, we then write
\begin{align}
P^{\pm}_S(x,p) \simeq M |x|^{\gamma} \Bigg[1 + \sum_{k = 1}^{K} \frac{h^{\pm}_{k/2}(p)}{x^k} + \mathcal{O}(x^{-K-1}) \Bigg],  \label{appB-expansion-inside}
\end{align}
where in this case \enquote{+} stands for $x > 0$ and \enquote{--} for $x < 0$.
We now plug the expansions \eqref{appB-expansion-outside} respectively \eqref{appB-expansion-inside} into the appropriate Equations \eqref{KFP-lattice-E} respectively \eqref{KFP-lattice-p}, keeping terms up to order $K = 2$. We find to lowest order in $\varepsilon$
\begin{align}
g^{\pm}_{1/2}(\alpha) = \tilde{g}^{\pm}_{1/2}, \qquad h^{\pm}_{1/2}(p) = - \frac{1}{\Omega} \frac{p}{1+p^2} + \tilde{h}^{\pm}_{1/2}\label{appB-inside-order1/2}.
\end{align}
Here $\tilde{g}$ and $\tilde{h}$ denote integration constants. The next order yields
\begin{align}
g^{\pm}_{1}(\alpha) &= \frac{(1+D)\sin(2\alpha)-D \cot(\alpha)}{2 D \Omega} - \frac{\alpha \beta}{\Omega} (1+ D\beta) + \tilde{g}^{\pm}_{1} \nonumber \\
h^{\pm}_{1}(p) &= \frac{D + (1-D)p^2}{\Omega^2 (1+p^2)^2} - \frac{p}{\Omega (1+p^2)} \tilde{h}^{\pm}_{1/2} + \tilde{h}^{\pm}_{1} \label{appB-inside-order1} .
\end{align}
Next, we want to match the two solutions inside and outside the strip across the boundary.
Replacing $x$ and $p$ by $\varepsilon$ and $\alpha$ we immediately find from the leading order term
\begin{align}
\gamma = \frac{\beta}{2}, \qquad M = \frac{N}{|2 \cos(\alpha)|^{\beta}} \simeq \frac{N}{2^{\beta}},
\end{align}
since the expansion Eq.~\eqref{appB-expansion-inside} is only valid for small $\alpha$.
For the sub-leading orders, we match all terms of the same order in $\varepsilon$ and $\alpha$.
In the upper half-plane, the matching corresponds to taking the limit $p \rightarrow +\infty$ for the solution inside the strip and again replacing $x$ and $p$ by $\varepsilon$ and $\alpha$,
\begin{align}
P^{\pm}_S(\sqrt{2 \varepsilon} \cos(\alpha), \sqrt{2 \varepsilon} \sin(\alpha)) \simeq N (2 \epsilon)^{\frac{\beta}{2}} |\cos(\alpha)|^{\beta} \Bigg[ 1 + \frac{\tilde{h}^{\pm}_{1/2}(p)}{\cos(\alpha)} (2 \varepsilon)^{-\frac{1}{2}} + \Big( \frac{\tilde{h}^{\pm}_{1}}{\cos^2(\alpha)} - \frac{1}{\Omega\sin(\alpha) \cos(\alpha)} \Big) (2\varepsilon)^{-1} + \mathcal{O}(\varepsilon^{-\frac{3}{2}}) \Bigg] . 
\end{align}
This should match the result from outside the strip for $\alpha \rightarrow 0, \pi$.
To order $\varepsilon^{-\frac{1}{2}}$, we then find
\begin{align}
\tilde{g}^{+}_{1/2} = \frac{\tilde{h}^{+}_{1/2}}{\sqrt{2}\cos(\alpha)} = \frac{\tilde{h}^{+}_{1/2}}{\sqrt{2}}, \quad &\alpha \rightarrow 0+ \nonumber \\
\tilde{g}^{+}_{1/2} = \frac{\tilde{h}^{-}_{1/2}}{\sqrt{2}\cos(\alpha)} = -\frac{\tilde{h}^{-}_{1/2}}{\sqrt{2}}, \quad &\alpha \rightarrow \pi- .
\end{align}
The terms of order $\varepsilon^{-1}$ give
\begin{align}
-\frac{\cot(\alpha)}{2\Omega} + \tilde{g}^{+}_{1} \simeq \frac{\tilde{h}^{+}_{1}}{2\cos^2(\alpha)} - \frac{1}{2\Omega\sin(\alpha)\cos(\alpha)}, \quad &\alpha \rightarrow 0+ \nonumber \\
-\frac{\cot(\alpha)}{2\Omega} - \frac{\pi}{\Omega} \beta (1+D\beta) + \tilde{g}^{+}_{1} \simeq \frac{\tilde{h}^{-}_{1}}{2\cos^2(\alpha)} - \frac{1}{2\Omega\sin(\alpha)\cos(\alpha)}, \quad &\alpha \rightarrow \pi-.
\end{align}
Since the cotangent tends to $+\infty$ for $\alpha \rightarrow 0+$ and to $-\infty$ for $\alpha \rightarrow \pi-$, the singular terms cancel and the last two conditions simplify to $\tilde{g}^{+}_{1} = \tilde{h}^{+}_{1}/2 = -\tilde{h}^{-}_{1}+  \pi \beta (1+D\beta)/\Omega$.
Similarly, we find in the lower half-plane, where we take the limit $p \rightarrow -\infty$ from inside the strip and the limits $\alpha \rightarrow 2\pi-$ respectively $\alpha \rightarrow \pi+$ from outside the strip, $\tilde{g}^{-}_{1/2} = \tilde{h}^{+}_{1/2}/\sqrt{2} = -\tilde{h}^{-}_{1/2}/\sqrt{2}$ and $\tilde{g}^{-}_{1} = \tilde{h}^{+}_{1}/2 = \tilde{h}^{-}_{1}/2 + \pi \beta (1+D\beta)/\Omega$.
From the lowest two orders in $\varepsilon$, we thus find
\begin{align}
\tilde{g}^{+}_{1/2} &= \tilde{g}^{-}_{1/2} = \tilde{h}^{+}_{1/2}/\sqrt{2} = -\tilde{h}^{-}_{1/2}/\sqrt{2} \equiv a_{1/2} \nonumber \\
\tilde{g}^{+}_{1} &= \tilde{g}^{-}_{1} = \tilde{h}^{+}_{1}/2 = \tilde{h}^{-}_{1}/2 + \frac{\pi}{\Omega} \beta (1+D\beta)  \equiv a_1 .
\end{align}
We now use the fact that the phase-space probability density has to be invariant under $(x,p) \rightarrow (-x,-p)$, or equivalently $\alpha \rightarrow \alpha + \pi$, we must have for $0 < \alpha < \pi$ outside the strip $P_P^{+}(\epsilon,\alpha) = P_P^{-}(\epsilon,\alpha+\pi)$.
To lowest order $\varepsilon^{-\frac{1}{2}}$ this condition is trivially fulfilled.
However, at order $\varepsilon^{-1}$, we find a condition on the (as yet unknown) exponent $\beta$,
\begin{align}
g^{-}_{1}(\alpha+\pi) - g^{+}_{1}(\alpha) = - \frac{\pi}{\Omega} \beta (1+D\beta) = 0.
\end{align}
Since the solution $\beta = 0$ is does not decay at large $\varepsilon$, we must have $\beta = -1/D$.
The two lowest orders do not provide any constraint on the value of $a_{1/2}$ and $a_1$.
To find the latter, we need to use the next order $\varepsilon^{-\frac{3}{2}}$, where we find outside the strip
\begin{align}
&g^{\pm}_{3/2}(\alpha) = \tilde{g}_{3/2}^{\pm}  \nonumber \\
& \; + a_{1/2}\frac{(2 + D)((2+3D)\sin(2\alpha) - 2 \alpha D) - 4 D \cot(\alpha)}{8 D \Omega} .
\end{align}
Inside the strip, we have for large $p$
\begin{align}
h^{\pm}_{3/2}(p) &\simeq - \frac{(2+D)a_{1/2}}{\sqrt{2} D} p^2 + \frac{3(1+D)}{D \Omega} p + \tilde{h}^{\pm}_{3/2} - \frac{ \pi (2+D)}{2 D \Omega}  + \frac{1 + D - 2 D a_1}{D \Omega} p^{-1} .
\end{align}
Since this coefficient is multiplied by $x^{-3}$, the first two terms are of order $\varepsilon^{-\frac{1}{2}} \alpha^2$ (note that we focus on the expansion around $\alpha = 0$ for simplicity, the case $\alpha = \pi$ can be examined in a similar manner) and have no correspondence outside the strip.
The next two terms are of order $\varepsilon^{-\frac{3}{2}}$ and should match the constant contribution from $g^{\pm}_{3/2}(\alpha)$.
Note that we get an extra contribution of order $\varepsilon^{-\frac{3}{2}}$ from $h^{\pm}_{1}(p)$, Eq.~\eqref{appB-inside-order1}.
Summing up, we have the following conditions
\begin{align}
\tilde{g}_{3/2}^{+} - \frac{a_{1/2} \cot(\alpha)}{2 \Omega} &\simeq \frac{1}{(\sqrt{2} \cos(\alpha))^3} \Big( \tilde{h}^{+}_{3/2} - \frac{ \pi (2+D)}{2 D \Omega} \Big) - \frac{a_{1/2}}{2 \Omega \cos^2(\alpha) \sin(\alpha)}, \qquad \alpha \rightarrow 0+ \nonumber \\
\tilde{g}_{3/2}^{+} - \frac{a_{1/2} \cot(\alpha)}{2 \Omega} - \frac{a_{1/2} \pi}{4 \Omega} &\simeq \frac{1}{(\sqrt{2} \cos(\alpha))^3} \Big( \tilde{h}^{-}_{3/2} - \frac{ \pi (2+D)}{2 D \Omega} \Big) + \frac{a_{1/2}}{2 \Omega \cos^2(\alpha) \sin(\alpha)}, \qquad \alpha \rightarrow \pi- \nonumber \\
\tilde{g}_{3/2}^{-} - \frac{a_{1/2} \cot(\alpha)}{2 \Omega} - \frac{a_{1/2} \pi}{4 \Omega} &\simeq \frac{1}{(\sqrt{2} \cos(\alpha))^3} \Big( \tilde{h}^{-}_{3/2} - \frac{ \pi (2+D)}{2 D \Omega} \Big) + \frac{a_{1/2}}{2 \Omega \cos^2(\alpha) \sin(\alpha)}, \qquad \alpha \rightarrow \pi+ \nonumber \\
\tilde{g}_{3/2}^{-} - \frac{a_{1/2} \cot(\alpha)}{2 \Omega} - \frac{a_{1/2} \pi}{2 \Omega} &\simeq \frac{1}{(\sqrt{2} \cos(\alpha))^3} \Big( \tilde{h}^{+}_{3/2} - \frac{ \pi (2+D)}{2 D \Omega} \Big) - \frac{a_{1/2}}{2 \Omega \cos^2(\alpha) \sin(\alpha)}, \qquad \alpha \rightarrow 2 \pi - .
\end{align}
The divergent terms once again cancel in the respective limits. The remaining system of equations,
\begin{align}
\tilde{g}_{3/2}^{+}  &\simeq \frac{1}{2 \sqrt{2}} \Big( \tilde{h}^{+}_{3/2} - \frac{ \pi (2+D)}{2 D \Omega} \Big), \qquad \alpha \rightarrow 0+ \nonumber \\
\tilde{g}_{3/2}^{+} - \frac{a_{1/2} \pi}{4 \Omega} &\simeq -\frac{1}{2 \sqrt{2}} \Big( \tilde{h}^{-}_{3/2} - \frac{ \pi (2+D)}{2 D \Omega} \Big), \qquad \alpha \rightarrow \pi- \nonumber \\
\tilde{g}_{3/2}^{-} - \frac{a_{1/2} \pi}{4 \Omega} &\simeq -\frac{1}{2 \sqrt{2}} \Big( \tilde{h}^{-}_{3/2} - \frac{ \pi (2+D)}{2 D \Omega} \Big), \qquad \alpha \rightarrow \pi+ \nonumber \\
\tilde{g}_{3/2}^{-} - \frac{a_{1/2} \pi}{2 \Omega} &\simeq \frac{1}{2 \sqrt{2}} \Big( \tilde{h}^{+}_{3/2} - \frac{ \pi (2+D)}{2 D \Omega} \Big), \qquad \alpha \rightarrow 2 \pi - ,
\end{align}
is only solvable for $a_{1/2} = -\sqrt{2}/D$.
Comparing this to the large-frequency expansion Eq.~\eqref{large-omega-large-e}, this precisely recovers the first correction to the leading order power-law.
The corresponding solution for the remaining coefficients reads
\begin{align}
\tilde{g}_{3/2}^{+} &\equiv a_{3/2} \nonumber \\
\tilde{g}_{3/2}^{-} &= a_{3/2} - \frac{\sqrt{2} \pi (2+D)}{4 D \Omega} \nonumber \\
\tilde{h}_{3/2}^{+} &= 2 \sqrt{2} a_{3/2} + \frac{\pi (2+D)}{2 D \Omega} \nonumber \\
\tilde{h}_{3/2}^{-} &= -2 \sqrt{2} a_{3/2} - \frac{\pi (2+D)}{2 D \Omega}.
\end{align}
A similar but lengthy argument for the order $\varepsilon^{-2}$ terms fixes the value of $a_1 = 1/D^2$, in agreement with Eq.~\eqref{large-omega-large-e}.

\section{Momentum-dependent diffusion coefficient - small $D$ \label{app:d1}}

In the presence of a the momentum-dependent diffusion coefficient $D_p = D_0 + D_1/(1+p^2/p_\text{c}^2)$, the rescaled Kramers-Fokker-Planck equation \eqref{KFP-lattice-D} reads
\begin{align}
\Bigg[ \Omega \bigg( - u \partial_z + z \partial_u \bigg) +  \partial_u \bigg(\frac{u}{1+D u^2} + \Big( 1 + \frac{\mathfrak{D}}{1 + D u^2} \Big) \partial_u \bigg) \Bigg] P_D(z,u) = 0, \label{KFP-lattice-D2}
\end{align}
where we defined $\mathfrak{D} = D_0/D_1$ (see Section \ref{sec:experimental}).
In the limit $D \rightarrow 0$ with $\mathfrak{D}$ fixed, the solution is a slightly modified Boltzmann-Gibbs form,
\begin{align}
P_\text{BG}(z,u) = \frac{1}{2 \pi (1+\mathfrak{D})} e^{-\frac{z^2+u^2}{2 (1 + \mathfrak{D})}} .
\end{align}
In complete analogy to Section \ref{sec:small-D}, we define an auxiliary function $g(z,u)$ via
\begin{align}
P_D(z,u) = P_\text{BG}(z,u) g(z,u),
\end{align}
which yields an equation for $g(z,u)$ similar to Eq.~\eqref{small-d-auxilary},
\begin{align}
\Big[\mathcal{L}_0 &+ D \mathcal{L}_1 + D^2 \mathcal{L}_2\Big] g(z,u) = 0 \label{appC-auxiliary} \\
\mathcal{L}_0 &= \kappa^2 \big[ \Omega(z \partial_{u} - u \partial_{z}) - u \partial_{u} + \kappa \partial_{u}^2 \big] \nonumber \\
\mathcal{L}_1 &= 2 \Omega \kappa^2 (z u^2 \partial_{u} - u^3 \partial_{z}) - 3 \kappa u^2 + u^4 - \big(2 \kappa + \kappa^2 \big) u^3 \partial_{u} + \big(\kappa^2 + \kappa^3 \big) u^2 \partial_{u}^2 \nonumber \\
& \qquad + 2 \big(\kappa^2 - \kappa^3 \big) u \partial_u \nonumber \\
\mathcal{L}_2 &= \Omega \kappa^2 (z u^4 \partial_{u} - u^5 \partial_{z}) - \kappa u^4 + u^6 - 2 \kappa u^5 \partial_{u} + \kappa^2 u^4 \partial_{u}^2 \nonumber ,
\end{align}
where we defined $\kappa = 1 + \mathfrak{D}$.
The structure of the individual operators is the same as for $\mathfrak{D} = 0$, so that we can still expect an expansion of the form
\begin{align}
g(z,u) = 1 + \sum_{n = 1}^{M} D^{n} \sum_{k=0}^{4 n} \sum_{l=0}^{4 n - k} a^{(n)}_{k,l} z^k u^l \label{appC-expansion},
\end{align}
for the solution.
However, since the individual operators are now proportional to powers of $\kappa=1+\mathfrak{D}$, the expansion will only work for $\mathfrak{D}$ of at most order $1$.
Plugging the expansion \eqref{appC-expansion} into Eq.~\eqref{appC-auxiliary} and solving for the coefficients, we find up to first order in $D$
\begin{align}
P^{(1)}_D(z,u) = \frac{e^{-\frac{z^2+u^2}{2}}}{2 \pi \kappa} \Bigg[ 1 + \frac{D}{4 (3+4\Omega^2)} \bigg[ \frac{3 u^4}{\kappa^2} + \frac{18 z^2}{\kappa} - 27 + \bigg( \frac{4 u^3 z}{\kappa^2} - \frac{12 u z}{\kappa} \bigg) \Omega + \bigg(\frac{3(u^2+z^2)^2}{\kappa^2} - 24 \bigg) \Omega^2 \bigg] \Bigg],
\end{align}
which for $\mathfrak{D} = 0$ ($\kappa = 1$) reduces to the previous result, Eq.~\eqref{small-d-1order}.
We can continue this expansion to higher orders, in particular, to second order in $D$, we find for the equipartition ratio Eq.~\eqref{equipart}
\begin{align}
\chi^{(2)} = 1 + \frac{6 \kappa}{3 + 4 \Omega^2} D^2 .
\end{align}
Since this expression increases linearly with $\kappa = 1 + \frac{D_1}{D_0}$, this shows that a nonzero $D_1$ enhances the deviations from equipartition.

\section{Momentum-dependent diffusion coefficient - large $\Omega$ \label{app:d1-omega}}

Just like for the small-$D$ expansion, we can also repeat the large $\Omega$ expansion of Section \ref{sec:large-freq} in the presence of a non-zero $D_1$.
Employing the notation of Appendix \ref{app:d1}, the equivalent to the Kramers-Fokker-Planck equation \eqref{KFP-lattice-E} including $D_1$ reads
\begin{align}
\Big[\Omega \partial_\alpha &+ \tilde{\mathcal{L}}_{\varepsilon,\alpha} \Big] P_P(\varepsilon,\alpha) = 0 \\
\text{with} \qquad \tilde{\mathcal{L}}_{\varepsilon,\alpha} &= \partial_\alpha \frac{\sin(\alpha)\cos(\alpha)}{1+2 \varepsilon \sin^2(\alpha)} + \partial_\varepsilon \frac{2 \varepsilon \sin^2(\alpha)}{1+2 \varepsilon \sin^2(\alpha)}  \nonumber \\
& + D \bigg[ 2 \partial_\alpha \sin(\alpha) \cos(\alpha) \partial_\varepsilon  + \frac{1}{2 \varepsilon} \partial_\alpha \cos^2(\alpha) \partial_\alpha + (\sin^2(\alpha) - \cos^2(\alpha)) \partial_\varepsilon + 2 \sin^2(\alpha) \partial_\varepsilon \varepsilon \partial_\varepsilon \bigg] \nonumber \\
& + D \mathfrak{D} \bigg[ \partial_\varepsilon \frac{2 \varepsilon \sin^2(\alpha)}{1 + 2 \varepsilon \sin^2(\alpha)} \partial_\varepsilon + \partial_\varepsilon \frac{\sin(\alpha) \cos(\alpha)}{1 + 2 \varepsilon \sin^2(\alpha)} \partial_\alpha + \partial_\alpha \frac{\sin(\alpha) \cos(\alpha)}{1 + 2 \varepsilon \sin^2(\alpha)} \partial_\varepsilon + \frac{1}{2 \varepsilon} \partial_\alpha \frac{2 \varepsilon \cos^2(\alpha)}{1 + 2 \varepsilon \sin^2(\alpha)} \partial_\alpha\bigg] . \nonumber
\end{align}
As discussed in Section \ref{sec:large-freq}, for $\Omega \gg 1$ we have to leading order $\partial_\alpha P_P(\varepsilon,\alpha) = 0$.
Then we can average the above equation over $\alpha$ to obtain an equation for the energy probability density, similar to Eq.~\eqref{energy-density-FP},
\begin{align}
\partial_\varepsilon \Bigg[ 1 - \frac{1}{\sqrt{1+2 \varepsilon}} + D \bigg( \varepsilon + \mathfrak{D} \Big(1 - \frac{1}{\sqrt{1+2 \varepsilon}} \Big) \bigg) \partial_\varepsilon \Bigg] P_\varepsilon(\varepsilon) = 0 .
\end{align}
The solution to this equation reads
\begin{align}
P_\varepsilon(\varepsilon) &= \frac{1}{Z_\varepsilon} \bigg(2 \mathfrak{D} + \Big(1 + 2 \varepsilon + \sqrt{1 + 2\varepsilon}\Big) \bigg)^{-\frac{1}{D}} \exp \Bigg[ \frac{2}{D} \frac{\text{artanh} \ \Big(\frac{1 + 2\sqrt{1+2\varepsilon}}{\sqrt{ 1 - 8 \mathfrak{D}}}\Big) - \text{artanh} \ \Big(\frac{3}{\sqrt{ 1 - 8 \mathfrak{D}}}\Big)}{\sqrt{1 - 8 \mathfrak{D}}} \Bigg]
\end{align}
for $\mathfrak{D} \neq 1/8$ and
\begin{align}
P_\varepsilon(\varepsilon) &= \frac{1}{Z_\varepsilon} \Big( 1 + 2 \sqrt{1+2\varepsilon} \Big)^{-\frac{2}{D}} \exp \bigg[ - \frac{2}{1 + 2 \sqrt{1+2\varepsilon}} \bigg]
\end{align}
for the special case $\mathfrak{D} = 1/8$.
In particular, we see that the asymptotic behavior $P_\varepsilon(\epsilon) \sim \varepsilon^{-\frac{1}{D}}$ is not changed by the introduction of $D_1$ respectively $\mathfrak{D}$.
Repeating the derivation of Section \ref{sec:detailed-balance} for the quantity $\phi$ follows the same lines as outlined there; we find that Eq.~\eqref{phi-result} is modified slightly due to the momentum-dependent diffusion coefficient,
\begin{align}
\phi &= \partial_p \frac{p}{1+p^2} + \partial_p \Big( D \Big(1 + \frac{\mathfrak{D}}{1+p^2} \Big) \partial_p \ln(P_0(x,p)) \Big) \label{phi-result-d1} ,
\end{align}
with $P_0(x,p) = P_\varepsilon((x^2+p^2)/2)/(2 \pi)$.
In particular, this expression now depends explicitly on $D$ and vanishes in the limit $D \rightarrow 0$ with $D_1 = \gamma p_c^2 D \mathfrak{D}$ finite,
\begin{align}
\phi \simeq - \frac{1 + p^4 - x^2 - \frac{1+x^2}{\sqrt{1+p^2+x^2}} + p^2 \Big( 2 - \frac{1}{\sqrt{1+x^2+p^2}} + x^2 \Big( 1 + \frac{1}{\sqrt{1+x^2+p^2}}\Big) \Big) }{2(1+p^2)^2} \frac{\gamma p_c^2}{D_1} D + \mathcal{O}(D^2) .
\end{align}

\end{widetext}

\bibliography{bib}

\begin{thebibliography}{42}%
\makeatletter
\providecommand \@ifxundefined [1]{%
 \@ifx{#1\undefined}
}%
\providecommand \@ifnum [1]{%
 \ifnum #1\expandafter \@firstoftwo
 \else \expandafter \@secondoftwo
 \fi
}%
\providecommand \@ifx [1]{%
 \ifx #1\expandafter \@firstoftwo
 \else \expandafter \@secondoftwo
 \fi
}%
\providecommand \natexlab [1]{#1}%
\providecommand \enquote  [1]{``#1''}%
\providecommand \bibnamefont  [1]{#1}%
\providecommand \bibfnamefont [1]{#1}%
\providecommand \citenamefont [1]{#1}%
\providecommand \href@noop [0]{\@secondoftwo}%
\providecommand \href [0]{\begingroup \@sanitize@url \@href}%
\providecommand \@href[1]{\@@startlink{#1}\@@href}%
\providecommand \@@href[1]{\endgroup#1\@@endlink}%
\providecommand \@sanitize@url [0]{\catcode `\\12\catcode `\$12\catcode
  `\&12\catcode `\#12\catcode `\^12\catcode `\_12\catcode `\%12\relax}%
\providecommand \@@startlink[1]{}%
\providecommand \@@endlink[0]{}%
\providecommand \url  [0]{\begingroup\@sanitize@url \@url }%
\providecommand \@url [1]{\endgroup\@href {#1}{\urlprefix }}%
\providecommand \urlprefix  [0]{URL }%
\providecommand \Eprint [0]{\href }%
\providecommand \doibase [0]{http://dx.doi.org/}%
\providecommand \selectlanguage [0]{\@gobble}%
\providecommand \bibinfo  [0]{\@secondoftwo}%
\providecommand \bibfield  [0]{\@secondoftwo}%
\providecommand \translation [1]{[#1]}%
\providecommand \BibitemOpen [0]{}%
\providecommand \bibitemStop [0]{}%
\providecommand \bibitemNoStop [0]{.\EOS\space}%
\providecommand \EOS [0]{\spacefactor3000\relax}%
\providecommand \BibitemShut  [1]{\csname bibitem#1\endcsname}%
\let\auto@bib@innerbib\@empty
%</preamble>
\bibitem [{\citenamefont {{L.D. Landau and E.M. Lifshitz}}(1980)}]{lan80}%
  \BibitemOpen
  \bibfield  {author} {\bibinfo {author} {\bibnamefont {{L.D. Landau and E.M.
  Lifshitz}}},\ }\href@noop {} {\emph {\bibinfo {title} {{Statistical
  Phyiscs}}}}\ (\bibinfo  {publisher} {Pergamon Press, Oxford},\ \bibinfo
  {year} {1980})\BibitemShut {NoStop}%
\bibitem [{\citenamefont {{C. Cohen-Tannoudji and W.D.
  Phillips}}(1990)}]{coh90}%
  \BibitemOpen
  \bibfield  {author} {\bibinfo {author} {\bibnamefont {{C. Cohen-Tannoudji and
  W.D. Phillips}}},\ }\bibfield  {title} {\enquote {\bibinfo {title} {{New
  mechanisms for laser cooling}},}\ }\href@noop {} {\bibfield  {journal}
  {\bibinfo  {journal} {Phys. Today}\ }\textbf {\bibinfo {volume} {43(10)}},\
  \bibinfo {pages} {33--40} (\bibinfo {year} {1990})}\BibitemShut {NoStop}%
\bibitem [{\citenamefont {{Y. Castin, J. Dalibard and C.
  Cohen-Tannoudji}}(1990)}]{cas90}%
  \BibitemOpen
  \bibfield  {author} {\bibinfo {author} {\bibnamefont {{Y. Castin, J. Dalibard
  and C. Cohen-Tannoudji}}},\ }\bibfield  {title} {\enquote {\bibinfo {title}
  {{The limits of Sisyphus cooling}},}\ }\href@noop {} {\bibfield  {journal}
  {\bibinfo  {journal} {Proceedings of the LIKE workshop}\ ,\ \bibinfo {pages}
  {5--24}} (\bibinfo {year} {1990})}\BibitemShut {NoStop}%
\bibitem [{\citenamefont {{F. Bardou, J.P. Bouchaud, O. Emile, A. Aspect and C.
  Cohen-Tannoudji}}(1994)}]{bar94}%
  \BibitemOpen
  \bibfield  {author} {\bibinfo {author} {\bibnamefont {{F. Bardou, J.P.
  Bouchaud, O. Emile, A. Aspect and C. Cohen-Tannoudji}}},\ }\bibfield  {title}
  {\enquote {\bibinfo {title} {{Subrecoil Laser Cooling and L{\'e}vy
  Flights}},}\ }\href@noop {} {\bibfield  {journal} {\bibinfo  {journal} {Phys.
  Rev. Lett.}\ }\textbf {\bibinfo {volume} {72}},\ \bibinfo {pages} {203}
  (\bibinfo {year} {1994})}\BibitemShut {NoStop}%
\bibitem [{\citenamefont {{H.J. Metcalf and P. van der
  Straten}}(1999)}]{met99}%
  \BibitemOpen
  \bibfield  {author} {\bibinfo {author} {\bibnamefont {{H.J. Metcalf and P.
  van der Straten}}},\ }\href@noop {} {\emph {\bibinfo {title} {{Laser Cooling
  and Trapping}}}}\ (\bibinfo  {publisher} {Springer, New York},\ \bibinfo
  {year} {1999})\BibitemShut {NoStop}%
\bibitem [{\citenamefont {{J. Dalibard and C. Cohen-Tannoudji}}(1989)}]{dal89}%
  \BibitemOpen
  \bibfield  {author} {\bibinfo {author} {\bibnamefont {{J. Dalibard and C.
  Cohen-Tannoudji}}},\ }\bibfield  {title} {\enquote {\bibinfo {title} {{Laser
  cooling below the Doppler limit by polarization gradients: simple theoretical
  models}},}\ }\href@noop {} {\bibfield  {journal} {\bibinfo  {journal} {J.
  Opt. Soc. Am. B}\ }\textbf {\bibinfo {volume} {6}},\ \bibinfo {pages}
  {2023--2045} (\bibinfo {year} {1989})}\BibitemShut {NoStop}%
\bibitem [{\citenamefont {{S. Marksteiner, K. Ellinger and P.
  Zoller}}(1996)}]{mar96}%
  \BibitemOpen
  \bibfield  {author} {\bibinfo {author} {\bibnamefont {{S. Marksteiner, K.
  Ellinger and P. Zoller}}},\ }\bibfield  {title} {\enquote {\bibinfo {title}
  {{Anomalous diffusion and L\'evy walks in optical lattices}},}\ }\href@noop
  {} {\bibfield  {journal} {\bibinfo  {journal} {Phys. Rev. A}\ }\textbf
  {\bibinfo {volume} {53}},\ \bibinfo {pages} {3409--3430} (\bibinfo {year}
  {1996})}\BibitemShut {NoStop}%
\bibitem [{\citenamefont {{E. Lutz}}(2003)}]{lut03}%
  \BibitemOpen
  \bibfield  {author} {\bibinfo {author} {\bibnamefont {{E. Lutz}}},\
  }\bibfield  {title} {\enquote {\bibinfo {title} {{Anomalous diffusion and
  Tsallis statistics in an optical lattice}},}\ }\href@noop {} {\bibfield
  {journal} {\bibinfo  {journal} {Phys. Rev. A}\ }\textbf {\bibinfo {volume}
  {67}},\ \bibinfo {pages} {051402} (\bibinfo {year} {2003})}\BibitemShut
  {NoStop}%
\bibitem [{\citenamefont {{P. Douglas, S. Bergamini and F.
  Renzoni}}(2006)}]{dou06}%
  \BibitemOpen
  \bibfield  {author} {\bibinfo {author} {\bibnamefont {{P. Douglas, S.
  Bergamini and F. Renzoni}}},\ }\bibfield  {title} {\enquote {\bibinfo {title}
  {{Tunable Tsallis Distributions in Dissipative Optical Lattices}},}\
  }\href@noop {} {\bibfield  {journal} {\bibinfo  {journal} {Phys. Rev. Lett.}\
  }\textbf {\bibinfo {volume} {96}},\ \bibinfo {pages} {110601} (\bibinfo
  {year} {2006})}\BibitemShut {NoStop}%
\bibitem [{\citenamefont {{H. Katori, S. Schlipf and H.
  Walther}}(1997)}]{kat97}%
  \BibitemOpen
  \bibfield  {author} {\bibinfo {author} {\bibnamefont {{H. Katori, S. Schlipf
  and H. Walther}}},\ }\bibfield  {title} {\enquote {\bibinfo {title}
  {{Anomalous Dynamics of a Single Ion in an Optical Lattice}},}\ }\href@noop
  {} {\bibfield  {journal} {\bibinfo  {journal} {Phys. Rev. Lett.}\ }\textbf
  {\bibinfo {volume} {79}},\ \bibinfo {pages} {2221--2224} (\bibinfo {year}
  {1997})}\BibitemShut {NoStop}%
\bibitem [{\citenamefont {{Y. Sagi, M. Brook, I. Almog and N.
  Davidson}}(2012)}]{sag12}%
  \BibitemOpen
  \bibfield  {author} {\bibinfo {author} {\bibnamefont {{Y. Sagi, M. Brook, I.
  Almog and N. Davidson}}},\ }\bibfield  {title} {\enquote {\bibinfo {title}
  {{Observation of Anomalous Diffusion and Fractional Self-Similarity in One
  Dimension}},}\ }\href@noop {} {\bibfield  {journal} {\bibinfo  {journal}
  {Phys. Rev. Lett.}\ }\textbf {\bibinfo {volume} {108}},\ \bibinfo {pages}
  {093002} (\bibinfo {year} {2012})}\BibitemShut {NoStop}%
\bibitem [{\citenamefont {{E. Lutz and F. Renzoni}}(2013)}]{lut13}%
  \BibitemOpen
  \bibfield  {author} {\bibinfo {author} {\bibnamefont {{E. Lutz and F.
  Renzoni}}},\ }\bibfield  {title} {\enquote {\bibinfo {title} {{Beyond
  Boltzmann-Gibbs statistical mechanics in optical lattices}},}\ }\href@noop {}
  {\bibfield  {journal} {\bibinfo  {journal} {Nature Phys.}\ }\textbf {\bibinfo
  {volume} {9}},\ \bibinfo {pages} {615} (\bibinfo {year} {2013})}\BibitemShut
  {NoStop}%
\bibitem [{\citenamefont {{A. Dechant, D.A. Kessler and E.
  Barkai}}(2015)}]{dec15}%
  \BibitemOpen
  \bibfield  {author} {\bibinfo {author} {\bibnamefont {{A. Dechant, D.A.
  Kessler and E. Barkai}}},\ }\bibfield  {title} {\enquote {\bibinfo {title}
  {{Deviations from Boltzmann-Gibbs equilibrium in confined optical
  lattices}},}\ }\href@noop {} {\bibfield  {journal} {\bibinfo  {journal}
  {Phys. Rev. Lett.}\ }\textbf {\bibinfo {volume} {115}},\ \bibinfo {pages}
  {173006} (\bibinfo {year} {2015})}\BibitemShut {NoStop}%
\bibitem [{\citenamefont {{W.T. Coffey, Yu.P. Kalmykov and J.T.
  Waldron}}(2004)}]{cof04}%
  \BibitemOpen
  \bibfield  {author} {\bibinfo {author} {\bibnamefont {{W.T. Coffey, Yu.P.
  Kalmykov and J.T. Waldron}}},\ }\href@noop {} {\emph {\bibinfo {title} {{The
  Langevin Equation}}}}\ (\bibinfo  {publisher} {World Scientific, Singapore},\
  \bibinfo {year} {2004})\BibitemShut {NoStop}%
\bibitem [{\citenamefont {{H. Risken}}(1986)}]{ris86}%
  \BibitemOpen
  \bibfield  {author} {\bibinfo {author} {\bibnamefont {{H. Risken}}},\
  }\href@noop {} {\emph {\bibinfo {title} {{The Fokker-Planck Equation}}}}\
  (\bibinfo  {publisher} {Springer, Berlin},\ \bibinfo {year}
  {1986})\BibitemShut {NoStop}%
\bibitem [{\citenamefont {{R.C. Tolman}}(1918)}]{tol18}%
  \BibitemOpen
  \bibfield  {author} {\bibinfo {author} {\bibnamefont {{R.C. Tolman}}},\
  }\bibfield  {title} {\enquote {\bibinfo {title} {{A General Theory of Energy
  Partition with Applications to Quantum Theory}},}\ }\href@noop {} {\bibfield
  {journal} {\bibinfo  {journal} {Phys. Rev.}\ }\textbf {\bibinfo {volume}
  {11}},\ \bibinfo {pages} {261} (\bibinfo {year} {1918})}\BibitemShut
  {NoStop}%
\bibitem [{\citenamefont {{K. Huang}}(1987)}]{hua87}%
  \BibitemOpen
  \bibfield  {author} {\bibinfo {author} {\bibnamefont {{K. Huang}}},\
  }\href@noop {} {\emph {\bibinfo {title} {{Statistical Mechanics}}}}\
  (\bibinfo  {publisher} {Wiley, New York},\ \bibinfo {year}
  {1987})\BibitemShut {NoStop}%
\bibitem [{\citenamefont {{C.W. Gardiner}}(19996)}]{gar96}%
  \BibitemOpen
  \bibfield  {author} {\bibinfo {author} {\bibnamefont {{C.W. Gardiner}}},\
  }\href@noop {} {\emph {\bibinfo {title} {{Handbook of Stochastic Methods}}}}\
  (\bibinfo  {publisher} {Springer, Berlin},\ \bibinfo {year}
  {19996})\BibitemShut {NoStop}%
\bibitem [{\citenamefont {{P.C. Holz, A. Dechant and E. Lutz}}(2015)}]{hol15}%
  \BibitemOpen
  \bibfield  {author} {\bibinfo {author} {\bibnamefont {{P.C. Holz, A. Dechant
  and E. Lutz}}},\ }\bibfield  {title} {\enquote {\bibinfo {title} {{Infinite
  density for cold atoms in shallow optical lattices}},}\ }\href@noop {}
  {\bibfield  {journal} {\bibinfo  {journal} {Europhys. Lett.}\ }\textbf
  {\bibinfo {volume} {109}},\ \bibinfo {pages} {23001} (\bibinfo {year}
  {2015})}\BibitemShut {NoStop}%
\bibitem [{\citenamefont {{D.A. Kessler and E. Barkai}}(2010)}]{kes10}%
  \BibitemOpen
  \bibfield  {author} {\bibinfo {author} {\bibnamefont {{D.A. Kessler and E.
  Barkai}}},\ }\bibfield  {title} {\enquote {\bibinfo {title} {{Infinite
  Covariant Density for Diffusion in Logarithmic Potentials and Optical
  Lattices}},}\ }\href@noop {} {\bibfield  {journal} {\bibinfo  {journal}
  {Phys. Rev. Lett.}\ }\textbf {\bibinfo {volume} {105}},\ \bibinfo {pages}
  {120602} (\bibinfo {year} {2010})}\BibitemShut {NoStop}%
\bibitem [{\citenamefont {{D.A. Kessler and E. Barkai}}(2012)}]{kes12}%
  \BibitemOpen
  \bibfield  {author} {\bibinfo {author} {\bibnamefont {{D.A. Kessler and E.
  Barkai}}},\ }\bibfield  {title} {\enquote {\bibinfo {title} {{Theory of
  Fractional L\'evy Kinetics for Cold Atoms Diffusing in Optical Lattices}},}\
  }\href@noop {} {\bibfield  {journal} {\bibinfo  {journal} {Phys. Rev. Lett.}\
  }\textbf {\bibinfo {volume} {108}},\ \bibinfo {pages} {230602} (\bibinfo
  {year} {2012})}\BibitemShut {NoStop}%
\bibitem [{\citenamefont {{A. Dechant and E. Lutz}}(2012)}]{dec12}%
  \BibitemOpen
  \bibfield  {author} {\bibinfo {author} {\bibnamefont {{A. Dechant and E.
  Lutz}}},\ }\bibfield  {title} {\enquote {\bibinfo {title} {{Anomalous spatial
  diffusion and multifractality in optical lattices}},}\ }\href@noop {}
  {\bibfield  {journal} {\bibinfo  {journal} {Phys. Rev. Lett.}\ }\textbf
  {\bibinfo {volume} {108}},\ \bibinfo {pages} {230601} (\bibinfo {year}
  {2012})}\BibitemShut {NoStop}%
\bibitem [{\citenamefont {{C. Tsallis}}(1988)}]{tsa88}%
  \BibitemOpen
  \bibfield  {author} {\bibinfo {author} {\bibnamefont {{C. Tsallis}}},\
  }\bibfield  {title} {\enquote {\bibinfo {title} {{Possible generalization of
  Boltzmann-Gibbs statistics}},}\ }\href@noop {} {\bibfield  {journal}
  {\bibinfo  {journal} {J. Stat. Phys.}\ }\textbf {\bibinfo {volume} {52}},\
  \bibinfo {pages} {479} (\bibinfo {year} {1988})}\BibitemShut {NoStop}%
\bibitem [{\citenamefont {{C. Tsallis, S.V.F. Levy, A.M.C. Souza and R.
  Maynard}}(1995)}]{tsa95}%
  \BibitemOpen
  \bibfield  {author} {\bibinfo {author} {\bibnamefont {{C. Tsallis, S.V.F.
  Levy, A.M.C. Souza and R. Maynard}}},\ }\bibfield  {title} {\enquote
  {\bibinfo {title} {{Statistical-Mechanical Foundation of the Ubiquity of
  L{\'e}vy Distributions in Nature}},}\ }\href@noop {} {\bibfield  {journal}
  {\bibinfo  {journal} {Phys. Rev. Lett.}\ }\textbf {\bibinfo {volume} {77}},\
  \bibinfo {pages} {5442} (\bibinfo {year} {1995})}\BibitemShut {NoStop}%
\bibitem [{\citenamefont {{O. Hirschberg, D. Mukamel and G.M.
  Sch{\"u}tz}}(2011)}]{hir11}%
  \BibitemOpen
  \bibfield  {author} {\bibinfo {author} {\bibnamefont {{O. Hirschberg, D.
  Mukamel and G.M. Sch{\"u}tz}}},\ }\bibfield  {title} {\enquote {\bibinfo
  {title} {{Approach to equilibrium of diffusion in a logarithmic
  potential}},}\ }\href@noop {} {\bibfield  {journal} {\bibinfo  {journal}
  {Phys. Rev. E}\ }\textbf {\bibinfo {volume} {84}},\ \bibinfo {pages} {041111}
  (\bibinfo {year} {2011})}\BibitemShut {NoStop}%
\bibitem [{\citenamefont {{A. Dechant, E. Lutz, D.A. Kessler and E.
  Barkai}}(2011)}]{dec11}%
  \BibitemOpen
  \bibfield  {author} {\bibinfo {author} {\bibnamefont {{A. Dechant, E. Lutz,
  D.A. Kessler and E. Barkai}}},\ }\bibfield  {title} {\enquote {\bibinfo
  {title} {{Fluctuations of time averages for Langevin dynamics in a binding
  force field}},}\ }\href@noop {} {\bibfield  {journal} {\bibinfo  {journal}
  {Phys. Rev. Lett.}\ }\textbf {\bibinfo {volume} {107}},\ \bibinfo {pages}
  {240603} (\bibinfo {year} {2011})}\BibitemShut {NoStop}%
\bibitem [{\citenamefont {{O. Hirschberg, D. Mukamel and G.M.
  Sch{\"u}tz}}(2012)}]{hir12}%
  \BibitemOpen
  \bibfield  {author} {\bibinfo {author} {\bibnamefont {{O. Hirschberg, D.
  Mukamel and G.M. Sch{\"u}tz}}},\ }\bibfield  {title} {\enquote {\bibinfo
  {title} {{Diffusion in a logarithmic potential: scaling and selection in the
  approach to equilibrium}},}\ }\href@noop {} {\bibfield  {journal} {\bibinfo
  {journal} {J. Stat. Mech.}\ }\textbf {\bibinfo {volume} {2}},\ \bibinfo
  {pages} {P02001} (\bibinfo {year} {2012})}\BibitemShut {NoStop}%
\bibitem [{\citenamefont {{G.S. Manning}}(1969)}]{man69}%
  \BibitemOpen
  \bibfield  {author} {\bibinfo {author} {\bibnamefont {{G.S. Manning}}},\
  }\bibfield  {title} {\enquote {\bibinfo {title} {{Limiting Laws and
  Counterion Condensation in Polyelectrolyte Solutions I. Colligative
  Properties}},}\ }\href@noop {} {\bibfield  {journal} {\bibinfo  {journal} {J.
  Chem. Phys.}\ }\textbf {\bibinfo {volume} {51}},\ \bibinfo {pages} {924}
  (\bibinfo {year} {1969})}\BibitemShut {NoStop}%
\bibitem [{\citenamefont {{A.J. Bray}}(2000)}]{bra00}%
  \BibitemOpen
  \bibfield  {author} {\bibinfo {author} {\bibnamefont {{A.J. Bray}}},\
  }\bibfield  {title} {\enquote {\bibinfo {title} {{Random walks in logarithmic
  and power-law potentials, nonuniversal persistence and vortex dynamics in the
  two-dimensional XY model}},}\ }\href@noop {} {\bibfield  {journal} {\bibinfo
  {journal} {Phys. Rev. E}\ }\textbf {\bibinfo {volume} {62}},\ \bibinfo
  {pages} {103--112} (\bibinfo {year} {2000})}\BibitemShut {NoStop}%
\bibitem [{\citenamefont {{H.C. Fogedby and R. Metzler}}(2007)}]{fog07}%
  \BibitemOpen
  \bibfield  {author} {\bibinfo {author} {\bibnamefont {{H.C. Fogedby and R.
  Metzler}}},\ }\bibfield  {title} {\enquote {\bibinfo {title} {{DNA Bubble
  Dynamics as a Quantum Coulomb Problem}},}\ }\href@noop {} {\bibfield
  {journal} {\bibinfo  {journal} {Phys. Rev. Lett.}\ }\textbf {\bibinfo
  {volume} {98}},\ \bibinfo {pages} {070601} (\bibinfo {year}
  {2007})}\BibitemShut {NoStop}%
\bibitem [{\citenamefont {{R.L. Stratonovich}}(1963)}]{str63}%
  \BibitemOpen
  \bibfield  {author} {\bibinfo {author} {\bibnamefont {{R.L. Stratonovich}}},\
  }\href@noop {} {\emph {\bibinfo {title} {{Topics in the Theory of Random
  Noise}}}}\ (\bibinfo  {publisher} {Gordon and Breach, New York},\ \bibinfo
  {year} {1963})\BibitemShut {NoStop}%
\bibitem [{\citenamefont {{R. Graham and H.
  Haken}}(1971{\natexlab{a}})}]{gra71}%
  \BibitemOpen
  \bibfield  {author} {\bibinfo {author} {\bibnamefont {{R. Graham and H.
  Haken}}},\ }\bibfield  {title} {\enquote {\bibinfo {title} {{Generalized
  thermodynamic potential for Markoff systems in detailed balance and far from
  thermal equilibrium}},}\ }\href@noop {} {\bibfield  {journal} {\bibinfo
  {journal} {Z. Physik}\ }\textbf {\bibinfo {volume} {243}},\ \bibinfo {pages}
  {289} (\bibinfo {year} {1971}{\natexlab{a}})}\BibitemShut {NoStop}%
\bibitem [{\citenamefont {{R. Graham and H.
  Haken}}(1971{\natexlab{b}})}]{gra71b}%
  \BibitemOpen
  \bibfield  {author} {\bibinfo {author} {\bibnamefont {{R. Graham and H.
  Haken}}},\ }\bibfield  {title} {\enquote {\bibinfo {title} {{Fluctuations and
  Stability of Stationary Non-Equilibrium Systems in Detailed Balance}},}\
  }\href@noop {} {\bibfield  {journal} {\bibinfo  {journal} {Z. Physik}\
  }\textbf {\bibinfo {volume} {245}},\ \bibinfo {pages} {141} (\bibinfo {year}
  {1971}{\natexlab{b}})}\BibitemShut {NoStop}%
\bibitem [{\citenamefont {{T. Chou, K. Mallick and R.K.P. Zia}}(2011)}]{cho11}%
  \BibitemOpen
  \bibfield  {author} {\bibinfo {author} {\bibnamefont {{T. Chou, K. Mallick
  and R.K.P. Zia}}},\ }\bibfield  {title} {\enquote {\bibinfo {title}
  {{Non-equilibrium statistical mechanics: from a paradigmatic model to
  biological transport}},}\ }\href@noop {} {\bibfield  {journal} {\bibinfo
  {journal} {Rep. Prog. Phys.}\ }\textbf {\bibinfo {volume} {74}},\ \bibinfo
  {pages} {116601} (\bibinfo {year} {2011})}\BibitemShut {NoStop}%
\bibitem [{\citenamefont {{M. Esposito and C. Van den Broeck}}(2010)}]{esp10}%
  \BibitemOpen
  \bibfield  {author} {\bibinfo {author} {\bibnamefont {{M. Esposito and C. Van
  den Broeck}}},\ }\bibfield  {title} {\enquote {\bibinfo {title} {{Three
  Detailed Fluctuation Theorems}},}\ }\href@noop {} {\bibfield  {journal}
  {\bibinfo  {journal} {Phys. Rev. Lett.}\ }\textbf {\bibinfo {volume} {104}},\
  \bibinfo {pages} {090601} (\bibinfo {year} {2010})}\BibitemShut {NoStop}%
\bibitem [{\citenamefont {{T. Tom{\'e} and M.J. de Oliveira}}(2010)}]{tom10}%
  \BibitemOpen
  \bibfield  {author} {\bibinfo {author} {\bibnamefont {{T. Tom{\'e} and M.J.
  de Oliveira}}},\ }\bibfield  {title} {\enquote {\bibinfo {title} {{Entropy
  production in irreversible systems described by a Fokker-Planck equation}},}\
  }\href@noop {} {\bibfield  {journal} {\bibinfo  {journal} {Phys. Rev. E}\
  }\textbf {\bibinfo {volume} {82}},\ \bibinfo {pages} {021120} (\bibinfo
  {year} {2010})}\BibitemShut {NoStop}%
\bibitem [{\citenamefont {{C. Kwon, J. Yeo, H.K. Lee and H.
  Park}}(2016)}]{kwo16}%
  \BibitemOpen
  \bibfield  {author} {\bibinfo {author} {\bibnamefont {{C. Kwon, J. Yeo, H.K.
  Lee and H. Park}}},\ }\bibfield  {title} {\enquote {\bibinfo {title}
  {{Unconventional entropy production in the presence of momentum-dependent
  forces}},}\ }\href@noop {} {\bibfield  {journal} {\bibinfo  {journal} {J.
  Korean Phys. Soc.}\ }\textbf {\bibinfo {volume} {68}},\ \bibinfo {pages}
  {633} (\bibinfo {year} {2016})}\BibitemShut {NoStop}%
\bibitem [{\citenamefont {{D.A. Steck}}(2003)}]{ste03}%
  \BibitemOpen
  \bibfield  {author} {\bibinfo {author} {\bibnamefont {{D.A. Steck}}},\
  }\bibfield  {title} {\enquote {\bibinfo {title} {Cesium d line data},}\
  }\href@noop {} {\bibfield  {journal} {\bibinfo  {journal} {Los Alamos
  National Laboratory (unpublished)}\ }\textbf {\bibinfo {volume} {124}}
  (\bibinfo {year} {2003})}\BibitemShut {NoStop}%
\bibitem [{\citenamefont {{L. Perotti, V. Alekeseev and H.
  Walther}}(2000)}]{per00}%
  \BibitemOpen
  \bibfield  {author} {\bibinfo {author} {\bibnamefont {{L. Perotti, V.
  Alekeseev and H. Walther}}},\ }\bibfield  {title} {\enquote {\bibinfo {title}
  {{Transport of a single ion in an optical lattice: spatial diffusion and
  energy}},}\ }\href@noop {} {\bibfield  {journal} {\bibinfo  {journal} {Opt.
  Comm.}\ }\textbf {\bibinfo {volume} {183}},\ \bibinfo {pages} {73--94}
  (\bibinfo {year} {2000})}\BibitemShut {NoStop}%
\bibitem [{Note1()}]{Note1}%
  \BibitemOpen
  \bibinfo {note} {The precise relation between $D$ and $U_0$ depends on the
  details of the atomic transition and the parametrization, however this only
  causes a slight change in the proportionality constant \cite
  {cas90,mar96}}\BibitemShut {NoStop}%
\bibitem [{\citenamefont {{R. Grimm, M. Weidem{\"u}ller and Y.B.
  Ovchinnikov}}(2000)}]{gri00}%
  \BibitemOpen
  \bibfield  {author} {\bibinfo {author} {\bibnamefont {{R. Grimm, M.
  Weidem{\"u}ller and Y.B. Ovchinnikov}}},\ }\bibfield  {title} {\enquote
  {\bibinfo {title} {{Optical Dipole Traps for Neutral Atoms}},}\ }\href@noop
  {} {\bibfield  {journal} {\bibinfo  {journal} {Adv. At. Mol. Opt. Phys.}\
  }\textbf {\bibinfo {volume} {42}},\ \bibinfo {pages} {95} (\bibinfo {year}
  {2000})}\BibitemShut {NoStop}%
\bibitem [{\citenamefont {{T.W. Hodapp, C. Gerz, C. Furtlehner, C.I. Westbrook,
  W.D. Phillips, and J. Dalibard}}(1995)}]{hod95}%
  \BibitemOpen
  \bibfield  {author} {\bibinfo {author} {\bibnamefont {{T.W. Hodapp, C. Gerz,
  C. Furtlehner, C.I. Westbrook, W.D. Phillips, and J. Dalibard}}},\ }\bibfield
   {title} {\enquote {\bibinfo {title} {{Three-dimensional spatial diffusion in
  optical molasses}},}\ }\href@noop {} {\bibfield  {journal} {\bibinfo
  {journal} {Appl. Phys. B: Lasers Opt. B.}\ }\textbf {\bibinfo {volume}
  {60}},\ \bibinfo {pages} {135--143} (\bibinfo {year} {1995})}\BibitemShut
  {NoStop}%
\end{thebibliography}%

\end{document}